\documentclass[12pt]{article}
\usepackage{amssymb}
\usepackage{color,graphicx}
\usepackage{amsmath}
\usepackage[v1compatible]{axodraw2}

\newcommand{\ep}{\varepsilon}
\newcommand{\bea}{\begin{eqnarray}}
\newcommand{\eea}{\end{eqnarray}}
\newcommand{\be}{\begin{equation}}
\newcommand{\ee}{\end{equation}}

\begin{document}

\begin{center}
  {\Large {\bf About calculation of massless and massive Feynman integrals
}}
\\ \vspace*{5mm} A.~V.~Kotikov
\end{center}

\begin{center}
Bogoliubov Laboratory of Theoretical Physics \\
Joint Institute for Nuclear Research\\
141980 Dubna, Russia
\end{center}

\begin{abstract}
  We report
  some results of calculations of massless and massive Feynman integrals
  particularly focusing on difference equations for coefficients of for their series expansions

\end{abstract}


\section{Introduction}

Calculation of Feynman integrals gives basic information about both experimentally investigated
processes and the characteristics of physical models.
The calculations of the matrix elements of the processes under study depend on the masses of
the particles involved in the interaction and, strictly speaking, require the calculation of
Feynman integrals including those with massive propagators. Depending on the kinematics of
the studied processes, the values of some masses can be neglected.

The study of the characteristics of physical models (for example, critical parameters, anomalous
dimensions of particles and operators) usually requires calculations of massless Feynman
integrals, which have a much simpler structure. They allow  one to obtain results for these
characteristics in high orders of perturbation theory.

Note that when calculating Feynman integrals, it is advisable to use analytical methods
whenever possible. The fact is that approximate methods for calculating Feynman integrals
do not always have sufficiently high accuracy. Moreover, the calculation of Feynman
integrals is ambiguous for numerical calculations both due to the singular nature of the
integrals themselves and (especially for gauge theories) with strong mutual contractions
between the contributions of different diagrams or even between parts of the same diagram.

Moreover, in many important situations it is necessary to know exactly the exact results.
For example, in the framework of renormalization group calculations in theories with high
internal symmetry, it is important to know \cite{Peterman:1978tb} about the vanishing of
$\beta$-functions in the order under consideration.
 
We note that when using dimensional regularization \cite{tHooft:1972tcz}, i.e. for an arbitrary dimension
of space, once found diagrams for any field model (or process) can be applied to other models
(and processes), since the main ones are scalar diagrams.
Consequently, the complexity of analytical calculations of Feynman integrals pays off for their
universality in application to various quantum-field models.

We also note the fact that the calculation of complicated diagrams may have some independent
interest. So the use of non-trivial identities such as the “uniqueness” relation
\cite{DEramo:1971hnd,Vasiliev:1981dg} can give information (see \cite{Kazakov:1984km,Kazakov:1983pk})
about the properties of some integrals and series that are not yet in the reference
literature. For example, calculations of the same Feynman integral carried out in Refs.
\cite{Kazakov:1984km} and \cite{Kotikov:1995cw} using
different calculation methods made it possible to find a previously unknown relation between
hypergeometric functions with arguments $1$ and $-1$. This relation has been proven only just
recently \cite{Kotikov:2018uat}.

Recently, many powerful original methods have appeared for calculating Feynman integrals (see, a
recent review in \cite{Kotikov:2018wxe}), which are often inferior in width to standard ones, i.e.
the $\alpha$-presentation and technique of Feynman parameters  (see, for example,
\cite{Peterman:1978tb,Ryder}), however, can significantly advance in the number of loops for a limited range of
quantities (or processes).



In this paper, we consider massless diagrams that contribute to the coefficient functions and anomalous dimensions
for the process of deep inelastic scattering (DIS) of leptons by hadrons.
Then we will show the main steps in calculating diagrams with massive propagators.
We also show that the moments of structure functions (SFs) and expansion coefficients (in inverse mass) have
many common features. The same or similar complex sums arise there and there, which can be conveniently
calculated using recursive relationships between them.

Such similarity is not strange.
First of all, the moments of the DIS structure functions
are not calculated directly.
The connection between deep inelastic scattering and elastic forward scattering is used, which is realized
through the optical theorem.
The forward elastic scattering amplitudes are determined by (scalar) diagrams depending on two momenta, $q$ and $p$
with the condition $p^2=0$.
Such diagrams can be expanded in series with respect to $(pq)/Q^2$ and, in particular, the structure of the
coefficients of these series is similar to the corresponding structure of the expansion coefficients for the
inverse mass in massive diagrams (here we consider the case of only one mass). This similarity of the structures
of the coefficients is not surprising, a similar similarity of the results of the calculation of these types of
diagrams was discussed long ago in Ref. \cite{Davydychev:1995mq}.

In Sections 2-4 we
demonstrate
the development of the multiloop calculation methods,
as applied to diagrams containing an arbitrary number of derivatives (or momenta) on lines.
Such diagrams arise when using the “gluing” method \cite{Chetyrkin:1982zq} and the method
"projectors" \cite{Gorishnii:1983su,Tkachov:1983st} for calculating the SF moments
of deep inelastic scattering of leptons at hadrons in QCD.

Contrary to original papers \cite{Kazakov:1986mu,Kotikov:1987mw} all calculations are carried out
in $p$-space. We note that
original results have been done in $x$-space for so-called dual diagrams (see, for example,
\cite{Kazakov:1986mu,Kotikov:1987mw,Kotikov:1995cw}). A dual diagram is obtained
from the initial one by replacement of all $p$ by $x$ with the rules of correspondence between the
graph and the integral, as in a $x$-space. The transition to the dual diagram is indicated by $\stackrel{d}{=}$.

Of course, the results of the integration
of the diagrams do not changed during the procedure.  However
the graphic representations 
are different. Shortly speaking, all loops (triangles, $n$-leg one-loop internal graphs)
should be replaced by the corresponding chains (three-leg vertices, $n$-leg vertices).
With the usage of the dual technique, the evaluation of the $\alpha_s$-corrections to the longitudinal
DIS structure function
has been done in \cite{Kazakov:1986mu,KK92}. All the calculations were done for the massless diagrams.
The extension of such calculations to the massive case were done in  \cite{Kotikov:1990zs}.
Some recent evaluations of the massive dual Feynman integrals can be found also in \cite{Henn:2014yza}.

In Section 3 and 4 we present several examples of calculations of massless diagrams. 
An introductions to the basic steps of the methods of ``gluing'' and ``projectors'' is given in Appendix B.
The calculation of massive diagrams is given in Sections 5 and 6. Additional rules are given for their efficient
calculation, examples of two- and three-point diagrams are considered. The recurrence relations for the coefficients
of decomposition in inverse mass are considered.
A short review of modern computing technology is given.

To calculate the most complex parts of massless and massive diagrams, recurrence relations for the
coefficients of their decomposition are used.
\footnote{Currently, there are also popular recursive relationships (see \cite{Tarasov:1996br,Lee:2012cn})
  for diagrams with different values of space, however, their consideration is above slope of the present paper.}
Solving these recursive relationships allows you to get accurate results for these most complex parts.
Obtaining these results is discussed in detail in Sections 3 and 6 for massless and massive diagrams, respectively.

A popular  property of maximal transcendentality is shown in Section 7.
It was introduced in  \cite{KL00}
for the Balitsky-Fadin-Kuraev-Lipatov (BFKL) kernel \cite{BFKL,next} in the  ${\mathcal N}=4$
Supersymmetric Yang-Mills (SYM) model \cite{BSSGSO}, is also applicable for the anomalous dimension matrices
of the twist-2 and twist-3 Wilson operators and for the coefficient functions of the
``deep-inelastic scattering''
in this model. The property gives a possibility to recover the results for the anomalous dimensions
\cite{KL,KoLiVe,KLOV} and  the coefficient functions
\cite{Bianchi:2013sta}
without any direct calculations by using the QCD corresponding values
\cite{VMV}.

The
very similar
property appears
also in the results of calculation of the large class of Feynman integrals, mostly 
for  so-called  master integrals \cite{Broadhurst:1987ei}.
The results for most of them can be reconstructed also without any direct calculations using a knowledge of
several terms in their inverse-mass expansion \cite{FleKoVe}. Note that the properties of the results are
related with the ones of
the amplitudes, form-factors and correlation functions (see \cite{Eden:2011we}-\cite{deLeeuw:2019qvz}
and references therein) studied recently  in the framework of the  ${\mathcal N}=4$ SYM.

In Section 6,
we demonstrate the existence of the propertiy of maximal transcendentality (or maximal complexity)
in  the
results of two-loop two- and three-point Feynman integrals (see also \cite{Kotikov:2010gf,Kotikov:2012ac}).

\section{Basic formulas}

Let us briefly consider the rules for calculation of massless diagrams.
All calculations are carried out in momentum space.

Following to Ref. \cite{Chetyrkin:1980pr}, we
introduce the traceless product  $q^{\mu_1 ... \mu_n}$ of the momentums connected with the usual
product $q^{\mu_1} ... q^{\mu_n}$ by the following equations
\bea
q^{\mu_1 ... \mu_n}  &=&  \hat{S} \, \sum_{p\geq 0} \, \frac{(-1)^p n! \Gamma(n-p+\lambda)}{2^{2p} (n-2p)!p!\Gamma(n+\lambda)} \,
g^{\mu_1 \mu_1} ... g^{\mu_{2p-1} \mu_{2p}} \, q^{2p} \, q^{\mu_{2p+1}} ... q^{\mu_n} \, ,
\nonumber \\
  q^{\mu_1} ... q^{\mu_n}  &=&  \hat{S} \, \sum_{p\geq 0} \,
  \frac{n! \Gamma(n-2p+1+\lambda)}{2^{2p} (n-2p)!p!\Gamma(n-p+1+\lambda)} \,
g^{\mu_1 \mu_1} ... g^{\mu_{2p-1} \mu_{2p}} \, q^{2p} \,  q^{\mu_{2p+1} ... \mu_n} \, ,
 \label{tProperty}
 \eea
 where $\hat{S}$ is a symmetrization on indeces $\mu_i$ ($i=1, ...,n$).

 We give also the simple but quite useful conditions:
\be
(q_1q_2)^{(n)}  \equiv q_1^{\mu_1 ... \mu_n} q_2^{\mu_1 ... \mu_n}  = q_1^{\mu_1} ... q_1^{\mu_n} q_2^{\mu_1 ... \mu_n}  = q_1^{\mu_1 ... \mu_n} q_2^{\mu_1} ... q_2^{\mu_n} \, ,
\label{TP}
\ee
which follow immediately from the  traceless-product definition: $g^{\mu_i \mu_j}  q^{\mu_1 ... \mu_i ... \mu_j ... \mu_n} =0$.

Propagators will be represented as
\bea
\frac{1}{(q^2)^{\alpha}} \equiv  \frac{1}{q^{2\alpha}} = \hspace{3mm} \raisebox{1mm}{{
\begin{axopicture}(70,30)(0,4)
  \SetWidth{1.0}
\Line(5,5)(65,5)
\Vertex(5,5){2}
\SetWidth{1.0}
\Vertex(65,5){2}
\Line(5,5)(-5,5)
\Line(65,5)(75,5)
\Text(33,7)[b]{}
\Text(33,-1)[t]{$\alpha$}
\Text(-3,-5)[b]{$\to$}
\Text(-3,-12)[b]{$q$}
\end{axopicture}
}}
\hspace{3mm}
,~~
\frac{q^{\mu}}{q^{2\alpha}}
= \hspace{3mm} \raisebox{1mm}{{
\begin{picture}(70,30)(0,4)
  \SetWidth{1.0}
\Line[arrow](5,5)(65,5)
\Vertex(5,5){2}
\SetWidth{1.0}
\Vertex(65,5){2}
\Line(5,5)(-5,5)
\Line(65,5)(75,5)
\Text(33,10)[b]{$\mu$}
\Text(33,-1)[t]{$\alpha$}
\Text(-3,-5)[b]{$\to$}
\Text(-3,-12)[b]{$q$}
\end{picture}
}}
\hspace{3mm}
,~~ \nonumber \\
\frac{q^{\mu_1} ... q^{\mu_n}}{q^{2\alpha}}
= \hspace{3mm} \raisebox{1mm}{{
\begin{picture}(70,30)(0,4)
  \SetWidth{1.0}
\Line[arrow,arrow](5,5)(65,5)
\Vertex(5,5){2}
\SetWidth{1.0}
\Vertex(65,5){2}
\Line(5,5)(-5,5)
\Line(65,5)(75,5)
\Text(33,10)[b]{$n$}
\Text(33,-1)[t]{$\alpha$}
\Text(-3,-5)[b]{$\to$}
\Text(-3,-12)[b]{$q$}
\end{picture}
}}
\hspace{3mm}
,~~
\frac{q^{\mu_1 ... \mu_n}}{q^{2\alpha}}
\hspace{3mm} \raisebox{1mm}{{
\begin{picture}(70,30)(0,4)
  \SetWidth{1.0}
\Line[arrow,arrow](5,5)(65,5)
\Vertex(5,5){2}
\SetWidth{1.0}
\Vertex(65,5){2}
\Line(5,5)(-5,5)
\Line(65,5)(75,5)
\Text(33,10)[b]{$(n)$}
\Text(33,-1)[t]{$\alpha$}
\Text(-3,-5)[b]{$\to$}
\Text(-3,-12)[b]{$q$}
\end{picture}
}}
\hspace{3mm}
.~~
\label{Def}
\eea

The use of the  traceless product  $q^{\mu_1 ... \mu_n}$ makes it possible to ignore terms of the
type $g^{\mu_i \mu_j}$ that arise upon
integration: they can be readily recovered from the general structure of the traceless product. Therefore,
in the process of integration it is only necessary to follow the coefficient of the leading term
$q^{\mu_1} ... q^{\mu_n}$. 

Everywhere in the paper on arguments $k, k_1, k_2 ...$  integration is carried out in $d = 4-2\ep$-space.
Hereafter in Sections 3, 4 and 5
the labels $k, k_1, k_2 ...$ denote integnal momentums. The symbols  $q, q_1, q_2 ...$ and $p, p_1, p_2 ...$
denote the external mometums with the conditions $p^2=0, p_1^2=0, p_2^2=0 ...$, respectively.\\

The following formulas hold.
\footnote{
  A traceless product can also be used to calculate complicated diagrams in another way.
  A propagator of complicated diagram can be expanded in a series of two other propagators, having
  traceless products in their numerators. This technique  \cite{Chetyrkin:1980pr,Kotikov:1995cw}
  is very effective for evaluating 
  complicated scalar diagrams with propagators having arbitrary degrees (see Refs.
  \cite{Broadhurst:1996yc,Kotikov:1989nm} and the review \cite{Teber:2016unz}.}
\\

{\bf A.}~~For simple chain:
\bea
&&\frac{q^{\mu_1} ... q^{\mu_n}}{q^{2\alpha_1}} \, \frac{q^{\nu_1} ... q^{\nu_m}}{q^{2\alpha_2}} =  \frac{q^{\mu_1} ... q^{\mu_n}q^{\nu_1} ... q^{\nu_m}}{q^{2(\alpha_1+\alpha_2)}}
\, , 
\nonumber \\
&& \hspace{-3cm} \mbox{or graphically} \nonumber \\
&&\raisebox{1mm}{{
\begin{picture}(90,30)(0,4)
  \SetWidth{1.0}
\Line[arrow](5,5)(40,5)
\Line[arrow](40,5)(85,5)
\Vertex(5,5){2}
\Vertex(5,5){2}
\Line(5,5)(-5,5)
\Line(85,5)(95,5)
\Vertex(40,5){2}
\Vertex(85,5){2}
\Text(15,10)[b]{$n$}
\Text(65,10)[b]{$m$}
\Text(15,-1)[t]{$\alpha_1$}
\Text(65,-1)[t]{$\alpha_2$}
\Text(-3,-5)[b]{$\to$}
\Text(-3,-12)[b]{$q$}
\end{picture}
}}
\hspace{3mm}
=
\hspace{3mm} \raisebox{1mm}{{
\begin{picture}(70,30)(0,4)
  \SetWidth{1.0}
\Line[arrow,arrow](5,5)(65,5)
\Vertex(5,5){2}
\SetWidth{1.0}
\Vertex(65,5){2}
\Line(5,5)(-5,5)
\Line(65,5)(75,5)
\Text(33,10)[b]{$n+m$}
\Text(33,-1)[t]{$\alpha_1+\alpha_2$}
\Text(-3,-5)[b]{$\to$}
\Text(-3,-12)[b]{$q$}
\end{picture}
}}
\hspace{3mm}
,
\label{chain}
\eea
i.e. the product of propagators is equivalent to a new propagator with an index equal to the sum of
the indices of the original propagators.  The number of momentums in the product in the numerator
is equal to the sum of the products of the impulses in the original propagators.\\

{\bf B.}~~ A simple loop can be integrated as
\be
\int \frac{Dk \, k^{\mu_1} ... k^{\mu_n}}{(q-k)^{2\alpha_1}k^{2\alpha_2}} = \frac{1}{(4\pi)^{d/2}} \, \frac{q^{\mu_1} ... q^{\mu_n}}{q^{2(\alpha_1+\alpha_2-d/2)}} \,
  A^{0,n}(\alpha_1,\alpha_2) + ... \, ,
\nonumber
\ee
where
we
neglect the terms of the order $g^{\mu_i \mu_j}$. Here
\be
Dk = \frac{d^dk}{(2\pi)^d} \, .
 \label{Measure}
\ee
is usual integration in Euclidean measure and
\be
A^{n,m}(\alpha,\beta) = \frac{a_n(\alpha)a_m(\beta)}{a_{n+m}(\alpha+\beta-d/2)},~~ a_n(\alpha)=\frac{\Gamma(\tilde{\alpha}+n)}{\Gamma(\alpha)},~~ \tilde{\alpha}=\frac{d}{2}-\alpha \, .
\label{Anm}
\ee

It is convenient to rewrite the equation graphically as
\bea
\raisebox{1mm}{{
    \begin{axopicture}(90,10)(0,4)
  \SetWidth{1.0}
\Arc(45,-7)(40,20,160)
\Arc[arrow](45,17)(40,200,340)
\Vertex(5,5){2}
\Vertex(85,5){2}
\Line(5,5)(-5,5)
\Line(85,5)(95,5)
\Text(45,-16)[b]{$n$}
\Text(45,27)[t]{$\alpha_1$}
\Text(45,-29)[t]{$\alpha_2$}
\Text(-3,-5)[b]{$\to$}
\Text(-3,-12)[b]{$q$}
\end{axopicture}
}}
\hspace{3mm}
=  \frac{1}{(4\pi)^{d/2}} \, A^{0,n}(\alpha_1,\alpha_2) \,
\hspace{3mm} \raisebox{1mm}{{
\begin{picture}(70,30)(0,4)
  \SetWidth{1.0}
\Line[arrow,arrow](5,5)(65,5)
\Vertex(5,5){2}
\SetWidth{1.0}
\Vertex(65,5){2}
\Line(5,5)(-5,5)
\Line(65,5)(75,5)
\Text(33,10)[b]{$n$}
\Text(33,-1)[t]{$\scriptstyle \alpha_1+\alpha_2-d/2$}
\Text(-3,-5)[b]{$\to$}
\Text(-3,-12)[b]{$q$}
\end{picture}
}}
\hspace{3mm}
 + ... \, .
\label{loop}
 \eea

\vskip 1.5cm

For the corresponding traceless product eq. (\ref{loop}) becomes to be exact, i.e.
\be
\int \frac{Dk \, k^{\mu_1 ... \mu_n}}{(q-k)^{2\alpha_1}k^{2\alpha_2}} = \frac{1}{(4\pi)^{d/2}} \, \frac{q^{\mu_1 ... \mu_n}}{q^{2(\alpha_1+\alpha_2-d/2)}} \,
A^{0,n}(\alpha_1,\alpha_2)  \, , \nonumber
\ee
or graphically\\
\be
\raisebox{1mm}{{
    \begin{axopicture}(90,10)(0,4)
  \SetWidth{1.0}
\Arc(45,-7)(40,20,160)
\Arc[arrow](45,17)(40,200,340)
\Vertex(5,5){2}
\Vertex(85,5){2}
\Line(5,5)(-5,5)
\Line(85,5)(95,5)
\Text(45,-16)[b]{$(n)$}
\Text(45,27)[t]{$\alpha_1$}
\Text(45,-29)[t]{$\alpha_2$}
\Text(-3,-5)[b]{$\to$}
\Text(-3,-12)[b]{$q$}
\end{axopicture}
}}
\hspace{3mm}
=  \frac{1}{(4\pi)^{d/2}} \, A^{0,n}(\alpha_1,\alpha_2) \,
\hspace{3mm} \raisebox{1mm}{{
\begin{picture}(70,30)(0,4)
  \SetWidth{1.0}
\Line[arrow,arrow](5,5)(65,5)
\Vertex(5,5){2}
\SetWidth{1.0}
\Vertex(65,5){2}
\Line(5,5)(-5,5)
\Line(65,5)(75,5)
\Text(33,10)[b]{$(n)$}
\Text(33,-1)[t]{$\scriptstyle \alpha_1+\alpha_2-D/2$}
\Text(-3,-5)[b]{$\to$}
\Text(-3,-12)[b]{$p$}
\end{picture}
}}
\hspace{3mm}
.
\label{tloop}
 \ee

\vskip 1.2cm

As it was noted already, indeed, 
we use the transless product $q^{\mu_1 ... \mu_n}$ in the r.h.s. but really we need only the first term,
i.e. $q^{\mu_1} ... q^{\mu_n}$, because the rest (i.e. the terms containing $g^{\mu_i \mu_j}$) is exactly
reconstracted from the exact form of traceless production. 

Such property can be used in another way:
a convenient representation of the results (\ref{loop}) and (\ref{tloop}) can be obtained also
by using an extra lightlike momentum $u$ (i.e. with $u^2=0$) and considering the product
$(uk)^n=u_{\mu_1} ... u_{\mu_n} \,  k^{\mu_1} ... k^{\mu_n} = u_{\mu_1} ... u_{\mu_n} \,  k^{\mu_1 ... \mu_n}$, because   $u^2=0$.

The results have the following form
\bea
\int \frac{Dk \, (uk)^{n} }{(q-k)^{2\alpha_1}k^{2\alpha_2}} = \frac{1}{(4\pi)^{d/2}} \, \frac{(uq)^{n} }{q^{2(\alpha_1+\alpha_2-d/2)}} \,
  A^{0,n}(\alpha_1,\alpha_2)  \, ,    \nonumber
\eea
or graphically\\
\be (u_{\mu_1} ... u_{\mu_n}) \, \raisebox{1mm}{{
    \begin{axopicture}(90,10)(0,4)
  \SetWidth{1.0}
\Arc(45,-7)(40,20,160)
\Arc[arrow](45,17)(40,200,340)
\Vertex(5,5){2}
\Vertex(85,5){2}
\Line(5,5)(-5,5)
\Line(85,5)(95,5)
\Text(45,-16)[b]{$n$}
\Text(45,27)[t]{$\alpha_1$}
\Text(45,-29)[t]{$\alpha_2$}
\Text(-3,-5)[b]{$\to$}
\Text(-3,-12)[b]{$q$}
\end{axopicture}
}}
\hspace{3mm}
=  \frac{1}{(4\pi)^{d/2}} \, A^{0,n}(\alpha_1,\alpha_2) \, (u_{\mu_1} ... u_{\mu_n}) \,
\hspace{3mm} \raisebox{1mm}{{
\begin{picture}(70,30)(0,4)
  \SetWidth{1.0}
\Line[arrow,arrow](5,5)(65,5)
\Vertex(5,5){2}
\SetWidth{1.0}
\Vertex(65,5){2}
\Line(5,5)(-5,5)
\Line(65,5)(75,5)
\Text(33,10)[b]{$n$}
\Text(33,-1)[t]{$\scriptstyle \alpha_1+\alpha_2-D/2$}
\Text(-3,-5)[b]{$\to$}
\Text(-3,-12)[b]{$q$}
\end{picture}
}}
\hspace{3mm}
\, .
\label{uloop}
 \ee

 \vskip 1.2cm
 
 Note that we use everything $\mu_i$ belonging to the traceless product, i.e. we consider the case of a scalar
 diagrams with a traceless product. In real theories such as QCD, there are still other  indices $\lambda_j$,
 which correspond to the numerators of propagators. In such cases, we can no longer neglect the terms
 $g^{\lambda_i\lambda_j}$ and $g^{\mu_i\lambda_j}$ and, therefore, the integration rules are complicated.
 They can be found in Refs. \cite{Kazakov:1986mu,Kotikov:1987mw}.

So, all diagrams, which can be expressed
as combinations of loops and chains can be evaluated immediately.
However, starting already with the two-loop level, there are diagrams, which cannot be expressed
as combinations of loops and chains (simplest example is shown below in Fig.1).
For these
cases there are additional rules, which will be shown only graphically with a purpose to
increase an illustation power.\\

{\bf C.}~~
When $\sum \alpha_i=d$, there is so-called uniqueness ratio \cite{DEramo:1971hnd,Vasiliev:1981dg,Kazakov:1984km}
for the triangle with indices
$\alpha_i$ ($i=1,2,3$)
\vskip 1 cm
\be
\raisebox{1mm}{{
    \begin{axopicture}(90,10)(0,4)
  \SetWidth{1.0}
\Line(5,5)(45,45)
\Line[arrow](5,5)(85,5)
\Line(45,45)(85,5)
\Vertex(5,5){2}
\Vertex(85,5){2}
\Vertex(45,45){2}
\Line(5,5)(-5,5)
\Line(85,5)(90,5)
\Line(45,45)(55,60)
\Text(70,7)[b]{$n$}
\Text(35,25)[t]{$\scriptstyle \alpha_2$}
\Text(45,-2)[t]{$\scriptstyle \alpha_1+n$}
\Text(55,25)[t]{$\scriptstyle \alpha_3$}
\Text(-3,-5)[b]{$\to$}
\Text(-3,-12)[b]{$\scriptstyle q_2-q_1$}
\Text(93,-5)[b]{$\to$}
\Text(93,-12)[b]{$\scriptstyle q_1-q_3$}
\Text(67,50)[b]{$\to$}
\Text(67,45)[b]{$\scriptstyle q_3-q_2$}
\end{axopicture}
}}
\hspace{3mm}
\overset{\sum \alpha_i=d}{=}
\frac{1}{(4\pi)^{d/2}} \, \sum_{m=0}^{n} \, C_n^m \, A^{n-m,m}(\alpha_2,\alpha_3) \,
\hspace{3mm} 
\raisebox{1mm}{{
    \begin{axopicture}(90,10)(0,4)
  \SetWidth{1.0}
\Line[arrow](5,5)(45,25)
\Line[arrow](45,25)(85,5)
\Line[arrow](45,25)(45,45)
\Vertex(5,5){2}
\Vertex(85,5){2}
\Vertex(45,45){2}
\Line(5,5)(-5,5)
\Line(85,5)(90,5)
\Line(45,45)(55,60)
\Text(20,20)[b]{$\scriptstyle m$}
\Text(70,20)[b]{$\scriptstyle n-m$}
\Text(20,5)[t]{$\scriptstyle \tilde{\alpha}_3+m$}
\Text(65,5)[t]{$\scriptstyle \tilde{\alpha}_2+n-m$}
\Text(55,37)[t]{$\scriptstyle \tilde{\alpha}_1$}
\Text(-3,-5)[b]{$\to$}
\Text(-3,-12)[b]{$\scriptstyle q_2-q_1$}
\Text(93,-5)[b]{$\to$}
\Text(93,-12)[b]{$\scriptstyle q_1-q_3$}
\Text(67,50)[b]{$\to$}
\Text(67,45)[b]{$\scriptstyle q_3-q_2$}
\end{axopicture}
}}
\hspace{3mm} \, ,
\label{TreUni}
\ee
\vskip 1 cm
\hspace{-8mm}
where
\be
C_n^m=\frac{n!}{m!(n-m)!} \, .
\label{Cnm}
\ee

The results (\ref{TreUni}) can be exactlly obtained in the following way: perform
an inversion $q_i \to 1/q_i$ $(i=1,2,3)$, $k \to 1/k$
in the subintegral expression and in the integral measure. The inversion
keeps angles between momentums. After the  inversion, one propagator is cancelled because $\sum \alpha_i =d$ and
the l.h.s. becomes to be equal to
a loop. Evaluating it using the rule (\ref{loop}) and returning after it to the initial momentums, we recover the rule (\ref{TreUni}). An extension of the rule (\ref{TreUni}) to the
case with two traceless products can be found in \cite{Kotikov:1987mw}.\\

{\bf D.}~~ For any triangle with indices
$\alpha_i$ ($i=1,2,3$) there is the following relation, which is based on
integration by parts (IBP) procedure \cite{Vasiliev:1981dg,Chetyrkin:1981qh}
%
\vskip 1cm

\bea
&&(d-2\alpha_1-\alpha_2-\alpha_3+n+m+k) \hspace{0.5cm}
\raisebox{1mm}{{
    \begin{axopicture}(90,10)(0,4)
  \SetWidth{1.0}
\Line[arrow](5,5)(45,45)
\Line[arrow](5,5)(85,5)
\Line[arrow](45,45)(85,5)
\Vertex(5,5){2}
\Vertex(85,5){2}
\Vertex(45,45){2}
\Line(5,5)(-5,5)
\Line(85,5)(90,5)
\Line(45,45)(55,60)
\Text(45,10)[b]{$n$}
\Text(20,30)[b]{$m$}
\Text(70,30)[b]{$k$}
\Text(35,25)[t]{$\scriptstyle \alpha_2$}
\Text(45,-2)[t]{$\scriptstyle \alpha_1$}
\Text(55,25)[t]{$\scriptstyle \alpha_3$}
\Text(-3,-5)[b]{$\to$}
\Text(-3,-12)[b]{$\scriptstyle q_2-q_1$}
\Text(93,-5)[b]{$\to$}
\Text(93,-12)[b]{$\scriptstyle q_1-q_3$}
\Text(70,50)[b]{$\to$}
\Text(70,40)[b]{$\scriptstyle q_3-q_2$}
\end{axopicture}
}}
\hspace{3mm} \nonumber \\
&&\nonumber \\
&&\nonumber \\
&&\nonumber \\
&&= \alpha_2 \biggl[ \, \hspace{0.5cm} \,
  \raisebox{1mm}{{
    \begin{axopicture}(90,10)(0,4)
  \SetWidth{1.0}
\Line[arrow](5,5)(45,45)
\Line[arrow](5,5)(85,5)
\Line[arrow](45,45)(85,5)
\Vertex(5,5){2}
\Vertex(85,5){2}
\Vertex(45,45){2}
\Line(5,5)(-5,5)
\Line(85,5)(90,5)
\Line(45,45)(55,60)
\Text(45,10)[b]{$n$}
\Text(20,30)[b]{$m$}
\Text(70,30)[b]{$k$}
\Text(35,25)[t]{$\scriptstyle \alpha_2+1$}
\Text(45,-2)[t]{$\scriptstyle \alpha_1-1$}
\Text(55,25)[t]{$\scriptstyle \alpha_3$}
\Text(-3,-5)[b]{$\to$}
\Text(-3,-12)[b]{$\scriptstyle q_2-q_1$}
\Text(93,-5)[b]{$\to$}
\Text(93,-12)[b]{$\scriptstyle q_1-q_3$}
\Text(70,50)[b]{$\to$}
\Text(70,40)[b]{$\scriptstyle q_3-q_2$}
\end{axopicture}
}}
\hspace{3mm}
- (q_2-q_1)^2 \times
  \raisebox{1mm}{{
    \begin{axopicture}(90,10)(0,4)
  \SetWidth{1.0}
\Line[arrow](5,5)(45,45)
\Line[arrow](5,5)(85,5)
\Line[arrow](45,45)(85,5)
\Vertex(5,5){2}
\Vertex(85,5){2}
\Vertex(45,45){2}
\Line(5,5)(-5,5)
\Line(85,5)(90,5)
\Line(45,45)(55,60)
\Text(45,10)[b]{$n$}
\Text(20,30)[b]{$m$}
\Text(70,30)[b]{$k$}
\Text(35,25)[t]{$\scriptstyle \alpha_2+1$}
\Text(45,-2)[t]{$\scriptstyle \alpha_1$}
\Text(55,25)[t]{$\scriptstyle \alpha_3$}
\Text(-3,-5)[b]{$\to$}
\Text(-3,-12)[b]{$\scriptstyle q_2-q_1$}
\Text(93,-5)[b]{$\to$}
\Text(93,-12)[b]{$\scriptstyle q_1-q_3$}
\Text(70,50)[b]{$\to$}
\Text(70,40)[b]{$\scriptstyle q_3-q_2$}
\end{axopicture}
}}
  \hspace{3mm}
\nonumber \\
&&\nonumber \\
&&\nonumber \\
&&\nonumber \\
&&  +
m (q_2-q_1)^{\mu_m} \times
\raisebox{1mm}{{
    \begin{axopicture}(90,10)(0,4)
  \SetWidth{1.0}
\Line[arrow](5,5)(45,45)
\Line[arrow](5,5)(85,5)
\Line[arrow](45,45)(85,5)
\Vertex(5,5){2}
\Vertex(85,5){2}
\Vertex(45,45){2}
\Line(5,5)(-5,5)
\Line(85,5)(90,5)
\Line(45,45)(55,60)
\Text(45,10)[b]{$n$}
\Text(20,30)[b]{$\scriptstyle m-1$}
\Text(70,30)[b]{$k$}
\Text(35,25)[t]{$\scriptstyle \alpha_2+1$}
\Text(45,-2)[t]{$\scriptstyle \alpha_1-1$}
\Text(55,25)[t]{$\scriptstyle \alpha_3$}
\Text(-3,-5)[b]{$\to$}
\Text(-3,-12)[b]{$\scriptstyle q_2-q_1$}
\Text(93,-5)[b]{$\to$}
\Text(93,-12)[b]{$\scriptstyle q_1-q_3$}
\Text(70,50)[b]{$\to$}
\Text(70,40)[b]{$\scriptstyle q_3-q_2$}
\end{axopicture}
}}
\hspace{5mm}
\Biggr] + \alpha_3
\, \biggl[\alpha_2 \leftrightarrow \alpha_3, m \leftrightarrow k \biggr] \, .
\label{TreIBP}
\eea

\vskip 1cm

Eq. (\ref{TreIBP}) can been obtained by introducing the factor $(\partial/\partial k_{\mu}) \, (k-q_1)^{\mu}$ to the subintegral expression of the triangle, shown below as $[...]$,
and using the integration by parts procedure as follows:
\bea
&& d \int Dk \, \bigl[ ...\bigr] = \int Dk \, \left(\frac{\partial}{\partial k_{\mu}} \, (k-q_1)^{\mu}\right) \,  \bigl[ ...\bigr]      =
\int Dk \,  \frac{\partial}{\partial k_{\mu}} \, \left((k-q_1)^{\mu} \,  \bigl[ ...\bigr] \right) \nonumber \\&&
- \int Dk \, (k-q_1)^{\mu} \,
\frac{\partial}{\partial k_{\mu}} \, \left( \bigl[ ...\bigr]\right) \, .
\label{IBPpro}
\eea
The first term in the r.h.s. becomes to be zero because it can be represented as a surphase integral on the infinite surphase. Evalutiong the second  term in the r.h.s.
we  reproduce Eq. (\ref{TreIBP}).

As it is possible to see from Eqs. (\ref{TreIBP}) and (\ref{IBPpro}) the line with the index $\alpha_1$ is distingulished. The contributions of the other lines are
same. So, we will call below the line with the index $\alpha_1$ as a ``distingulished line''. It is clear that a various choices of the distingulished line produce
different tipes of the IBP relations.

Using equation (\ref{TreIBP}) allows you to change the indices of the line diagrams by an integer. One can
also change line indices using the point group of transformations  \cite{Vasiliev:1981dg,Gorishnii:1984te}.
The elements of the group are:
a) the transition to impulse presentation,
b) conformal inversion transformation $p \to p' =p/p^2$, c) a special series of transformations that makes it possible to make one of the vertices unique, and then apply relation (3) to it. An extension of the group of transformations
for diagrams with the traceless product can be found in Ref. \cite{Kotikov:1987mw}.

\section{Basic massless two-loop integrals}

The general topology of the two-loop two-point diagram, which cannot be expressed as a combination of loops and chanins is shown on Fig.1.

\begin {figure} [htbp]
\centerline{
\begin{picture}(150,100)(0,4)
  \SetWidth{0.7}
\Arc[arrow](90,50)(60,0,90)
\Arc[arrow](90,50)(60,90,180)
\Arc[arrow](90,50)(60,180,270)
\Arc[arrow](90,50)(60,270,360)
\Line(30,50)(20,50)
\Line(150,50)(160,50)
\Line[arrow](90,-10)(90,110)
\Vertex(30,50){2}
\Vertex(150,50){2}
\Vertex(90,110){2}
\Vertex(90,-10){2}
  \SetWidth{1}
\Text(45,95)[b]{$n_1$}
\Text(135,95)[b]{$n_2$}
\Text(45,20)[b]{$n_4$}
\Text(135,20)[b]{$n_3$}
\Text(80,50)[t]{$n_5$}
\Text(45,77)[t]{$\alpha_1$}
\Text(135,75)[t]{$\alpha_2$}
\Text(45,0)[t]{$\alpha_4$}
\Text(135,0)[t]{$\alpha_3$}
\Text(100,50)[t]{$\alpha_5$}
\Text(10,40)[b]{$\to$}
\Text(10,30)[b]{$p$}
\end{picture}
}
\vskip 0.5cm
\caption{The diagram which cannot be expressed as a combination of loops and chains.
}
\label{sunsetMMm}
\end{figure}
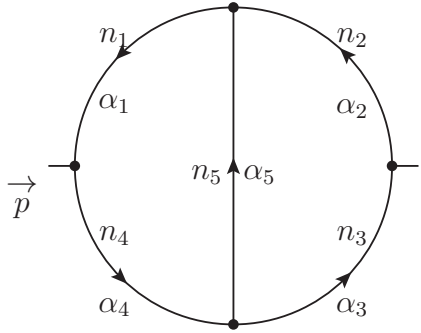

\vskip 0.5cm


Below in the present analysis we will concentrate mostly on two particular cases, which can be taken from 
the diagram shown in Fig. 1, for $\alpha_3+n_3=\alpha$, $n_3=n$,
$\alpha_j (j\neq 3)=1$, $n_j (j\neq 3)=0$ (we denote $I_1(\alpha,n)$) and for  $\alpha_5+n_5=\alpha$,
$n_5=n$, $\alpha_j (j\neq 5)=1$, $n_j (j\neq 5)=0$ (we denote $I_2(\alpha,n)$)\\
\vskip 0.5cm
\be
I_1(\alpha,n) \, =
 \hspace{3mm}
\raisebox{1mm}{{
    \begin{axopicture}(90,10)(0,4)
  \SetWidth{1.0}
\Arc(45,-7)(40,20,90)
\Arc(45,-7)(40,90,160)
\Line(45,-25)(45,35)
\Arc(45,17)(40,200,270)
\Arc[arrow](45,17)(40,270,340)
\Vertex(45,-23){2}
\Vertex(45,33){2}
\Vertex(5,5){2}
\Vertex(85,5){2}
\Line(5,5)(-5,5)
\Line(85,5)(95,5)
\Text(63,-10)[b]{$n$}
\Text(75,-22)[t]{$\alpha$}
\Text(-3,-5)[b]{$\to$}
\Text(-3,-12)[b]{$q$}
\end{axopicture}
}}
\hspace{3mm},~~
I_2(\alpha,n) \, =
 \hspace{3mm}
\raisebox{1mm}{{
    \begin{axopicture}(90,10)(0,4)
  \SetWidth{1.0}
\Arc(45,-7)(40,20,90)
\Arc(45,-7)(40,90,160)
\Line[arrow](45,-25)(45,35)
\Arc(45,17)(40,200,270)
\Arc(45,17)(40,270,340)
\Vertex(45,-23){2}
\Vertex(45,33){2}
\Vertex(5,5){2}
\Vertex(85,5){2}
\Line(5,5)(-5,5)
\Line(85,5)(95,5)
\Text(37,5)[t]{$n$}
\Text(59,5)[t]{$\alpha$}
\Text(-3,-5)[b]{$\to$}
\Text(-3,-12)[b]{$q$}
\end{axopicture}
}}
\hspace{3mm} .
\label{I1+2}
\ee
\vskip 1cm

It is convenient to calculate the  diagrams $I_1(\alpha,n)$ and $I_2(\alpha,n)$ using
the functional relations \footnote{The functional relations were obtained in
  \cite{Kazakov:1983pk,Kazakov:1984bw} by application of IBP realtions with various distinguished lines.}
similar to those obtained in \cite{Kazakov:1983pk,Kazakov:1984bw},
which reduces the amount of computations.

Repeating analysis done in \cite{Kazakov:1983pk}, we obtain the following   functional relations:
\bea
&&I_1(\alpha,n) = \frac{1}{q^2} \, I_1(\alpha-1,n)
- \frac{1}{2\ep} \, \Bigl( 2I_{1,1}(\alpha,n) + I_{2,1}(\alpha,n)
\Bigr) \, ,
  \label{I1alpha}\\
  &&I_2(\alpha,n) =
  \frac{2}{n+d-2-2\alpha} \, I_{2,1}(\alpha,n)
  - \frac{n+2d-4-2\alpha}{n+d-2-2\alpha} \, \frac{1}{q^2} \, I_2(\alpha-1,n)  \, ,
  \label{I2alpha}
  \eea
  where the inhomogeneus terms are
\bea
&&I_{11}(\alpha,n) =
  \raisebox{1mm}{{
    \begin{axopicture}(90,10)(0,4)
  \SetWidth{1.0}
\CArc(25,5)(20,0,180)
\CArc(25,5)(20,180,360)
 \Arc(65,5)(20,0,180)
  \Arc[arrow](65,5)(20,180,360)
  \Vertex(5,5){2}
  \Vertex(45,5){2}
\Vertex(85,5){2}
\Line(5,5)(-5,5)
\Line(85,5)(95,5)
\Text(65,-10)[b]{$n$}
\Text(65,-20)[t]{$\alpha$}
\Text(25,-20)[t]{$2$}
\Text(-3,-5)[b]{$\to$}
\Text(-3,-12)[b]{$q$}
\end{axopicture}
}}
  \hspace{3mm}
  - \frac{1}{q^2} \hspace{3mm}
  \raisebox{1mm}{{
    \begin{axopicture}(90,10)(0,4)
  \SetWidth{1.0}
\Arc(45,-7)(40,20,160)
\Arc(5,-35)(40,20,85)
\Arc(45,17)(40,200,270)
\Arc[arrow](45,17)(40,270,340)
\Vertex(5,5){2}
\Vertex(85,5){2}
\Vertex(43,-23){2}
\Line(5,5)(-5,5)
\Line(85,5)(95,5)
\Text(65,-10)[b]{$n$}
\Text(65,-25)[t]{$\alpha-1$}
\Text(33,7)[t]{$2$}
\Text(-3,-5)[b]{$\to$}
\Text(-3,-12)[b]{$q$}
\end{axopicture}
  }}
\hspace{3mm},
\label{I11alpha}\\
  \nonumber \\ &&
\nonumber \\ &&
\nonumber \\ &&
I_{21}(\alpha,n) =
\frac{1}{q^2}  \hspace{0.3cm}
\raisebox{1mm}{{
    \begin{axopicture}(90,10)(0,4)
  \SetWidth{1.0}
\Arc(45,-7)(40,20,160)
\Arc(71,-23)(30,65,177)
\Arc(45,17)(40,200,270)
\Arc[arrow](45,17)(40,270,340)
\Vertex(5,5){2}
\Vertex(85,5){2}
\Vertex(41,-23){2}
\Line(5,5)(-5,5)
\Line(85,5)(95,5)
\Text(65,-10)[b]{$n$}
\Text(67,-24)[t]{$\scriptstyle \alpha-1$}
\Text(43,7)[t]{$2$}
\Text(-3,-5)[b]{$\to$}
\Text(-3,-12)[b]{$q$}
\end{axopicture}
  }} 
  \nonumber 
  \hspace{3mm}
  - \hspace{3mm}
  \raisebox{1mm}{{
    \begin{axopicture}(90,10)(0,4)
  \SetWidth{1.0}
\Arc(45,-7)(40,20,160)
\Arc(71,-23)(30,65,177)
\Arc(45,17)(40,200,270)
\Arc[arrow](45,17)(40,270,340)
\Vertex(5,5){2}
\Vertex(85,5){2}
\Vertex(41,-23){2}
\Line(5,5)(-5,5)
\Line(85,5)(95,5)
\Text(65,-10)[b]{$n$}
\Text(67,-24)[t]{$\alpha$}
\Text(43,27)[t]{$2$}
\Text(-3,-5)[b]{$\to$}
\Text(-3,-12)[b]{$q$}
\end{axopicture}
  }}
  \nonumber 
  \hspace{3mm} .
\label{I11alpha}\\
  \eea
\vskip 1cm

We see
that the inhomogeneus terms in the functional equations (\ref{I1alpha}) and (\ref{I2alpha}), i.e.
the diagrams $I_{11}(\alpha,n)$ and $I_{21}(\alpha,n)$, can be represented as combinations of loops and chains and,
thus, they can be evaluated by rules
(\ref{chain}) and (\ref{loop}).\\

We would like to note that for the massless two-point diagrams the subject of the study is so-called coefficient functions, which expression
$C_{i ...} (\alpha,n)$ $(i=1,2)$ for the considered diagrams $I_{i ...} (\alpha,n)$ can be introduced in the following form
\be
I_{i ...} (\alpha,n) = \frac{1}{(4\pi)^{d}} \, C_{i ...} (\alpha,n) \, \frac{q^{\mu_1 ... \mu_n}}{q^{2(\alpha + 2\ep)}} \, .
  \label{Cialpha}
  \ee

  The result (\ref{Cialpha}) contains the fact that we consider the two-loop diagrams. In general, $L$-loop diagram
  $I_L (\alpha_1, ..., \alpha_N,n)$
  having propagators with indices $\alpha_i$ $(i=1,...,N)$ and one traceless product of momentums can be respresented as
  \be
  I_L (\alpha_1, ..., \alpha_N,n)= \frac{1}{(4\pi)^{dL/2}} \, C_L (\alpha_1, ..., \alpha_N,,n) \,
  \frac{q^{\mu_1 ... \mu_n}}{q^{2(\overline{\alpha} - d/2*\ep)}} \, ,
  \label{Cialpha.ge}
  \ee
where $\overline{\alpha}=\sum_1^N \alpha_i$.

The results for $C_{i ...} (\alpha,n)$ can be obtained directly from rules
(\ref{chain}) and (\ref{loop}). They have the following form:
\bea
&&C_{1,1} (\alpha,n) = A(2,1) \Bigl(A^{0,n}(1,\alpha)-A^{0,n}(1,\alpha+\ep)\Bigr)
\, , \label{C11alpha} \\
&&C_{2,1} (\alpha,n) = A^{0,n}(2,\alpha-1) A^{0,n}(1,\alpha+\ep)-A^{0,n}(1,\alpha)A^{0,n}(2,\alpha+\ep) \, , \label{C21alpha}
\eea
where
\be
A(\alpha_1,\alpha_2) = A^{0,0}(\alpha_1,\alpha_2) \, .
\label{A00}
\ee
and the result for $A^{n,m}(\alpha_1,\alpha_2)$ is given in (\ref{Anm}).

Thus, the coefficient functions $C_{i,1} (\alpha,n)$ $(i=1,2)$ can be represented as combinations of $\Gamma$-functions.

\subsection{$I_1(0,n)$ and $I_2(0,n)$}

The  diagrams $I_1(0,n)$ and $I_2(0,n)$ can be considered a s natural boundary conditions for the functional
equations (\ref{I1alpha}) and  (\ref{I2alpha}). Moreover, in
a sence, they
have a special property: their
results
can be obtained with help of Eqs.
(\ref{chain}) and (\ref{loop}) but with an additional resummatiomn.

Indeed,
expanding for $I_1(0,n)$ and $I_2(0,n)$, respectively, the corresponding
products of momentums in the following way:
\bea
\prod^n_{i=1} \, (q-k_2)^{\mu_i} = \sum_{k=0}^{n}\, C_n^k \, (-1)^k \, \prod^k_{i=1} k_2^{\mu_i} \, \prod^{n-k}_{j=1} q^{\mu_j} \hspace{1cm} \mbox{for  $I_1(0,n)$} \, ,
\label{I1.0} \\
\prod^n_{i=1} \, (k_1-k_2)^{\mu_i} = \sum_{k=0}^{n}\, C_n^k (-1)^k \, \prod^k_{i=1} k_2^{\mu_i} \, \prod^{n-k}_{j=1} k_1^{\mu_j} \hspace{1cm} \mbox{for  $I_2(0,n)$} \, ,
\label{I2.0}
\eea
we see that the diagrams $I_1(0,n)$ and $I_2(0,n)$ can be represented as
\bea
&&I_1(0,n) =  \sum_{k=0}^{n}\, C_n^k \, (-1)^k \, \prod^{n-k}_{j=1} q^{\mu_j} \, \overline{I}_1(0,k) \label{I1.1} \, , \\
&&I_2(0,n) =  \sum_{k=0}^{n}\, C_n^k \, (-1)^k \, \overline{I}_2(n-k,k) \label{I2.1} \, ,
\label{I2.02}
\eea
where
\vskip 0.5cm
\bea
\overline{I}_{1}(0,k) =
  \raisebox{1mm}{{
    \begin{axopicture}(90,10)(0,4)
  \SetWidth{1.0}
\Arc(45,-7)(40,20,90)
\Arc(45,-7)(40,90,160)
\Arc[arrow](85,45)(40,197,270)
\Arc(45,17)(40,200,270)
\Arc(45,17)(40,270,340)
\Vertex(5,5){2}
\Vertex(85,5){2}
\Vertex(48,33){2}
\Line(5,5)(-5,5)
\Line(85,5)(95,5)
\Text(67,13)[b]{$\scriptstyle k$}
\Text(-3,-5)[b]{$\to$}
\Text(-3,-12)[b]{$q$}
\end{axopicture}
  }}\hspace{0.3cm},
  \hspace{0.2cm}
\overline{I}_{2}(n-k,k) =
  \raisebox{1mm}{{
    \begin{axopicture}(90,10)(0,4)
  \SetWidth{1.0}
\CArc(25,5)(20,0,180)
\CArc[arrow](25,5)(20,180,360)
 \Arc(65,5)(20,0,180)
  \Arc[arrow](65,5)(20,180,360)
  \Vertex(5,5){2}
  \Vertex(45,5){2}
\Vertex(85,5){2}
\Line(5,5)(-5,5)
\Line(85,5)(95,5)
\Text(65,-10)[b]{$\scriptstyle k$}
\Text(25,-10)[b]{$\scriptstyle n-k$}
\Text(-3,-5)[b]{$\to$}
\Text(-3,-12)[b]{$q$}
\end{axopicture}
}} 
  \hspace{3mm} .
\label{I11alpha}
\eea
\vskip 1cm
So, the diagrams $I_1(0,n)$ and $I_2(0,n)$ 
can be expressed as combination of loops and chains and, thus, their coefficient functions 
can be calculated using rules (\ref{chain}) and (\ref{loop}).

So, we have for the coefficient functions $C_1(0,n)$ and $C_2(0,n)$:
\bea
C_1(0,n) = \sum_{k=0}^{n}\, C_n^k \, (-1)^k \, A^{0,k}(1,1) \,  A^{0,k}(1,1+\ep) \, , \label{C1.0} \\
C_2(0,n) = \sum_{k=0}^{n}\, C_n^k \, (-1)^k \, A^{0,k}(1,1) \,  A^{0,n-k}(1,1) \, . \label{C2.0} 
\eea

As we noted already at the beginning of the subsection, the
results for $C_i(0,n)$ ($i=1,2$) are very important because they give a possibility to recover all
results $C_i(m,n)$ using Eqs. (\ref{I1alpha}) and  (\ref{I2alpha}) with $\alpha =m$. But the calculation of
$C_i(0,n)$ needs performing additional series (see Eqs. (\ref{C1.0}) and (\ref{C2.0})). So,
we present these calculations in some
details, separetely for $I_1(0,n)$ and $I_2(0,n)$
in the following two subsections.

\subsection{$I_1(0,n)$}

Now we consider $C_1(0,n)$ form Eq.(\ref{C1.0}), which can be represented as
\be
C_1(0,n) = \frac{N_2}{2\ep^2} \, \sum_{k=0}^{n}\, C_n^k \, (-1)^k \,
\frac{\Gamma(k+1-\ep)\Gamma(1-3\ep)}{\Gamma(k+2-3\ep)\Gamma(1-\ep)} \, \frac{1}{k+1-2\ep} \, .
\label{C1.0.1}
\ee

It is conveninet to rewrite
the last term in the r.h.s. as
\be
\frac{1}{k+1-2\ep} = \frac{1}{k+1} + \frac{2\ep}{(k+1)(k+1-2\ep)} \, ,
\label{fac}
\ee
that leads to the following form for $C_1(0,n)$:
\be
C_1(0,n) = \overline{C}_1(0,n) +2\ep \, \tilde{C}_1(0,n)
\label{C1.0.2}
\ee
with
\bea
&&\overline{C}_1(0,n) = \frac{N_2}{2\ep^2} \, \sum_{k=0}^{n}\, C_n^k \, (-1)^k \,
\frac{\Gamma(k+1-\ep)\Gamma(1-3\ep)}{\Gamma(k+2-3\ep)\Gamma(1-\ep)} \, \frac{1}{k+1} \, ,
\label{oC1.0} \\
&&\tilde{C}_1(0,n) = \frac{N_2}{2\ep^2} \, \sum_{k=0}^{n}\, C_n^k \, (-1)^k \,
\frac{\Gamma(k+1-\ep)\Gamma(1-3\ep)}{\Gamma(k+2-3\ep)\Gamma(1-\ep)} \, \frac{1}{(k+1)(k+1-2\ep)} \, ,
\label{tC1.0}
\eea
where the normalization $N_2$ and the factors $K_1$ and $K_2$ are
\be
N_2=\Gamma^2(1+\ep)K_1K_2,~~ K_1=\frac{\Gamma^2(1-\ep)}{\Gamma(1-2\ep)},~~
K_1=\frac{\Gamma(1-\ep)\Gamma(1-2\ep)\Gamma(1+2\ep)}{\Gamma^2(1+\ep)\Gamma(1-3\ep)} \, .
\label{N2}
\ee

The result for $\overline{C}_1(0,n)$ can be found exactly as
\bea
&&\frac{\overline{C}_1(0,n)}{N_2/(2\ep^2)} = \sum_{k=1}^{n+1} \, \frac{(-1)^{k+1} n!}{k!((n-k+1)!} \, 
\frac{\Gamma(k-\ep)\Gamma(1-3\ep)}{\Gamma(k+1-3\ep)\Gamma(1-\ep)} \, \nonumber \\
&&= -\frac{1}{n+1} \, \Bigl[ {}_2F_1\bigl(-(n+1),-\ep;1-3\ep;1\bigr)-1\Bigr] \, \frac{\Gamma(-ep)}{\Gamma(1-ep)}
\nonumber \\
&&= \frac{1}{(n+1)\ep} \, \Bigl[B(n+2,-2,-3)-1\Bigr]=  \frac{1}{(n+1)\ep} \, \Bigl[\frac{n+1-2\ep}{n+1-3\ep} \,
  B(n+1,-2,-3)-1\Bigr] \, ,
\label{oC1.0.1}
\eea
where
\be
B(n+1,a,b)=\frac{\Gamma(n+1+a\ep)\Gamma(1+b\ep)}{\Gamma(1+a\ep)\Gamma(n+1+b\ep)} \, .
\label{B}
\ee

The result for $B(n+1,a,b)$
can be easily evaluated using the expansion of the corresponding $\Gamma$-functions \cite{FleKoVe}
\bea
&&\frac{\Gamma(n+1+a\ep)}{n! \Gamma(1+a\ep)} = exp\left[-\sum_{m=1}^{\infty} \, \frac{(-a\ep)^m}{m} \, S_m(n) \right], \label{Gamma.n} \\
&&\Gamma(1+a\ep) =  exp\left[-\gamma a\ep + \sum_{m=1}^{\infty} \, \frac{(-a\ep)^m}{m} \, \zeta_m \right], \label{Gamma.1}
\eea
where $\gamma$ is Euler's constant.

Indeed, we have
\bea
&&B(n+1,a,b)=
exp\left[-\sum_{m=1}^{\infty} \, \frac{(-\ep)^m}{m} \, \bigl[a^m-b^m\bigr] \, S_m(n) \right] = 1+(a-b)\ep \, S_1(n) \nonumber \\
&&+ \frac{(a-b)\ep^2}{2} \, \left[(a-b)S^2_1(n)-(a+b)S_2(n)\right] +  \frac{(a-b)\ep^3}{3!} \nonumber \\
&&\times \left[(a-b)^2S^3_1(n)-3(a^2-b^2)S_1(n)S_2(n) +2(a^2+ab+b^2)S_3(n)\right] + O(\ep^4) \, .
\label{Bab}
\eea

Hereafter we use the following nested sums:
\be
S_{\pm i}(n)=\sum^n_{m=1} \, \frac{(-1)^m}{m^i},~~ S_{\pm i,j}(n)=\sum^n_{m=1} \, \frac{(-1)^m}{m^i} S_{\pm j}(m), 
\label{Sin}
\ee
and $\zeta(n) = S_n(\infty)$ - Riemann zeta-function.

The result for $\tilde{C}_1(0,n)$ can be expressed as
\be
\frac{\tilde{C}_1(0,n)}{N_2/(2\ep^2)} = \sum_{k=0}^{n} \,  C_n^k \, (-1)^k \,  \frac{1}{(k+1)^3} \,
\left(1+ \ep \left[2S_1(k)+ \frac{5}{k+1}\right] \right) + O(\ep^2) \, .
\label{tC1.0.1}
\ee

To evaluate the r.h.s it is convenient to calculate firstly the series $\phi_1(a,n)$ and $\Phi_1(a,n)$
\bea
\phi_1(a,n) = \sum_{k=0}^{n} \,  C_n^k \, (-1)^k \,  \frac{1}{k+1-a},~~
\Phi_1(a,n) = \sum_{k=0}^{n} \,  C_n^k \, (-1)^k \,  \frac{S_1(k)}{k+1-a},
\nonumber
\eea
differentiate them several times with respect of $a$ and put $a=0$. Indeed,
\be
\frac{\tilde{C}_1(0,n)}{N_2/(2\ep^2)} = \phi^{(2)}_1(n) + \ep \Bigl(2\Phi^{(2)}_1(n) + 5 \phi^{(3)}_1(n)\Bigr)
 + O(\ep^2) \, ,
\label{tC1.0.2}
\ee
where
\bea
\phi^{(m)}_1(n) = \frac{1}{m!} \, \frac{\partial^m}{(\partial a)^m} \, \phi_1(a,n) \biggl{|}_{a=0},~~
\Phi^{(m)}_1(n) = \frac{1}{m!} \, \frac{\partial^m}{(\partial a)^m} \, \Phi_1(a,n) \biggl{|}_{a=0} \, .
\nonumber
\eea

The series $\phi_1(a,n)$ can be calculated directly as
\be
\phi_1(a,n) = \frac{1}{1-a} \, {}_2F_1(-n,1-a,2-a) = \frac{\Gamma(1-a)\Gamma(n+1)}{\Gamma(n+2-a)}
\label{phi1a}
\ee
and we have
\bea
&&\phi_1(n,0) = \frac{1}{n+1},~~ \phi_1^{(1)}(n) = \frac{1}{n+1} \, S_1(n+1),~~\nonumber \\
&&\phi_1^{(2)}(n) = \frac{1}{2(n+1)} \, \Bigl(S^2_1(n+1)+S_2(n+1)\Bigr),~~ \nonumber \\
&&\phi_1^{(3)}(n) = \frac{1}{6(n+1)} \, \Bigl(S^3_1(n+1)+3S_1(n+1)S_2(n+1)+2S_3(n+1)\Bigr) \, .
\label{phi1m}
\eea

Evaluating the series
$\Phi_1(a,n)$
is quite difficut, and we will show this in a separate subsection.
The calculation of more complicated series can be found in a famous paper \cite{Vermaseren:1998uu}. 

\subsubsection{$\Phi_1(a,n)$}

It is convenient
to consider the recursive relation
between $\Phi_1(a,n)$ and $\Phi_1(a,n-1)$.
Indeed, we can represented $\Phi_1(a,n)$ in the following form
\be
\Phi_1(a,n) = \sum_{k=0}^{n-1} \, \frac{(-1)^k n!}{k!(n-1-k)!} \,  \frac{S_1(k)}{(n-k)(k+1-a)} +
(-1)^n \, \frac{S_1(n)}{n+1-a} \, .
\label{Phi1a}
\ee

Taking partial fraction
\be
\frac{1}{(n-k)(k+1-a)} = \frac{1}{n+1-a} \left[\frac{1}{n-k} + \frac{1}{k+1-a}\right] \, ,
\label{ParFra}
\ee
we can rewrite (\ref{Phi1a}) in the following way
\be
\Phi_1(a,n) =  \frac{1}{n+1-a} \, \left[n \Phi_1(a,n) +
\sum_{k=0}^{n} \, \frac{(-1)^k n!}{k!(n-k)!} \, S_1(k) \right] \,.
\label{Phi1a1}
\ee

The series in the r.h.s. can be calculated usng results for $\hat{\Phi}_1(\alpha,\beta)$, studied in Appendix A, as
\bea
&&\sum_{k=0}^{n} \, \frac{(-1)^k n!}{k!(n-k)!} \, S_1(k) = \frac{\partial}{(\partial x)} \,
\sum_{k=0}^{n} \, \frac{(-1)^k n! \Gamma(1-x)}{\Gamma(k+1-x)(n-k)!} \biggr{|}_{x=0} \nonumber \\
&& = n! \,  \frac{\partial}{(\partial x)} \, \left\{ \Gamma(1-x) \, \hat{\Phi}_1(\alpha=0,\beta=1-x) \right\}
\biggr{|}_{x=0} =
\frac{\partial}{(\partial x)} \, \frac{x}{x-n} \biggr{|}_{x=0} = - \frac{1}{n} \, .
\nonumber
\eea

It is convenient to introduce the new function $\overline{\Phi}_1(a,n)$ as
\be
\Phi_1(a,n) = \frac{\Gamma(1-a)\Gamma(n+1)}{\Gamma(n+2-a)} \, \overline{\Phi}_1(a,n) \, ,
\label{oPhi1a}
\ee
which has the following relation:
\be
\overline{\Phi}_1(a,n) = \overline{\Phi}_1(a,n-1) - \frac{\Gamma(n+1-a)}{n!\Gamma(1-a)} \, \frac{1}{n} \, .
\label{oPhi1a.1}
\ee

The last relation
can be solved as
\be
\overline{\Phi}_1(a,n) = \overline{\Phi}_1(a,0) - \sum_{m=1}^n \, \frac{\Gamma(m+1-a)}{m!\Gamma(1-a)} \, \frac{1}{m} \, .
\label{oPhi1a.2}
\ee
Since $\Phi_1(a,n=0)=0$, then $\overline{\Phi}_1(a,0)=0$ and we find finally
\be
\Phi_1(a,n) = - \frac{\Gamma(1-a)\Gamma(n+1)}{\Gamma(n+2-a)} \,
\sum_{m=1}^n \, \frac{\Gamma(m+1-a)}{m!\Gamma(1-a)} \, \frac{1}{m} \, 
\label{Phi1a.1}
\ee
and for $a=0$
\be
\Phi_1(0,n) = - \frac{1}{n+1} \,\sum_{m=1}^n \, \frac{1}{m} = - \frac{S_1(n)}{n+1} \, . 
\label{Phi1a.2}
\ee

Note that the case $\Phi_1(0,n)$ can be evaluated directly in the way similar to the one (\ref{oC1.0.1})
for $\overline{C}_1(0,n)$
and there is a full agreement of such calculations with (\ref{Phi1a.2}).

After little algebra, we have
\bea
&&\Phi^{(1)}_1(n) = - \frac{1}{n+1} \,\sum_{m=1}^n \, \frac{1}{m} \, \bigl(S_1(n+1) -S_1(m)\bigr) \nonumber \\
&&= - \frac{1}{n+1} \, \left[\frac{1}{2} \, \bigl(S^2_1(n) -S_2(n)\bigr) + \frac{S_1(n)}{n+1} \right] \, ,
\label{Phi1n1} \\
&&\Phi^{(2)}_1(n) = - \frac{1}{2(n+1)} \,\sum_{m=1}^n \, \frac{1}{m} \, \left[\bigl(S_1(n+1) -S_1(m)\bigr)^2
+ S_2(n+1) -S_2(m) \right] \nonumber \\
&&= - \frac{1}{2(n+1)} \, \biggl[\frac{1}{3} \, \bigl(S^3_1(n) -3S_1(n)S_2(n) -4S_3(n)\bigr) + 2S_{2,1}(n) \nonumber \\
&& +\frac{1}{n+1} \, \bigl(S^2_1(n) -S_2(n)\bigr) + \frac{2S_1(n)}{(n+1)^2} \biggr]
\label{Phi1n1} \, ,
\eea
where we used the formulas from the part C of Appendix A.

We would like to note about an appearence the nested sum $S_{2,1}(n)$, which cannot be obtained from expansions
of products of $\Gamma$-functions.

\subsubsection{$C_1(0,n)$}

Taking the results for $\phi^{(2)}_1(n)$, $\phi^{(3)}_1(n)$ and $\Phi^{(2)}_1(n)$ together, we can obtain the following
results for $\tilde{C}_1(0,n)$
\be
\frac{\tilde{C}_1(0,n)}{N_2/(2\ep^2)} = \frac{\ep}{n+1-2\ep} \, \left[\frac{1}{2} C(n) + \ep D(n) \right] + O(\ep^3),
\label{tC1.0.1}
\ee
where
\bea
&&\hat{C}(n) = S^2_1(n) +S_2(n) + \frac{2S_1(n)}{n+1} + \frac{2}{(n+1)^2} = S^2_1(n+1) +S_2(n+1) \, ,
\label{Cn} \\
&&\hat{D}(n) = \overline{D}(n) +  \frac{1}{2(n+1)}\bigl(S^2_1(n) +5S_2(n)\bigr) + \frac{S_1(n)}{(n+1)^2} +\frac{3}{(n+1)^3} 
\label{Dn}
\eea
and
\be
\overline{D}(n) = \frac{1}{2} \Bigl(S^3_1(n) + 7S_1(n)S_2(n) +6S_3(n) -4 S_{2,1}(n) \Bigr) \, . 
\label{Dn}
\ee

We note that there is another representation for $\tilde{C}_1(0,n)$:
\be
\frac{\tilde{C}_1(0,n)}{N_2/(2\ep^2)} = \frac{\ep}{n+1} \,
\left[\frac{1}{2} C(n) + \ep \overline{D}(n+1) \right] + O(\ep^3) \, .
\label{tC1.0.1}
\ee

Taking the results for $\overline{C}_1(0,n)$ and $\tilde{C}_1(0,n)$ of (\ref{tC1.0.1})
together, we have the final result for $C_1(0,n)$:
\bea
&&C_1(0,n) =  \frac{N_2}{2(n+1)} \, \Biggl[\frac{1}{\ep^2} \, S_1(n+1) + \frac{1}{2\ep} \,\bigl(3S^2_1(n+1) +7S_2(n+1)\bigr)
  \nonumber \\
  &&+ \frac{7}{6} \,S^3_1(n+1) +  \frac{19}{2} \, S_1(n+1)S_2(n+1) + \frac{37}{3} \,S_3(n+1) -4 S_{2,1}(n+1) \Biggr]
\, 
\label{C1.0.2}
\eea
with the normalization factor $N_2$ was defined in (\ref{N2}).

Since $K_2/K_1 \sim O(\ep^3)$ (see Eq. (\ref{6zeta3}) below), we can replace in the r.h.s. the normalization $N_2$ by $N_1$
\be
N_1 = \Gamma^2(1+\ep) \, K_1^2
\label{N1}
\ee
and put the normalization $N_1$ in the definition of $\mu^2_g$-scale of $g$-scheme \cite{Broadhurst:1999xk},
which relates with the usual $\overline{MS}$
one as $\mu^2_g = K_1 \mu^2_{\overline{MS}}$ (see discussions in Ref. \cite{Kotikov:2019bqo}).

\subsection{$I_2(0,n)$}

The diagram $I_2(0,n)$ s zero for odd $n$ values. So, we can calculate it firstly at the even  $n$ values and
recover its genetal form at the end.

So, now we consider $C_2(0,n)$ from Eq.(\ref{C2.0}), which can be represented as
\be
C_2(0,n) = \frac{N_1}{\ep^2} \, \sum_{k=0}^{n}\, C_n^k \, (-1)^k \,
\frac{\Gamma(k+1-\ep)\Gamma(n-k+1-\ep)\Gamma^2(1-2\ep)}{\Gamma(k+2-3\ep)\Gamma(n-k+2-2\ep)\Gamma^2(1-\ep)} 
\, ,
\label{C2.0.1}
\ee
where the normalization factor $N_1$ is given in Eq. (\ref{N1}).

The last part of r.h.s. can be written as
(for even $n$)
\bea
&&\frac{B(k+1,-1,-2) B(n-k+1,-1,-2)}{(k+1-2\ep)(n-k+1-2\ep)} =
\frac{2}{n+2-4\ep} \, \frac{1}{k+1-2\ep} \nonumber \\
&&\times \, B(k+1,-1,-2) B(n-k+1,-1,-2) \, ,
\label{sym}
\eea
because
\be
\frac{1}{(k+1-2\ep)(n-k+1-2\ep)} = \frac{2}{n+2-4\ep} \, \left[\frac{1}{k+1-2\ep} +
  \frac{1}{n-k+1-2\ep} \right]
\label{sym.1}
\ee
Moreover, with the required accuracy $O(\ep^0)$, we can rewrite the product\\ $B(k+1,-1,-2) B(n-k+1,-1,-2)$ in the
following way:
\bea
&&B(k+1,-1,-2) B(n-k+1,-1,-2)= B(k+1,-1,-2) \nonumber \\
&& + B(n-k+1,-1,-2) -1 + \ep^2 S_1(k)S_1(n-k) + O(\ep^2)\, ,
\label{expa}
\eea
because
\be
B(k+1,-1,-2) = 1+\ep S_1(n)+ \frac{\ep^2}{2} \, \bigl( S_1^2(n) -3 S_2(n) \bigr) + O(\ep^2)\, .
\label{expa.1}
\ee

Then,  $C_2(0,n)$ can be splitted to the four different parts:
\be
C_2(0,n) = \frac{2}{n+2-4\ep} \, \biggl[C_2^{(1)}(0,n)+ C_2^{(2)}(0,n) - C_2^{(3)}(0,n) + \ep^2 C_2^{(4)}(0,n) \biggr]
\, ,
\label{C2.0.2}
\ee
where
\bea
&&C_2^{(1)}(0,n) = \frac{N_1}{\ep^2} \, \sum_{k=0}^n \, (-1)^k \, C_N^k \, \frac{B(k+1,-1,-2)}{k+1-2\ep}, \label{C21.0.a} \\
&&C_2^{(2)}(0,n) = \frac{N_1}{\ep^2} \, \sum_{k=0}^n \, (-1)^k \, C_N^k \, \frac{B(n-k+1,-1,-2)}{k+1-2\ep}, \label{C22.0.a} \\
&&C_2^{(3)}(0,n) = \frac{N_1}{\ep^2} \, \sum_{k=0}^n \, (-1)^k \, C_N^k \, \frac{1}{k+1-2\ep}, \label{C23.0.a} \\
&&C_2^{(4)}(0,n) =
\sum_{k=0}^n \, (-1)^k \, C_N^k \, \frac{S_1(k)S_1(n-k)}{k+1} \, . \label{C24.0.a}
\eea

The parts $C_2^{(1)}(0,n)$ and $C_2^{(3)}(0,n)$ can be evaluated directly. Indeed,
\bea
&&\frac{C_2^{(1)}(0,n)}{N_1/\ep^2} = \sum_{k=0}^n \, (-1)^k \, C_N^k \,
\frac{\Gamma(k+1-\ep)\Gamma(1-2\ep)}{\Gamma(k+2-2\ep)\Gamma(1-\ep)}= \frac{1}{1-2\ep} \,
             {}_2F_1\bigl(-n,1-\ep;2-2\ep;1\bigr) 
   \nonumber \\
&&= \frac{\Gamma(n+1-\ep)\Gamma(1-2\ep)}{\Gamma(n+2-2\ep)\Gamma(1-\ep)} \,
   =  \frac{B(n+1,-1,-2)}{(n+1-2\ep)} \, , \nonumber \\
             &&\frac{C_2^{(3)}(0,n)}{N_1/\ep^2} 
             = \frac{1}{1-2\ep} \, {}_2F_1\bigl(-n,1-2\ep;2-2\ep;1\bigr) =
             \frac{\Gamma(n+1)\Gamma(1-2\ep)}{\Gamma(n+2-2\ep)}
             = \frac{B(n+1,0,-2)}{(n+1-2\ep)} \, .
 \nonumber   \eea

As in the case of $C_1(0,n)$, using Eq. (\ref{fac})
it is convenient to split the part $C_2^{(2)}(0,n)$ in two parts
\be
C_2^{(2)}(0,n) = \overline{C}_2^{(2)}(0,n) +2\ep \, \tilde{C}_2^{(2)}(0,n) \, ,
\label{C22.0.2a}
\ee
where
\bea
&&\overline{C}_2^{(2)}(0,n) = \frac{N_1}{\ep^2} \, \sum_{k=0}^{n}\, C_n^k \, (-1)^k \,
\frac{B(n-k+1,-1,-2)}{k+1}
\, ,
\label{oC22.0.1a} \\
&&\tilde{C}_2^{(2)}(0,n) = \frac{N_1}{\ep^2} \, \sum_{k=0}^{n}\, C_n^k \, (-1)^k \, \frac{B(n-k+1,-1,-2)}{(k+1)(k+1-2\ep)}
\, .
\label{tC22.0.1a}
\eea

As in the case of
$\overline{C}_1(0,n)$, the part $\overline{C}_2^{(2)}(0,n)$ can be found exactly:
\bea
&&\frac{\overline{C}_2^{(2)}(0,n)}{N_2/\ep^2} = \sum_{k=1}^{n+1} \, \frac{(-1)^{k+1} n!}{k!((n-k+1)!} \, 
\frac{\Gamma(n-k+2-\ep)\Gamma(1-2\ep)}{\Gamma(n-k+3-2\ep)\Gamma(1-\ep)} \, \nonumber \\
&&= -\frac{1}{n+1} \, \frac{\Gamma(n+2-\ep)\Gamma(1-2\ep)}{\Gamma(n+3-2\ep)\Gamma(1-\ep)} \,
\Bigl[ {}_2F_1\bigl(-(n+1),\ep-(n+1);2\ep-(n+1);1\bigr)-1\Bigr]  \nonumber \\
&&= \frac{B(n+2,-1,-2)}{(n+1)} \, \Bigl[1- \frac{\Gamma(\ep-(n+1))\Gamma(n+1-\ep)}{\Gamma(\ep)\Gamma(-\ep)} \Bigr]
\, .
\label{oC22.0.2a}
\eea

Since
\be
\frac{\Gamma(\ep-(n+1))}{\Gamma(-\ep)} = \frac{(-1)^{n+1}\Gamma(1-\ep)}{\Gamma(n+2-\ep)} \, ,
\label{rati.1a}
\ee
then $\overline{C}_2^{(2)}(0,n)$ has the following form for even $n$ values
\bea
\frac{\overline{C}_2^{(2)}(0,n)}{N_2/\ep^2} &=& \frac{B(n+2,-1,-2)}{(n+1)} \, \Bigl[1-
  \frac{(-1)^{n+1} \ep}{n+1-\ep}\Bigr] \nonumber \\
&&= \frac{B(n+2,-1,-2)}{(n+1)} \, \frac{n+1-2 \ep}{n+1-\ep}
= \frac{B(n+1,-1,-2)}{(n+1)} \, .
\label{oC22.0.3a}
\eea

The result for $\tilde{C}_2^{(2)}(0,n)$ can be expressed as
\be
\frac{\tilde{C}_2^{(2)}(0,n)}{N_1/\ep^2} = \sum_{k=0}^{n} \,  C_n^k \, (-1)^k \,  \frac{1}{(k+1)^2} \,
\left(1+ \ep \left[ S_1(n-k)+ \frac{2}{k+1}\right] \right) + O(\ep^2) \, .
\label{tC122.0.1a}
\ee

The last part $C_2^{(4)}(0,n)$ is
splitted in two parts because $S_1(m)=S_1(m+1)-1/(m+1)$:
\bea
&&C_2^{(4)}(0,n) = \overline{C}_2^{(4)}(0,n) - \tilde{C}_2^{(4)}(0,n),
\label{C24.0.2a} \\
&&\overline{C}_2^{(4)}(0,n) =
\sum_{k=0}^n \, (-1)^k \, C_N^k \, \frac{S_1(k+1)S_1(n-k)}{k+1},  \label{oC24.0.1a} \\
&&\tilde{C}_2^{(4)}(0,n) = \sum_{k=0}^n \, (-1)^k \, C_N^k \, \frac{S_1(n-k)}{(k+1)^2} \, . \label{tC24.0.1a}
\eea

It is easy to show that the part $ \overline{C}_2^{(4)}(0,n)$ is zero at even $n$ values. Indeed,
\be
\overline{C}_2^{(4)}(0,n) = - \sum_{k=1}^{n+1} \, \frac{(-1)^k n!S_1(k)S_1(n-k+1)}{k!(n-k+1)!}
= - \sum_{k=0}^{n+1} \, \frac{(-1)^k n!S_1(k)S_1(n-k+1)}{k!(n-k+1)!} \, .
\label{oC24.0.2a}
\ee
After replacement $k \to n+1-k$ the last series obtains the additional factor $(-1)^{n+1}$ and, thus,
it is zero for even $n$ values. It is really the case: the results for $ \overline{C}_2^{(4)}(0,n)$
are exactly evaluated in the part B of Appendix A.\\

Now we return to the coefficient function  $C_2(0,n)$ given in Eq. (\ref{C2.0.2}).
It is convenient to split the result (\ref{C2.0.1})
into two parts:
\be
\frac{n+2-4\ep}{2} \, C_2(0,n) = \overline{C}_2(0,n) + 2\ep \, \tilde{C}_2(0,n) \, ,
\label{C2.0.1b}
\ee
where
\bea
&&\overline{C}_2(0,n) = C_2^{(1)}(0,n) + \overline{C}_2^{(2)}(0,n) - C_2^{(3)}(0,n) \, , \label{oC2.0.1b} \\
&&\tilde{C}_2(0,n) = \tilde{C}_2^{(2)}(0,n) - \frac{\ep}{2} \, \tilde{C}_2^{(4)}(0,n) \, . \label{tC2.0.1b}
\eea

The part $\overline{C}_2(0,n)$ can be evaluated exactly as
\be
\overline{C}_2(0,n) = \frac{1}{n+1-2\ep} \, \left[2\left(1-\frac{\ep}{n+1}\right) \, B(n+1,-1,-2) - B(n+1,0,-2)
  \right] \, .
\label{oC2.0.2b} 
\ee

Using expansions (\ref{Bab}) of $B(n+1,a,b)$ with respect of $\ep$, we have
\bea
&&\overline{C}_2(0,n) = \frac{1}{n+1-2\ep} \, \left[1-\frac{2\ep}{n+1} + \ep^2
  \left(S_2(n)-S_1^2(n)-\frac{2S_1(n)}{n+1}\right) \right] \nonumber \\
&&=\frac{1}{n+1} \, \left[1 + \ep^2
  \left(S_2(n)-S_1^2(n)-\frac{2S_1(n)}{n+1}\right) \right] \, .
\label{oC2.0.3b} 
\eea

The result for $\tilde{C}_2(0,n)$ can be expressed as
\be
\frac{\tilde{C}_2^{(2)}(0,n)}{N_1/\ep^2} = \sum_{k=0}^{n} \,  C_n^k \, (-1)^k \,  \frac{1}{(k+1)^2} \,
\left(1+ \ep \left[ \frac{1}{2} \, S_1(n-k)+ \frac{2}{k+1}\right] \right) + O(\ep^2) \, .
\label{tC122.0.1a}
\ee

To evaluate the r.h.s it is convenient to use the result (\ref{phi1a}) for $\phi_1(a,n)$, to
calculate
the series
$\Phi_2(a,n)$:
\be
\Phi_2(a,n) = \sum_{k=0}^{n} \,  C_n^k \, (-1)^k \,  \frac{S_1(n-k)}{k+1-a}
\label{Phi2a}
\ee
and to differentiate several times these series with respect of $a$ and put $a=0$. Indeed,
\be
\frac{\tilde{C}_2(0,n)}{N_1/\ep^2} = \phi^{(1)}_1(n) + \ep \Bigl(\frac{1}{2} \, \Phi^{(1)}_2(n)
+ 2 \phi^{(2)}_1(n)\Bigr)
 + O(\ep^2) \, ,
\label{tC1.0.2}
\ee
where
\be
\Phi^{(m)}_2(n) = \frac{1}{m!} \, \frac{\partial^m}{(\partial a)^m} \, \Phi_2(a,n) \biggl{|}_{a=0} \, .
\nonumber
\ee

We consider the evaluation of the series $\Phi_2(a,n)$ in the next subsection.

\subsubsection{$\Phi_2(a,n)$}

As in the case of $\Phi_1(a,n)$,
to obtain
the results for $\Phi_2(a,n)$
it is convenient
to consider the difference relation between $\Phi_2(a,n)$ and $\Phi_2(a,n-1)$.
Indeed, we can rewrite (\ref{Phi2a}) in the following way
\be
\Phi_2(a,n) = \sum_{k=0}^{n-1} \, \frac{(-1)^k n!}{k!(n-1-k)!} \,  \frac{S_1(n-k)}{(n-k)(k+1-a)} \, .
\label{Phi2a}
\ee

Taking partial fraction (\ref{ParFra})
we can represent
(\ref{Phi2a}) in the following way
\be
\Phi_2(a,n) = \frac{1}{n+1-a} \, \left[ \sum_{k=0}^{n-1} \, \frac{(-1)^k n!}{k!(n-1-k)!} \,  \frac{S_1(n-k)}{k+1-a}
  + \sum_{k=0}^{n-1} \, \frac{(-1)^k n!}{k!(n-k)!} \, S_1(n-k) \right]\, .
\label{Phi2a.1}
\ee

Taking $S_1(n-k)=S_1(n-k-1)+1/(n-k)$, we have
\bea
&&\Phi_2(a,n) =  \frac{1}{n+1-a} \, \left[n \Phi_1(a,n) +  \sum_{k=0}^{n-1} \, \frac{(-1)^k n!}{k!(n-k)!} \,
  \frac{1}{k+1-a} +
\sum_{k=0}^{n-1} \, \frac{(-1)^k n!}{k!(n-k)!} \, S_1(n-k) \right]
\nonumber \\
&&= \frac{1}{n+1-a} \, \left[n \Phi_1(a,n) + \phi_1(a,n) - \frac{(-1)^n}{n+1-a} 
   +
\sum_{k=0}^{n} \, \frac{(-1)^k n!}{k!(n-k)!} \, S_1(n-k) \right] \, .
\label{Phi2a.2}
\eea

The series in the r.h.s. can be calculated usng results for $\hat{\Phi}_1(\alpha,\beta)$ studied in Appendix A as
\bea
&&\sum_{k=0}^{n} \, \frac{(-1)^k n!}{k!(n-k)!} \, S_1(n-k) = \frac{\partial}{(\partial y)} \,
\sum_{k=0}^{n} \, \frac{(-1)^k n! \Gamma(1-y)}{k!\Gamma(n-k+1-y)} \biggr{|}_{y=0} \nonumber \\
&& = n! \,  \frac{\partial}{(\partial x)} \, \left\{ \Gamma(1-y) \, \hat{\Phi}_1(\alpha=-y,\beta=1) \right\}
\biggr{|}_{y=0} = (-1)^n \,
\frac{\partial}{(\partial x)} \, \frac{y}{y-n} \biggr{|}_{y=0} = - \frac{(-1)^n}{n} \, .
\nonumber
\eea

So, we have
\bea
\Phi_2(a,n) =  \frac{1}{n+1-a} \, \left[n \Phi_1(a,n) + \frac{\Gamma(1-a)\Gamma(n+1)}{\Gamma(n+2-a)}
  -(-1)^n \left(\frac{1}{n+1-a}+\frac{1}{n}\right)\right] \, .
\label{Phi2a.2}
\eea

It is convenient to introduce the new function $\overline{\Phi}_2(a,n)$ as
\be
\Phi_2(a,n) = \frac{\Gamma(1-a)\Gamma(n+1)}{\Gamma(n+2-a)} \, \overline{\Phi}_2(a,n) \, ,
\label{oPhi2a}
\ee
which has the following reccursion:
\be
\overline{\Phi}_2(a,n) = \overline{\Phi}_2(a,n-1) + \frac{1}{n+1-a} 
- (-1)^n \,\frac{\Gamma(n+1-a)}{n!\Gamma(1-a)} \, \left(\frac{1}{n+1-a} + \frac{1}{n}\right) \, .
\label{oPhi2a.1}
\ee
It can be solved as
\be
\overline{\Phi}_2(a,n) = \overline{\Phi}_2(a,0) + \sum_{m=1}^n \, \biggl[ \frac{1}{m+1-a} -(-1)^m \,
\frac{\Gamma(m+1-a)}{m!\Gamma(1-a)} \, \left(\frac{1}{m+1-a} + \frac{1}{m}\right)\biggr] \, .
\label{oPhi2a.2}
\ee

Since $\Phi_2(a,n=0)=0$, then $\overline{\Phi}_2(a,0)=0$ and we find finally
\be
\Phi_2(a,n) = \frac{\Gamma(1-a)\Gamma(n+1)}{\Gamma(n+2-a)} \,
\sum_{m=1}^n \, 
\biggl[ \frac{1}{m+1-a} -(-1)^m \,
\frac{\Gamma(m+1-a)}{m!\Gamma(1-a)} \, \left(\frac{1}{m+1-a} + \frac{1}{m}\right)\biggr] \, 
\label{Phi2a.3}
\ee
and for $a=0$
\be
\Phi_2(0,n) =  \frac{1}{n+1} \,
\sum_{m=1}^{n+1} \, \left[\frac{1}{m+1} -(-1)^m\left(\frac{1}{m+1}+ \frac{1}{m}\right)\right]
= \frac{1}{n+1} \, \left(S_1(n+1)+\frac{(-1)^{n+1}}{n+1}\right) \, .
\label{Phi2a.4}
\ee

We can calculate
$\Phi_2(0,n)$
directly in the way similar to the evaluation of $\Phi_1(0,n)$. The direct calculation
is a full agreement of such calculations with (\ref{Phi2a.4}).

We have for $\Phi^{(1)}_2(n)$:
\bea
&&\Phi^{(1)}_2(n) = - \frac{S_1(n+1)}{n+1} \, \sum_{m=1}^{n+1} \,
\left[\frac{1}{m+1} -(-1)^m\left(\frac{1}{m+1}+ \frac{1}{m}\right)\right] \nonumber \\
&&+ \frac{1}{n+1} \,  \, \sum_{m=1}^{n+1} \,
\left[\frac{1}{(m+1)^2} -(-1)^mS_1(m)\left(\frac{1}{m+1}+ \frac{1}{m}\right)\right] \, .
\label{Phi2a.4}
\eea

After little algebra, we obtain the final resuls
\be
\Phi^{(1)}_2(n) =  \frac{1}{n+1} \, \Bigl(S_2(n+1)+2S_{-2}(n+1)+S_1^2(n+1)\Bigr) \, .
\label{Phi1n1} 
\ee

We would like to note about an appearence the sum $S_{-2}(n)$, which cannot be obtained from expansions
of products of $\Gamma$-functions.

Taking the results for $\phi^{(1)}_1(n)$, $\phi^{(2)}_1(n)$
and $\Phi^{(2)}_2(n)$ together, we can obtain the following
results for $\tilde{C}_2(0,n)$
\be
\frac{\tilde{C}_2(0,n)}{N_1/\ep^2} = \frac{1}{n+1-2\ep} \, \left[S_1(n)+\frac{1}{n+1}
  +\ep \left(\frac{3}{2}S_1^2(n)+\frac{3}{2}S_2(n)+S_{-2}(n)+\frac{S_1(n)}{n+1}\right)\right]
   + O(\ep^2) \, .
\label{tC2.0.1b}
\ee

\subsubsection{$C_2(0,n)$}

Taking the results for $\overline{C}_2(0,n)$ and $\tilde{C}_2(0,n)$ together, we have for $C_2(0,n)$
\be
C_2(0,n) =  \frac{2N_1}{(n+2-4\ep)(n+1-2\ep)} \Biggl[\frac{1}{\ep^2} + \frac{2}{\ep} \, S_1(n) 
  + 2\,S^2_1(n) +  4\, S_2(n) + 2  S_{-2}(n) \Biggr] \, .
\label{C2.0.2b}
\ee

Changing the nested sums $S_i(n)$ to $S_i(n+1)$, we can simplify these results:
\bea
C_2(0,n) &=&  \frac{2N_1}{(n+2-4\ep)(n+1)} \Biggl[\frac{1}{\ep^2} + \frac{2}{\ep} \, S_1(n+1) 
  \nonumber \\
  &&+ 2\,S^2_1(n+1) +  4\, S_2(n+1) + 2  S_{-2}(n+1) \Biggr] \, \label{C1.0.3b} \\
&&=  \frac{2N_1}{(n+2)(n+1)} \Biggl[\frac{1}{\ep^2} + \frac{2}{\ep} \, \left(S_1(n+1)+\frac{1}{n+2}\right) 
  \nonumber \\
  &&+ 2\,S^2_1(n+1) +  4\, S_2(n+1) + 2  S_{-2}(n+1) + 8 \frac{S_1(n+1)}{n+2} + \frac{16}{(n+2)^2}\Biggr] \, .
\label{C2.0.4b}
\eea

We note, that
we can
put the normalization in the definition of $\mu^2_g$-scale of $g$-scheme \cite{Broadhurst:1999xk}, which
relates with the usual $\overline{MS}$
one as $\mu^2_g = K_1 \mu^2_{\overline{MS}}$.

Using the fact that the above result (\ref{C2.0.4b})
is obtained for even $n$ values (and $C_2(0,2m+1)=0$), we
can recover the
full resul for $C_2(0,n)$ in the following form:
\bea
C_2(0,n) &=&  \frac{(1+(-1)^n) \, N_1}{(n+2)(n+1)} \Biggl[\frac{1}{\ep^2} + \frac{2}{\ep} \, \left(S_1(n+1)+\frac{1}{n+2}\right) 
  \nonumber \\
  &&+ 2\,S^2_1(n+1) +  4\, S_2(n+1) + 2  S_{-2}(n+1) + 8 \frac{S_1(n+1)}{n+2} + \frac{16}{(n+2)^2}\Biggr] \, .
\label{C1.0.4b1}
\eea

\subsection{$I_1(1,n)$ and $I_2(1,n)$}

Now we evaluate the particular cases $I_1(1,n)$ and $I_2(1,n)$, which are important for the future studies.
Indeed, we will see that the integral are finite and have very compact form.

The results for the $C_{1,1}(1,n)$ and $C_{2,1}(1,n)$ are
\bea
&&C_{1,1}(1,n) = -\frac{N_2}{2\ep^2} \, \left[\frac{2K_1}{K_2} \, \frac{B(n+1,-1,-2)}{n+1-2\ep} -
  \frac{B(n+1,-2,-3)}{n+1-3\ep} \right] \, ,
\label{C11.1.re} \\
&&C_{2,1}(1,n) = -\frac{N_2}{2\ep^2} \, \frac{B(n+1,-1,-3)}{n+1-2\ep}  \, ,
\label{C21.1.re}
\eea
where the normalization factors $N_1$ and $N_2$ and also the factors $K_1$ and $K_2$ are given in
Eqs. (\ref{N1}) and (\ref{N2}), respectively.

Below we consider the results for $I_1(1,n)$ and $I_2(1,n)$ is the different subsections.

\subsubsection{$I_1(1,n)$}

Following to the case $C_{2}(0,n)$ (see Eq. (\ref{C2.0.1b})) it is convenient to split the result for
$C_{2}(1,n)$ into tow parts
\be
C_1(1,n) = \overline{C}_1(1,n) +2\ep \, \tilde{C}_{1}(0,n) \, ,
\label{oC0}
\ee
i.e. we
take together the coefficient functions $C_{1,1}(1,n)$ and $C_{2,1}(1,n)$ and also
the function $\overline{C}_1(0,n)$ of Eq. (\ref{oC1.0.1}) and to denote this contributon
as $\overline{C}_1(1,n)$. So, we have
\be
\overline{C}_1(1,n) = \overline{C}_1(0,n) - \frac{1}{2\ep} \, \left(2 C_{1,1}(1,n) - C_{2,1}(1,n)\right) \, .
\label{oC1}
\ee

Taking the results (\ref{C11.1.re}), (\ref{C21.1.re}) and (\ref{oC1.0.1}) we have for $\overline{C}_1(1,n)$
\bea
\frac{\overline{C}_1(1,n)}{N_2/(2\ep^2)} &=& \frac{1}{\ep} \, \biggl[ \frac{2K_1}{K_2} \, \frac{B(n+1,-1,-2)}{n+1-2\ep}
  \nonumber \\
&&  - \frac{B(n+1,-1,-3)}{n+1-2\ep} - \frac{2\ep B(n+1,-2,-3)}{(n+1)(n+1-3\ep)} - \frac{1}{n+1} \Biggr] \, .
\label{oC1.1}
\eea

Expanding the r.h.s. with
\be
\frac{K_1}{K_2} = \exp[6\zeta_3\ep^3] \, ,
\label{6zeta3}
\ee
we have
\be
\frac{\overline{C}_1(1,n)}{N_2/(2\ep^2)} = \frac{1}{n+1-2\ep} \, \left[-\hat{A}(n)\ep +
  \bigl(12\zeta_3 - \hat{B}(n)\bigr) \ep^2
  + )(\ep^3) \right] + O(\ep^3) \, ,
\label{oC1.2}
\ee
where
\bea
&&\hat{A}(n) = S^2_1(n) +S_2(n) + \frac{2S_1(n)}{n+1} + \frac{2}{(n+1)^2} = \hat{C}(n) \, ,
\label{An} \\
&&\hat{B}(n) = S^3_1(n)+5S_1(n)S_2(n)+4S_{3}(n)
+  \frac{1}{n+1}\bigl(S^2_1(n) +5S_2(n)\bigr) + \frac{2S_1(n)}{(n+1)^2} +\frac{6}{(n+1)^3} \, 
\label{Bn}
\eea
and we see that $\hat{A}(n)$ is equal to $\hat{C}(n)$ in (\ref{Cn}).

Taking together the results (\ref{tC1.0.1}) and (\ref{oC1.2}), we obtain finally
\be
C_1(1,n) = \frac{1}{n+1} \, \Bigl(S_3(n)+S_1(n)S_2(n)-S_{2,1}(n)+6\zeta_3 \Bigr) + O(\ep) \, .  \label{C1}
\ee

\subsubsection{$I_2(1,n)$}

Consider the contribution of $C_2(0,n)$ to $C_2(1,n)$. It has the form (see Eq.(\ref{I2alpha}) for $\alpha=1$)
\be
- \frac{n+2d-6}{n+d-4} \, C_2(0,n) = - \frac{n+2-4\ep}{n-2\ep} \, C_2(0,n) \, ,
\label{C2.1.a1}
\ee
where $C_2(0,n)$ is given above in (\ref{C1.0.4b1}). So, we have
\be
- \frac{n+2-4\ep}{n-2\ep} \, C_2(0,n) = - \frac{2N_1}{(n-2\ep)(n+1-2\ep)}
\Biggl[\frac{1}{\ep^2} + \frac{2}{\ep} \, S_1(n) 
  + 2\,S^2_1(n) +  4\, S_2(n) + 2  S_{-2}(n) \Biggr] \, .
\label{C2.1.a2}
\ee

Using the result for $C_{21}(1,n)$ of (\ref{C21.1.re}), we obtain the second contribution to $C_2(1,n)$ as
\be
- \frac{2}{n-2\ep} \, C_{21}(1,n) = \frac{2N_1}{(n-2\ep)(n+1-2\ep)}
\Biggl[\frac{1}{\ep^2} + \frac{2}{\ep} \, S_1(n) 
  + 2\,S^2_1(n) +  4\, S_2(n)  \Biggr] \, .
\label{C2.1.a3}
\ee

Since $K_1/K_2 \sim O(\ep^3)$ (see Eq. (\ref{6zeta3})) and, thus, $N_1/N_2 \sim O(\ep^3)$. So, the difference between
the normalizations  $N_1$ and $N_2$ in negligible for Eqs. (\ref{C2.1.a2}) and (\ref{C2.1.a3}).
Taking these two results together, we have for $C_2(1,n)$
\be
C_2(1,n) = - \frac{4S_{-2}(n)}{n(n+1)} + O(\ep) \, .
\label{C2.1.a4}
\ee

The results is convenient for $n\geq 1$.
\footnote{Really the result $C_1(1,n=0)=6\zeta_3$ can be obtained directly from Eq. (\ref{C2.1.a4})
  by using an analytic continuations of the sum $S_{-2}(n)$ (see Appendix C).}
For the case $n=0$ we can use the results for   $C_1(1,n=0)=6\zeta_3$, because $C_2(1,n=0)=C_1(1,n=0)$.

Using the fact that the above result (\ref{C2.1.a3}) is obtained for even $n$ values (and\\ $C_2(1,2m+1)=0$)
, we can recover the
full resul for $C_2(1,n)$ in the following form
\be
C_2(1,n) = (1+(-1)^n) \, \left(3 \delta^0_n \zeta_3 -(1-\delta^0_n) \frac{2S_{-2}(n)}{n(n+1)} \right) \, .
\label{C2.1.a5}
\ee

\section{Examples of calculation of some type of massless diagrams}

In the previous section, we studied the basic diagrams $I_1(\alpha,n)$ and $I_1(\alpha,n)$ with $\alpha=0$ and $\alpha=1$.
I would especially like to note the case  $\alpha=1$ where the coefficient functions of the integrals $I_1(1,n)$ and $I_1(1,n)$
have a very compact form.

In this section, we consider the expansion coefficients of scalar diagrams (we call them the "moments" of diagrams),
arising in the study of forward elastic scattering.
These moments are extracted from the initial diagrams by the method of "projectors"
\cite{Gorishnii:1983su,Tkachov:1983st}. Some basic diagrams and getting their points
are discussed in Appendix B.

Here we look at the diagrams $I_1(\alpha,n)$ and $I_1(\alpha,n)$ with $\alpha=n+1$ and $\alpha=n+2$.

\subsection{$I_1(n+1,n)$ and $I_2(n+1,n)$}

\vskip 1cm
\begin{figure}[htbp]
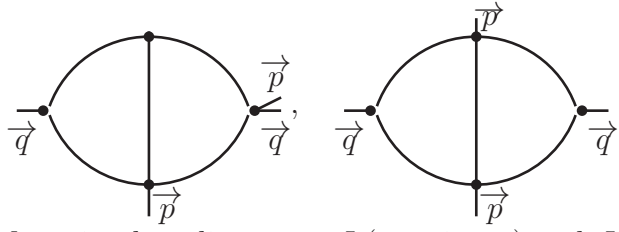

\centerline{
    \begin{axopicture}(90,10)(0,4)
  \SetWidth{1.0}
\Arc(45,-7)(40,20,90)
\Arc(45,-7)(40,90,160)
\Line(45,-25)(45,35)
\Arc(45,17)(40,200,340)
\Vertex(45,-23){2}
\Vertex(45,33){2}
\Vertex(5,5){2}
\Vertex(85,5){2}
\Line(5,5)(-5,5)
\Line(85,5)(95,5)
\Line(45,-25)(45,-35)
\Line(85,5)(95,10)
\Text(-3,-5)[b]{$\to$}
\Text(-3,-12)[b]{$q$}
\Text(93,-5)[b]{$\to$}
\Text(93,-12)[b]{$q$}
\Text(93,20)[b]{$\to$}
\Text(93,12)[b]{$p$}
\Text(52,-30)[b]{$\to$}
\Text(52,-38)[b]{$p$}
\end{axopicture}
\hspace{3mm}, \hspace{5mm}
    \begin{axopicture}(90,10)(0,4)
  \SetWidth{1.0}
\Arc(45,-7)(40,20,90)
\Arc(45,-7)(40,90,160)
\Line(45,-25)(45,35)
\Arc(45,17)(40,200,340)
\Vertex(45,-23){2}
\Vertex(45,33){2}
\Vertex(5,5){2}
\Vertex(85,5){2}
\Line(5,5)(-5,5)
\Line(85,5)(95,5)
\Line(45,33)(45,40)
\Line(45,-25)(45,-35)
\Text(-3,-5)[b]{$\to$}
\Text(-3,-12)[b]{$q$}
\Text(93,-5)[b]{$\to$}
\Text(93,-12)[b]{$q$}
\Text(50,40)[b]{$\to$}
\Text(50,35)[b]{$p$}
\Text(52,-30)[b]{$\to$}
\Text(52,-38)[b]{$p$}
\end{axopicture}
\hspace{3mm}
}
\vskip 1cm
\caption{Most simplest diagrams: $J_1(\alpha=1,q,p)$ and $J_2(\alpha=1,q,p)$.}
\label{sunsetMMm}
\end{figure}

\vskip 0.5cm

Let us first consider the two simplest diagrams: $J_1(\alpha=1,q,p)$ and $J_2(\alpha=1,q,p)$,
shown in Fig. 2. As it was discussed already,
using the method of ``projectors'' (see Appendix B) we can study the so-called moments
$J_i(\alpha=1,n)$ $(i=1,2)$. 
The moments of the diagrams shown in Fig. 2 are
represented by the diagrams shown in Eq. (\ref{I1+2})
for $\alpha = n+1$, i.e. $J_i(\alpha=1,n)=I_i(n+1,n)$

To evaluate the diagrams it is convenient to use the so-called momentum transformation. To show it's
effectivity it is useful to consider more complicated diagrams $I_i(\alpha+n,n)$ $(i=1,2)$, which have the
coefficient functions $C_i(\alpha+n,n)$
\be
I_{i} (\alpha+n,n) = \frac{1}{(4\pi)^{d}} \, C_{i} (\alpha+n,n) \, \frac{q^{\mu_1 ... \mu_n}}{q^{2(\alpha +n + 2\ep)}} \, ,
  \label{Cialpha+n}
  \ee
  which is similar to (\ref{Cialpha}).

  Now we introduce the following Fourier transforms \cite{Kotikov:2018wxe}
  \bea
 && \int \, d^dp \, e^{ipx} \, \frac{1}{p^{2\alpha}} = 2^{2\tilde{\alpha}} \, \pi^{d/2} \, a_0(\alpha) \,
  \frac{1}{x^{2\tilde{\alpha}}} \, ,~~ (\tilde{\alpha}=\frac{d}{2}-\alpha, \mbox{see Eq. (\ref{Anm})}) \, , \label{FT} \\
 &&\int \, d^dp \, e^{ipx} \, \frac{p^{\mu_1} ... p^{\mu_n}}{p^{2(\alpha+n)}} = (-i)^n \, 2^{2\tilde{\alpha}} \, \pi^{d/2} \, a_n(\alpha+n) \,
  \frac{p^{\mu_1} ... p^{\mu_n}}{x^{2\tilde{\alpha}}} + ... \, ,\label{FT} 
 \label{FTn}
  \eea
where
we
neglect the terms of the order $g^{\mu_i \mu_j}$.

The  Fourier transform (\ref{FT}) is the usual one (see, for example, the recent review \cite{Kotikov:2018wxe})
and the Fourier transform (\ref{FTn})
can be obtained from (\ref{FT}) with the help of projector
$\partial^{\mu_1}/(\partial x)^{\mu_1} ... \partial^{\mu_n}/(\partial x)^{\mu_n}$. 

Applijing  the Fourier transforms to l.h.s. we will come to the following diagrams in $x$-space:
\vskip 1cm
\bea
\raisebox{1mm}
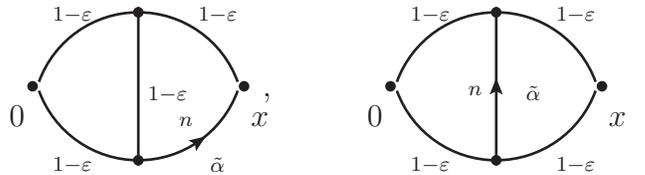
{{
    \begin{axopicture}(90,10)(0,4)
  \SetWidth{1.0}
\Arc(45,-7)(40,20,90)
\Arc(45,-7)(40,90,160)
\Line(45,-25)(45,35)
\Arc(45,17)(40,200,270)
\Arc[arrow](45,17)(40,270,340)
\Vertex(45,-23){2}
\Vertex(45,33){2}
\Vertex(5,5){2}
\Vertex(85,5){2}
\Text(63,-10)[b]{$\scriptstyle n$}
\Text(75,-22)[t]{$\scriptstyle \tilde{\alpha}$}
\Text(75,35)[t]{$\scriptstyle 1-\ep$}
\Text(20,-22)[t]{$\scriptstyle 1-\ep$}
\Text(20,35)[t]{$\scriptstyle 1-\ep$}
\Text(56,5)[t]{$\scriptstyle 1-\ep$}
\Text(-1,-10)[b]{$0$}
\Text(91,-10)[b]{$x$}
\end{axopicture}
}}
\, , \hspace{3mm} \hspace{9mm}
\raisebox{1mm}{{
    \begin{axopicture}(90,10)(0,4)
  \SetWidth{1.0}
\Arc(45,-7)(40,20,90)
\Arc(45,-7)(40,90,160)
\Line[arrow](45,-25)(45,35)
\Arc(45,17)(40,200,270)
\Arc(45,17)(40,270,340)
\Vertex(45,-23){2}
\Vertex(45,33){2}
\Vertex(5,5){2}
\Vertex(85,5){2}
\Text(37,5)[t]{$\scriptstyle n$}
\Text(75,-22)[t]{$\scriptstyle 1-\ep$}
\Text(75,35)[t]{$\scriptstyle 1-\ep$}
\Text(20,-22)[t]{$\scriptstyle 1-\ep$}
\Text(20,35)[t]{$\scriptstyle 1-\ep$}
\Text(59,5)[t]{$\scriptstyle \tilde{\alpha}$}
\Text(-1,-10)[b]{$0$}
\Text(91,-10)[b]{$x$}
\end{axopicture}
}}
\hspace{3mm}
 \, .
\nonumber
\eea
\vskip 1cm

If we replace $x$ and all internal coordinats by the corresponding moment $q$ and the internal moments we
obtain the corresponding diagram in the momentum space. We will call them as $\overline{I}_{i} (\tilde{\alpha},n)$.
Such procedure is called by ``dual transformation'' (see discussions in Ref. \cite{Kotikov:2018wxe}) and
denoted as $\stackrel{d}{=}$ (see aslo Introduction). So, we have
\vskip 1cm
\bea
\raisebox{1mm}
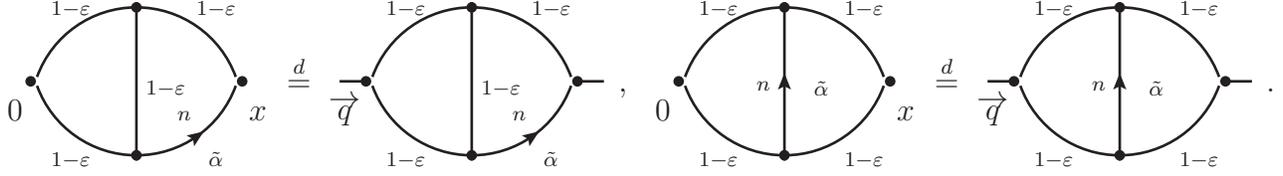
{{
    \begin{axopicture}(90,10)(0,4)
  \SetWidth{1.0}
\Arc(45,-7)(40,20,90)
\Arc(45,-7)(40,90,160)
\Line(45,-25)(45,35)
\Arc(45,17)(40,200,270)
\Arc[arrow](45,17)(40,270,340)
\Vertex(45,-23){2}
\Vertex(45,33){2}
\Vertex(5,5){2}
\Vertex(85,5){2}
\Text(63,-10)[b]{$\scriptstyle n$}
\Text(75,-22)[t]{$\scriptstyle \tilde{\alpha}$}
\Text(75,35)[t]{$\scriptstyle 1-\ep$}
\Text(20,-22)[t]{$\scriptstyle 1-\ep$}
\Text(20,35)[t]{$\scriptstyle 1-\ep$}
\Text(56,5)[t]{$\scriptstyle 1-\ep$}
\Text(-1,-10)[b]{$0$}
\Text(91,-10)[b]{$x$}
\end{axopicture}
}}
\hspace{3mm} \stackrel{d}{=}
\hspace{3mm}
\raisebox{1mm}{{
    \begin{axopicture}(90,10)(0,4)
  \SetWidth{1.0}
\Arc(45,-7)(40,20,90)
\Arc(45,-7)(40,90,160)
\Line(45,-25)(45,35)
\Arc(45,17)(40,200,270)
\Arc[arrow](45,17)(40,270,340)
\Vertex(45,-23){2}
\Vertex(45,33){2}
\Vertex(5,5){2}
\Vertex(85,5){2}
\Line(5,5)(-5,5)
\Line(85,5)(95,5)
\Text(63,-10)[b]{$\scriptstyle n$}
\Text(75,-22)[t]{$\scriptstyle \tilde{\alpha}$}
\Text(75,35)[t]{$\scriptstyle 1-\ep$}
\Text(20,-22)[t]{$\scriptstyle 1-\ep$}
\Text(20,35)[t]{$\scriptstyle 1-\ep$}
\Text(56,5)[t]{$\scriptstyle 1-\ep$}
\Text(-3,-5)[b]{$\to$}
\Text(-3,-12)[b]{$q$}
\end{axopicture}
}}
\hspace{3mm}
\, ,
\hspace{3mm}
\raisebox{1mm}{{
    \begin{axopicture}(90,10)(0,4)
  \SetWidth{1.0}
\Arc(45,-7)(40,20,90)
\Arc(45,-7)(40,90,160)
\Line[arrow](45,-25)(45,35)
\Arc(45,17)(40,200,270)
\Arc(45,17)(40,270,340)
\Vertex(45,-23){2}
\Vertex(45,33){2}
\Vertex(5,5){2}
\Vertex(85,5){2}
\Text(37,5)[t]{$\scriptstyle n$}
\Text(75,-22)[t]{$\scriptstyle 1-\ep$}
\Text(75,35)[t]{$\scriptstyle 1-\ep$}
\Text(20,-22)[t]{$\scriptstyle 1-\ep$}
\Text(20,35)[t]{$\scriptstyle 1-\ep$}
\Text(59,5)[t]{$\scriptstyle \tilde{\alpha}$}
\Text(-1,-10)[b]{$0$}
\Text(91,-10)[b]{$x$}
\end{axopicture}
}}
\hspace{3mm} \stackrel{d}{=}
\hspace{3mm}
\raisebox{1mm}{{
    \begin{axopicture}(90,10)(0,4)
  \SetWidth{1.0}
\Arc(45,-7)(40,20,90)
\Arc(45,-7)(40,90,160)
\Line[arrow](45,-25)(45,35)
\Arc(45,17)(40,200,270)
\Arc(45,17)(40,270,340)
\Vertex(45,-23){2}
\Vertex(45,33){2}
\Vertex(5,5){2}
\Vertex(85,5){2}
\Line(5,5)(-5,5)
\Line(85,5)(95,5)
\Text(37,5)[t]{$\scriptstyle n$}
\Text(75,-22)[t]{$\scriptstyle 1-\ep$}
\Text(75,35)[t]{$\scriptstyle 1-\ep$}
\Text(20,-22)[t]{$\scriptstyle 1-\ep$}
\Text(20,35)[t]{$\scriptstyle 1-\ep$}
\Text(59,5)[t]{$\scriptstyle \tilde{\alpha}$}
\Text(-3,-5)[b]{$\to$}
\Text(-3,-12)[b]{$q$}
\end{axopicture}
}}
\hspace{3mm}
\, .
\nonumber
\eea
\vskip 1cm

Lets to denote the coefficient functions of the last integral as $\overline{C}_i(\tilde{\alpha},n)$, i.e.
\be
\overline{I}_{i} (\tilde{\alpha},n) = \frac{1}{(4\pi)^{d}} \, \overline{C}_{i} (\tilde{\alpha},n) \, \frac{q^{\mu_1 ... \mu_n}}{q^{2(\tilde{\alpha}  - 2\ep)}} \, .
  \label{oCialpha+n}
  \ee

  Taking the Fourier transforms for the both side of Eq. (\ref{Cialpha+n}) (see  Ref. \cite{Kotikov:2018wxe}),
  we obtain the relations between the
  coefficient functions $C_{i} (\alpha+n,n)$ and $\overline{C}_{i} (\tilde{\alpha},n)$ 
  \be
  C_{i} (\alpha+n,n) = K(\alpha+n,n) \, \overline{C}_{i} (\tilde{\alpha},n) \, ,
  \label{Cialpha+n.1}
  \ee
  where
  \be
  K(\alpha+n,n) = \frac{a_0^4(1)a_n(\alpha+n)}{a_n(\alpha+n+2\ep)} =
  \frac{\Gamma^4(1-\ep)\Gamma(\tilde{\alpha})\Gamma(\alpha+n+2\ep)}{\Gamma(\alpha+n)\Gamma(\tilde{\alpha}-2\ep)} \, .
\label{Kalpha+n}
  \ee

  Now return to the case $\alpha=1$. 
  With the accuracy $O(\ep)$,
  \be
  \overline{C}_{i} (1-\ep,n) \approx C_{i} (1,n), ~~
  K(n+1,n) =\frac{\Gamma^5(1-\ep)\Gamma(n+1+2\ep)}{n!\Gamma(1-3\ep)} \approx 1 \, ,
\label{oCialpha+n.1}
\ee
where the sign $\approx$ means equality up to $0(\ep^0)$.

Thus, we have from (\ref{Cialpha+n.1})
\be
C_{i} (n+1,n) = C_{i} (1,n) + O(\ep) \, ,
\label{Cialpha+n.1}
\ee
i.e.
\bea
&&C_1(n+1,n) = \frac{1}{n+1} \, \Bigl(S_3(n)+S_1(n)S_2(n)-S_{2,1}(n)+6\zeta_3 \Bigr) + O(\ep), \label{C1.n+1} \\
&&C_2(n+1,n) = (1+(-1)^n) \, \left(3 \delta^0_n \zeta_3 -(1-\delta^0_n) \frac{2S_{-2}(n)}{n(n+1)}+ O(\ep)  \right) \, .
\label{C2.n+1}
\eea
\vskip 0.3cm

\subsection{$I_1(n+2,n)$ and $I_2(n+2,n)$}

Consider the more complex diagrams $\overline{J}_1(\alpha=1,q,p)$ and $\overline{J}_2(\alpha=1,q,p)$ shown in Fig. 3.
\vskip 1cm
\begin{figure}[htbp]
\centerline{
    \begin{axopicture}(90,10)(0,4)
  \SetWidth{1.0}
\Arc(45,-7)(40,20,90)
\Arc(45,-7)(40,90,160)
\Line(45,-25)(45,35)
\Arc(45,17)(40,200,340)
\Vertex(45,-23){2}
\Vertex(45,33){2}
\Vertex(5,5){2}
\Vertex(85,5){2}
\Vertex(65,-18){2}
\Line(5,5)(-5,5)
\Line(85,5)(95,5)
\Line(65,-18)(65,-28)
\Line(85,5)(95,10)
\Text(-3,-5)[b]{$\to$}
\Text(-3,-12)[b]{$q$}
\Text(93,-5)[b]{$\to$}
\Text(93,-12)[b]{$q$}
\Text(93,20)[b]{$\to$}
\Text(93,12)[b]{$p$}
\Text(72,-25)[b]{$\to$}
\Text(72,-32)[b]{$p$}
\end{axopicture}
\hspace{3mm}, \hspace{5mm}
    \begin{axopicture}(90,10)(0,4)
  \SetWidth{1.0}
\Arc(45,-7)(40,20,90)
\Arc(45,-7)(40,90,160)
\Line(45,-25)(45,35)
\Arc(45,17)(40,200,340)
\Vertex(45,-23){2}
\Vertex(45,33){2}
\Vertex(5,5){2}
\Vertex(85,5){2}
\Vertex(45,10){2}
\Line(5,5)(-5,5)
\Line(85,5)(95,5)
\Line(45,33)(45,40)
\Line(45,10)(55,10)
\Text(-3,-5)[b]{$\to$}
\Text(-3,-12)[b]{$q$}
\Text(93,-5)[b]{$\to$}
\Text(93,-12)[b]{$q$}
\Text(50,40)[b]{$\to$}
\Text(50,35)[b]{$p$}
\Text(57,5)[b]{$\to$}
\Text(57,-2)[b]{$p$}
\end{axopicture}
\hspace{3mm}
}
\vskip 1cm
\caption{The diagrams $\overline{J}_1(\alpha=1,q,p)$ and $\overline{J}_2(\alpha=1,q,p)$.}
\label{sunsetMMm}
\end{figure}

\vskip 0.5cm

Their moments $\overline{J}_i(\alpha=1,n)$ $(i=1,2)$
are represented by the diagrams in Eq. (\ref{I1+2}) with $\alpha=n+2$, i.e. by the diagrams
$I_i(n+2,n)$.
Their
coefficient functions $C_i(n+2,n)$ $(i=1,2)$ can be expressed throught the ones $C_i(n+1,n)$ given in eqs. (\ref{C1.n+1}) and (\ref{C2.n+1})
and $C_{i1}(n+2,n)$ given in eqs. (\ref{C11alpha}) and (\ref{C21alpha}) at $\alpha=n+2$.

Performing calculations, we have
the results for the $C_{11}(1,n)$ and $C_{21}(1,n)$ are
\bea
&&\frac{C_{11}(n+2,n)}{N_2/(2\ep^2)} =
\frac{2K_1}{K_2} \, \frac{B(n+1,1,0)}{n+1} -
  \frac{B(n+1,2,1)}{n+1+\ep} = \frac{2K_1}{K_2} \, \frac{B(n+2,1,0)}{n+1+\ep} -
  \frac{B(n+2,2,1)}{n+1+2\ep}
\, ,
\label{C11n+2.1} \\
&&\frac{C_{2,1}(n+2,n)}{N_2/(2\ep^2)} =
\frac{B(n+1,1,0)}{n+1+\ep}  \, B(n+1,2,1) - 3 \frac{B(n+1,1,0)}{n+1}  \, B(n+2,2,1) \, ,
\label{C21n+2.1}
\eea
where the normalization factors $N_1$ and $N_2$ and also the factors $K_1$ and $K_2$ are given in
Eqs. (\ref{N1}) and (\ref{N2}), respectively.

Since
\be
 B(n+2,2,1) =  B(n+2,2,1) \, \frac{n+1+2\ep}{n+1+\ep}~~\mbox{and}~~ B(n+1,1,0) B(n+1,2,1)=B(n+1,2,0) \, ,
 \label{Bn+2}
 \ee
 then
 \be
 \frac{C_{2,1}(n+2,n)}{N_2/(2\ep^2)} = -2 \,
 \frac{B(n+1,1,0)}{n+1+\ep}  \, \frac{n+1+3\ep}{n+1} = -2 \,
 \frac{B(n+2,1,0)}{n+1+\ep}  \, \frac{n+1+3\ep}{n+1+2\ep} \, .
\label{C21n+2.2}
\ee

Expanding all $B(n+2,...)$, we have
\bea
&&(n+1)\frac{C_{11}(n+2,n)}{N_2/(2\ep^2)} =
1+ \ep S_1(n+1) + \ep^2 \left[\frac{1}{2} \bigl(S_1^2(n+1)+S_2(n+1)\bigr) - \frac{1}{(n+1)^2}\right]
\nonumber \\
&&+ \ep^3\Biggl[
    \frac{1}{6} \bigl(S_1^3(n+1)+3S_1(n+1)S_2(n+1)-10S_3(n+1)\bigr) + 12\zeta_3
  - \frac{2S_2(n+1)}{n+1} \nonumber \\
  &&  - \frac{2S_1(n+1)}{(n+1)^2} + \frac{6}{(n+1)^3}\Biggr]
  + O(\ep^4) 
\, ,
\label{C11n+2.3} \\
&&(n+1)\frac{C_{21}(n+2,n)}{N_2/(2\ep^2)} = -2 \Biggl(
1+ 2\ep S_1(n+1) + 2\ep^2 \left[S_1^2(n+1)-S_2(n+1) - \frac{1}{(n+1)^2}\right]
\nonumber \\
&&+ 2\ep^3 \left[\frac{2}{3} \bigl(S_1^3(n+1)-3S_1(n+1)S_2(n+1)+2S_3(n+1)\bigr)
  - \frac{2S_1(n+1)}{(n+1)^2} + \frac{3}{(n+1)^3}\right] \Biggr)\nonumber \\
&&+ O(\ep^4)
\, .
\label{C21n+2.3}
\eea

Taking the results (\ref{C11n+2.3}) and (\ref{C21n+2.3}) together with the results for $C_i(n+1,n)$
($i=1,2$) in Eq. (\ref{C1.n+1}) and (\ref{C2.n+1}), we have the final results
\bea
&&(n+1) \frac{C_1(n+2,n)}{N_2} = \frac{1}{2\ep^2} \, S_1(n+1) +\frac{1}{4\ep} \,
\left[3 S_1^2(n+1) - 5 S_2(n+1)\right] + \frac{7}{12} \, S_1^3(n+1) \nonumber \\
&& - \frac{5}{4} \, S_1(n+1)S_2(n+1) +  \frac{19}{6} \, S_3(n+1)
- 2  S_{2,1}(n+1) - \frac{S_1(n+1)}{(n+1)^2} \, , \label{C1.n+2} \\
&&(n+1) \frac{C_2(n+2,n)}{N_2} = \frac{1+(-1)^n}{n+2} \, \Biggl(\frac{1}{\ep^2} + \frac{2}{\ep} \,
\left[S_1(n+1)  - \frac{1}{n+2} \right]  \nonumber \\
&&+ 2  S_1^2(n+1) -2 S_2(n+1) - 4 \, \frac{S_1(n+1)}{n+2} + \frac{1}{(n+2)^2} \Biggr) \, , \label{C2.n+2}
\eea
where we added the additional factor $(1+(-1)^n)/2$ to the coefficient function $C_2(n+2,n)$.

Consideration of the more complicated examples can be found in Refs. \cite{Kazakov:1986mu,Kotikov:1987mw}.

We only note here that for diagrams containing several propagators depending on the momentum $p$ (see, for example,
$\hat{J}_1(\alpha,\beta,q,p)$ in Appendix B), their moments
will contain the sum of two-point diagrams (see the moment $\hat{J}_1(\alpha,\beta,n)$ in Appendix B).

Calculation of moments of this type is a much more serious problem compared to $I_i(n+1,n)$ and
$I_i(n+2,n)$ with $(i=1,2)$.
Nevertheless, as was shown in Ref. \cite{Kotikov:1987mw}, it is almost always possible to separate the
contributions of complex integrals and
complex sums. Moreover, $\ep$-singularities remain only in the simplest parts, where
it is usually possible to sum in all orders
in $\ep$.



\section{Calculation of massive Feynman integrals}

Feynman integrals with massive propagators are significantly more complex objects compared to the
massless case. The basic rules for calculating such diagrams are discussed in Section 2, which are
supplemented by new ones containing directly massive propagators. These additional rules are discussed
in the next subsection.

\subsection{Basic formulas}
Let us briefly consider the rules for calculating diagrams
with the massive propagators.

Propagator with mass $M$ will be represented as
\be
\frac{1}{(q^2+M^2)^{\alpha}}
= \hspace{3mm} \raisebox{1mm}{{
\begin{axopicture}(70,30)(0,4)
  \SetWidth{2.0}
\Line(5,5)(65,5)
\SetWidth{1.0}
\Vertex(5,5){2}
\SetWidth{1.0}
\Vertex(65,5){2}
\Line(5,5)(-5,5)
\Line(65,5)(75,5)
\Text(33,10)[b]{$M$}
\Text(33,-1)[t]{$\alpha$}
\Text(-3,-5)[b]{$\to$}
\Text(-3,-12)[b]{$q$}
\end{axopicture}
}}
\hspace{3mm}
.~~
\label{DefM}
\ee

The following formulas hold.\\

{\bf A.}~~For simple chain of two massive propagators with the same mass:
\bea
&&\frac{1}{(q^2+M^2)^{\alpha_1}} \, \frac{1}{(q^2+M^2)^{\alpha_2}} =  \frac{1}{(q^2+M^2)^{(\alpha_1+\alpha_2)}}
\, , 
\nonumber \\
&& \hspace{-4cm} \mbox{or graphically} \nonumber \\
&&\raisebox{1mm}{{
\begin{picture}(90,30)(0,4)
  \SetWidth{1.0}
\Vertex(5,5){2}
\Vertex(5,5){2}
\Line(5,5)(-5,5)
\Line(85,5)(95,5)
\Vertex(40,5){2}
\Vertex(85,5){2}
\SetWidth{2.0}
\Line(5,5)(40,5)
\Line(40,5)(85,5)
 \SetWidth{1.0}
\Text(15,10)[b]{$M$}
\Text(65,10)[b]{$M$}
\Text(15,-1)[t]{$\alpha_1$}
\Text(65,-1)[t]{$\alpha_2$}
\Text(-3,-5)[b]{$\to$}
\Text(-3,-12)[b]{$q$}
\end{picture}
}}
\hspace{3mm}
=
\hspace{3mm} \raisebox{1mm}{{
\begin{picture}(70,30)(0,4)
  \SetWidth{2.0}
  \Line(5,5)(65,5)
  \SetWidth{1.0}
\Vertex(5,5){2}
\SetWidth{1.0}
\Vertex(65,5){2}
\Line(5,5)(-5,5)
\Line(65,5)(75,5)
\Text(33,10)[b]{$M$}
\Text(33,-1)[t]{$\alpha_1+\alpha_2$}
\Text(-3,-5)[b]{$\to$}
\Text(-3,-12)[b]{$q$}
\end{picture}
}}
\hspace{3mm}
,
\label{chainM}
\eea
i.e. the product of propagators with the same mass $M$ is equivalent to a new propagator with the mass $M$
and an
index equal to the sum of the indices of the original propagators.\\

{\bf B.}~~ Massive tadpole is integrated as 
\be
\int \frac{Dk }{k^{2\alpha_1}(k^2+M^2)^{\alpha_2}} = \frac{1}{(4\pi)^{d/2}} \,
\frac{R(\alpha_1,\alpha_2)}{M^{2(\alpha_1+\alpha_2-d/2)}} \, ,
\nonumber
\ee
where
\be
R(\alpha,\beta) = \frac{\Gamma(d/2-\alpha_1)\Gamma(\alpha_1+\alpha_2-d/2)}{\Gamma(d/2)\Gamma(\alpha_2)}
 \, .
\label{R}
\ee

{\bf C.}~~ A simple loop of two massive propagators with masses $M_1$ and $M_1$ can be represented
as hypergeometric function, which can be calculated in a general form, for example, by Feynman-parameter
method.
It is very conveninet, using this approach to represent the loop as an integral of a propagator
with the ``effective mass'' $\mu$ \cite{FleKoVe,Kotikov:1990kg,Kniehl:2005bc,Kniehl:2005yc}:
\bea
&&(4\pi)^{d/2} \times \int \frac{Dk }{[(q-k)^2+M_1^2]^{\alpha_1}[k^2+M_2^2]^{\alpha_2}} \nonumber \\
&&=
\frac{\Gamma(\alpha_1+\alpha_2-d/2)}{\Gamma(\alpha_1)\Gamma(\alpha_2)} \,
\int_0^1 \, \frac{ds \, s^{\alpha_1-1} \, (1-s)^{\alpha_2-1} }{[s(1-s)q^2+M_1^2s + M_2^2(1-s)]^{\alpha_1+\alpha_2-d/2}}
\nonumber \\
&&= 
\frac{\Gamma(\alpha_1+\alpha_2-d/2)}{\Gamma(\alpha_1)\Gamma(\alpha_2)} \,
\int_0^1 \, \frac{ds}{s^{1-\tilde{\alpha}_2} \, (1-s)^{1-\tilde{\alpha}_1} } \,
\frac{1}{[q^2+\mu^2]^{\alpha_1+\alpha_2-d/2}},~~
\left(\mu^2 = \frac{M_1^2}{1-s} +  \frac{M_2^2}{s}\right) \, .
\nonumber
\eea

It is useful
to rewrite the equation graphically as
\vskip 0.5cm
\bea
\raisebox{1mm}{{
    \begin{axopicture}(90,10)(0,4)
  \SetWidth{2.0}
\Arc(45,-7)(40,20,160)
\Arc(45,17)(40,200,340)
 \SetWidth{1.0}
\Vertex(5,5){2}
\Vertex(85,5){2}
\Line(5,5)(-5,5)
\Line(85,5)(95,5)
\Text(45,-16)[b]{$M_2$}
\Text(45,40)[b]{$M_1$}
\Text(45,27)[t]{$\alpha_1$}
\Text(45,-29)[t]{$\alpha_2$}
\Text(-3,-5)[b]{$\to$}
\Text(-3,-12)[b]{$q$}
\end{axopicture}
}}
\hspace{3mm}
= \frac{1}{(4\pi)^{d/2}} \,
\frac{\Gamma(\alpha_1+\alpha_2-d/2)}{\Gamma(\alpha_1)\Gamma(\alpha_2)} \,
\int_0^1 \, \frac{ds}{s^{1-\tilde{\alpha}_2} \, (1-s)^{1-\tilde{\alpha}_1} } \,
\hspace{3mm} \raisebox{1mm}{{
\begin{picture}(70,30)(0,4)
  \SetWidth{2.0}
\Line(5,5)(65,5)
\SetWidth{1.0}
\Vertex(5,5){2}
\Vertex(65,5){2}
\Line(5,5)(-5,5)
\Line(65,5)(75,5)
\Text(33,10)[b]{$\mu$}
\Text(33,-1)[t]{$\scriptstyle \alpha_1+\alpha_2-d/2$}
\Text(-3,-5)[b]{$\to$}
\Text(-3,-12)[b]{$q$}
\end{picture}
}}
\hspace{3mm}
\, .
\label{loopM}
 \eea

\vskip 1.2cm

{\bf D.}~~ For any triangle with indices
$\alpha_i$ ($i=1,2,3$) and masses $M_i$  there is the following relation, which is based on
integration by parts (IBP) procedure \cite{Chetyrkin:1981qh,Vasiliev:1981dg,Kotikov:1990kg}
%
\vskip 1cm

\bea
&&(d-2\alpha_1-\alpha_2-\alpha_3) \hspace{0.5cm}
\raisebox{1mm}{{
    \begin{axopicture}(90,10)(0,4)
  \SetWidth{2.0}
\Line(5,5)(45,45)
\Line(5,5)(85,5)
\Line(45,45)(85,5)
\SetWidth{1.0}
\Vertex(5,5){2}
\Vertex(85,5){2}
\Vertex(45,45){2}
\Line(5,5)(-5,5)
\Line(85,5)(90,5)
\Line(45,45)(55,60)
\Text(45,10)[b]{$\scriptstyle M_1$}
\Text(20,30)[b]{$\scriptstyle M_2$}
\Text(70,30)[b]{$\scriptstyle M_3$}
\Text(35,25)[t]{$\scriptstyle \alpha_2$}
\Text(45,-2)[t]{$\scriptstyle \alpha_1$}
\Text(55,25)[t]{$\scriptstyle \alpha_3$}
\Text(-3,-5)[b]{$\to$}
\Text(-3,-12)[b]{$\scriptstyle q_2-q_1$}
\Text(93,-5)[b]{$\to$}
\Text(93,-12)[b]{$\scriptstyle q_1-q_3$}
\Text(70,50)[b]{$\to$}
\Text(70,40)[b]{$\scriptstyle q_3-q_2$}
\end{axopicture}
}}
\hspace{3mm} \nonumber \\
&&\nonumber \\
&&\nonumber \\
&&\nonumber \\
&&= \alpha_2 \biggl[ \, \hspace{0.5cm} \,
  \raisebox{1mm}{{
    \begin{axopicture}(90,10)(0,4)
  \SetWidth{2.0}
  \Line(5,5)(45,45)
\Line(5,5)(85,5)
\Line(45,45)(85,5)
\SetWidth{1.0}
\Vertex(5,5){2}
\Vertex(85,5){2}
\Vertex(45,45){2}
\Line(5,5)(-5,5)
\Line(85,5)(90,5)
\Line(45,45)(55,60)
\Text(45,10)[b]{$\scriptstyle M_1$}
\Text(20,30)[b]{$\scriptstyle M_2$}
\Text(70,30)[b]{$\scriptstyle M_3$}
\Text(35,25)[t]{$\scriptstyle \alpha_2+1$}
\Text(45,-2)[t]{$\scriptstyle \alpha_1-1$}
\Text(55,25)[t]{$\scriptstyle \alpha_3$}
\Text(-3,-5)[b]{$\to$}
\Text(-3,-12)[b]{$\scriptstyle q_2-q_1$}
\Text(93,-5)[b]{$\to$}
\Text(93,-12)[b]{$\scriptstyle q_1-q_3$}
\Text(70,50)[b]{$\to$}
\Text(70,40)[b]{$\scriptstyle q_3-q_2$}
\end{axopicture}
}}
\hspace{3mm}
- \biggl[(q_2-q_1)^2 +M_1^2 +M_2^2 \biggr] \times
  \raisebox{1mm}{{
    \begin{axopicture}(90,10)(0,4)
  \SetWidth{2.0}
\Line(5,5)(45,45)
\Line(5,5)(85,5)
\Line(45,45)(85,5)
 \SetWidth{1.0}
\Vertex(5,5){2}
\Vertex(85,5){2}
\Vertex(45,45){2}
\Line(5,5)(-5,5)
\Line(85,5)(90,5)
\Line(45,45)(55,60)
\Text(45,10)[b]{$\scriptstyle M_1$}
\Text(20,30)[b]{$\scriptstyle M_2$}
\Text(70,30)[b]{$\scriptstyle M_3$}
\Text(35,25)[t]{$\scriptstyle \alpha_2+1$}
\Text(45,-2)[t]{$\scriptstyle \alpha_1$}
\Text(55,25)[t]{$\scriptstyle \alpha_3$}
\Text(-3,-5)[b]{$\to$}
\Text(-3,-12)[b]{$\scriptstyle q_2-q_1$}
\Text(93,-5)[b]{$\to$}
\Text(93,-12)[b]{$\scriptstyle q_1-q_3$}
\Text(70,50)[b]{$\to$}
\Text(70,40)[b]{$\scriptstyle q_3-q_2$}
\end{axopicture}
}} 
\hspace{3mm}
  \hspace{3mm}
\Biggr]
\nonumber \\
&&\nonumber \\
&&\nonumber \\
&&\nonumber \\
&&
+ \alpha_3
\, \biggl[\alpha_2 \leftrightarrow \alpha_3, M_2 \leftrightarrow M_3 \biggr] 
-2M_1^2 \alpha_1 
\times
\raisebox{1mm}{{
    \begin{axopicture}(90,10)(0,4)
  \SetWidth{2.0}
\Line(5,5)(45,45)
\Line(5,5)(85,5)
\Line(45,45)(85,5)
 \SetWidth{1.0}
\Vertex(5,5){2}
\Vertex(85,5){2}
\Vertex(45,45){2}
\Line(5,5)(-5,5)
\Line(85,5)(90,5)
\Line(45,45)(55,60)
\Text(45,10)[b]{$\scriptstyle M_1$}
\Text(20,30)[b]{$\scriptstyle M_2$}
\Text(70,30)[b]{$\scriptstyle M_3$}
\Text(35,25)[t]{$\scriptstyle \alpha_2$}
\Text(45,-2)[t]{$\scriptstyle \alpha_1+1$}
\Text(55,25)[t]{$\scriptstyle \alpha_3$}
\Text(-3,-5)[b]{$\to$}
\Text(-3,-12)[b]{$\scriptstyle q_2-q_1$}
\Text(93,-5)[b]{$\to$}
\Text(93,-12)[b]{$\scriptstyle q_1-q_3$}
\Text(70,50)[b]{$\to$}
\Text(70,40)[b]{$\scriptstyle q_3-q_2$}
\end{axopicture}
}}
\hspace{5mm}
\, .
\label{TreIBPM}
\eea

\vskip 1cm

As it was in the massless case (see Eq. (\ref{TreIBP})),
Eq. (\ref{TreIBPM}) can been obtained by introducing the factor $(\partial/\partial k_{\mu}) \, (k-q_1)^{\mu}$ to the subintegral expression of the triangle
and using the integration by parts procedure as in Eq. (\ref{IBPpro}).

As it was in the massless case,
the line with the index $\alpha_1$ is distingulished. The contributions of the other lines are
same. So, we will call below the line with the index $\alpha_1$ as a ``distingulished line''. It is clear that a various choices of the distingulished line produce
different tipes of the IBP relations.


\subsection{Basic massive two-loop integrals}

The general topology of the two-loop two-point diagram, which cannot be expressed as a combination of loops and chanins is shown on Fig.4.
\begin{figure}[htbp]
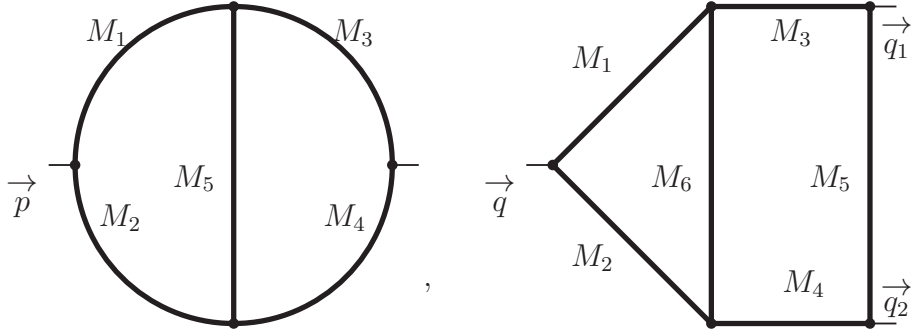

\centerline{
\begin{axopicture}(150,110)(0,4)
  \SetWidth{2.0}
\Arc(90,50)(60,0,90)
\Arc(90,50)(60,90,180)
\Arc(90,50)(60,180,270)
\Arc(90,50)(60,270,360)
\Line(90,-10)(90,110)
\SetWidth{0.7}
\Line(30,50)(20,50)
\Line(150,50)(160,50)
\Vertex(30,50){2}
\Vertex(150,50){2}
\Vertex(90,110){2}
\Vertex(90,-10){2}
  \SetWidth{1}
\Text(42,95)[b]{$M_1$}
\Text(135,95)[b]{$M_3$}
\Text(47,25)[b]{$M_2$}
\Text(132,25)[b]{$M_4$}
\Text(75,50)[t]{$M_5$}
\Text(10,40)[b]{$\to$}
\Text(10,30)[b]{$p$}
\end{axopicture}
%
%
~~ ,~~~
\begin{axopicture}(150,110)(0,4)
  \SetWidth{2.0}
\Line(30,50)(90,110)
\Line(30,50)(90,-10)
\Line(90,-10)(90,110)
\Line(90,110)(150,110)
\Line(90,-10)(150,-10)
\Line(150,-10)(150,110)
\SetWidth{0.7}
\Line(30,50)(20,50)
\Line(150,110)(160,110)
\Line(150,-10)(160,-10)
\Vertex(30,50){2}
\Vertex(150,110){2}
\Vertex(150,-10){2}
\Vertex(90,110){2}
\Vertex(90,-10){2}
  \SetWidth{1}
\Text(45,85)[b]{$M_1$}
\Text(120,95)[b]{$M_3$}
\Text(45,10)[b]{$M_2$}
\Text(125,0)[b]{$M_4$}
\Text(135,50)[t]{$M_5$}
\Text(75,50)[t]{$M_6$}
\Text(10,40)[b]{$\to$}
\Text(160,100)[b]{$\to$}
\Text(160,5)[t]{$\to$}
\Text(10,30)[b]{$q$}
\Text(160,90)[b]{$q_1$}
\Text(160,0)[t]{$q_2$}
\end{axopicture}
}
\vspace{5mm}
\caption{Two-loop two-point diagram $I(M_1,M_2,M_3,M_4,M_5)$ and three-point diagram $P(M_1,M_2,M_3,M_4,M_5,M_6)$ with $q_1^2=q_2^2=0$.
}
\label{sunsetMMm}
\end{figure}

\vskip 0.5cm


Below in the present analysis we will concentrate mostly on two-loop two-point and three-point
diagrams,
which can be taken from 
the diagram shown in Fig. 4.
We will call them as:
\bea
&&I_j = I(M_j=M \neq 0, M_p=0, p \neq j),~~ I_{ij} = I(M_i=M_j=M \neq 0, M_p=0, p \neq i \neq j), \nonumber \\
&&I_{ijs} = I(M_i=M_j=M_s=M \neq 0, M_p=0, p\neq i \neq j \neq s), \nonumber \\
&&I_{ijst} = I(M_i=M_j=M_s=M_t=M \neq 0, M_p=0, p\neq i \neq j \neq s\neq t), \label{Iijst} \\
&&P_j = P(M_j=M \neq 0, M_p=0, p\neq j),~~ P_{ij} = P(M_i=M_j=M \neq 0, M_p=0, p\neq i \neq j), \nonumber \\
&&P_{ijs} = P(M_i=M_j=M_s=M \neq 0, M_p=0, p\neq i \neq j \neq s), \nonumber \\
&&P_{ijst} = P(M_i=M_j=M_s=M_t=M \neq 0, M_p=0, p\neq i \neq j \neq s\neq t) \, . \label{Iijst}
\eea

Application of the IBP procedure \cite{Chetyrkin:1981qh}
to loop internal momenta leads to relations between  various Feynman integrals and,  therefore, to the necessity of
calculating only some of them, which in a sense are independent.
These independent diagrams (which were chosen completely arbitrarily, of course) are called master integrals
\cite{Broadhurst:1987ei}.

Applying the IBP procedure \cite{Chetyrkin:1981qh} to the 
master-integrals themselves leads to differential equations 
\cite{Kotikov:1990zs,Kotikov:1990kg} for them with
the inhomogeneous terms containing less complex diagrams.
\footnote{The ``less complex diagrams''  usually contain less number of
propagators and sometimes they can be represented as diagrams with less
number of loops and with some ``effective masses'' (see, for example, 
\cite{FleKoVe,Kniehl:2005bc,Kniehl:2005yc,Fleischer:1999hp} and references therein).}
Applying the IBP procedure to diagrams in inhomogeneous terms leads to new differential equations 
for them with new inhomogeneous terms 
containing even more less complex diagrams ($\equiv$ less$^2$ complex ones).
By repeating the procedure several times, in the last step we can obtain 
inhomogeneous terms containing mainly
tadpoles, which can be easily calculated  in-turn (see also the discussions in Section 7 below).

Solving the corresponding differential equations
in this last step, diagrams for the inhomogeneous terms of the differential equations in the previous step can be reproduced.
Repeating the procedure several times, me can get the results for the original Feynman diagram.

\section{Evaluation of series}

As the first example, we consider $P_{126}$ integral, whous graphical representation has the following form
\vskip 0.5cm
\be
P_{126} \, =
 \hspace{3mm}
\raisebox{1mm}
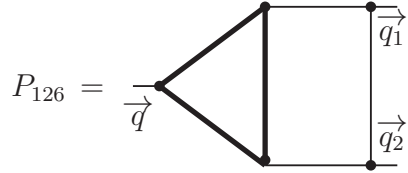
{{
    \begin{axopicture}(90,10)(0,4)
  \SetWidth{2.0}
  \Line(5,5)(45,35)
  \Line(5,5)(45,-25)
\Line(45,-25)(45,35)
\SetWidth{0.7}
\Line(45,35)(85,35)
\Line(45,-25)(85,-25)
\Line(85,35)(85,-25)
\Vertex(45,35){2}
\Vertex(45,-23){2}
\Vertex(5,5){2}
\Vertex(85,35){2}
\Vertex(85,-25){2}
\Line(5,5)(-5,5)
\Line(85,35)(95,35)
\Line(85,-25)(95,-25)
\Text(-3,-5)[b]{$\to$}
\Text(-3,-12)[b]{$q$}
\Text(93,28)[b]{$\to$}
\Text(93,20)[b]{$q_1$}
\Text(93,-12)[b]{$\to$}
\Text(93,-20)[b]{$q_2$}
\end{axopicture}
}}
\hspace{3mm}
\nonumber
\ee
\vskip 1cm
This integral was calculated in Ref.  \cite{Fleischer:1997bw} and presented also
in \cite{FleKoVe,Kotikov:2007vr}.

It has the following series representation:
\footnote{In Minkowski space all $\ln^n x \to \ln^n (-x)$ and there is an imaginary part which
  is important for cuts of the diagram. Such property exists also in other examples (see subsection 6.3 below).}
\bea
 P_{125} &=& - \frac{\hat{N}}{(q^{2})^2} \,
\sum_{n=1} \, \frac{(-x)^n}{2n} \, \Biggl[
 \frac{(n!)^2}{(2n)!} \, \biggl\{
 \frac{1}{\ep^2} - \frac{1}{\ep} \Bigl(S_1 + \ln x) + 5\overline{S}_1S_1 - \frac{1}{2} \Bigl(\overline{S}_1^2 + 15 S_1^2 + S_2 \Bigr)
\nonumber \\
&& + \frac{2}{n} S_1 - S_1 \ln x + \frac{1}{2} \ln^2 x \biggr\} 
 + n F(n) \Biggr] ,~~ \left(x=\frac{q^2}{M^2}\right) \, ,
\label{P126}
\end{eqnarray}
where
\be
S_i=S_i(n-1),~~\overline{S}_i=S_i(n-1)\, . 
\label{Si(n-1)}
\ee
Here the normalization 
$\hat{N}={(\overline{\mu}^2/m^{2})}^{2\varepsilon}$, 
where $\overline{\mu}=4\pi e^{-\gamma_E} \mu$ is in the standard
$\overline{MS}$-scheme and $\gamma_E$ is the Euler constant.

Eq. (\ref{P126}) contains the function $F(n)$:
\be
F(n)= F_1(n) - F_2(n) + F_3(n) - F_4(n) ,
\nonumber
\ee
with
\bea
&&F_1(n)= \sum_{m=1}^{n-1} \, \frac{\Gamma(n)\Gamma(n-m)}{m\Gamma(2n-m)} \, \left[\Psi(n)+\Psi(n-m)- \Psi(2n-m)-\Psi(1)\right], \nonumber \\
&&F_2(n)= \sum_{k=1}^{\infty} \, \frac{\Gamma^2(n+k)}{k!\Gamma(2n+k)} \, \frac{1}{k^2},~~
F_3(n)= \sum_{m=1}^{n-1} \, \frac{\Gamma(n)\Gamma(n-m)}{\Gamma(2n-m)} \, \frac{1}{m^2}, \nonumber \\
&&F_4(n)= \sum_{m=1}^{n-1} \, \frac{\Gamma(n+m)}{m!\Gamma(n)} \, \frac{1}{m^2},
\label{Fin}
\eea

Our purpose here to show the calculatio of the series $F_i(n)$. The most important point is the following: if some series has the
following form
\be
\tilde{F}(n)= \sum_{m=1}^{n-1} \, \tilde{f}(n,m),
\nonumber
\ee
then it is important to reprsent it as
\be
\tilde{F}(n)= \sum_{m=1}^{n-1} \, \hat{f}(m),
\nonumber
\ee
where the new function $\hat{f}(m)$ is $n$-independent.

Technically, the series $F_i(n)$ can be evaluated by considering a connection between $F_i(n+1)$ and $F_i(n)$, which can be expressed
as a difference equation of the following type:
\be
\tilde{C}(n+1) \, \tilde{F}(n+1) - \tilde{C}(n) \, \tilde{F}(n) = \tilde{F}^{(1)}(n)
\nonumber
\ee
with some coefficient functions $\tilde{C}(n)$. Usually the new function $\tilde{F}^{(1)}(n)$ can be summed by standard formulas from
techbooks. If it is not the case, then it is necessary to repear the above procedure for the function $\tilde{F}^{(1)}(n)$, then we will
come to new function, e.g., $\tilde{F}^{(2)}(n)$. So, repeating the above procedure for the functions $\tilde{F}^{(i)}(n)$, we will
come to the one $\tilde{F}^{(i+1)}(n)$, which can be summed by standard formulas from techbooks.

For all the above functions $F_i(n)$ $(i=1,2,3,4)$ we should repeat the procedure twicely. Is it not so convenient. More simple to
consider the new functions $F_i(n,a)$:
\bea
&&F_1(n,a)= \sum_{m=1}^{n-1} \, \frac{\Gamma(n+a)\Gamma(n-m+a)}{m\Gamma(2n-m+a)\Gamma(1+a)},~~
F_2(n,a)= \sum_{k=1}^{\infty} \, \frac{\Gamma^2(n+k)}{k!\Gamma(2n+k)} \, \frac{1}{k-a}, \nonumber \\
&&F_3(n,a)= \sum_{m=1}^{n-1} \, \frac{\Gamma(n)\Gamma(n-m)}{\Gamma(2n-m)} \, \frac{1}{m-a},~~
F_4(n,a)= \sum_{m=1}^{n-1} \, \frac{\Gamma(n+m)}{m!\Gamma(n)} \, \frac{1}{m-a},
\nonumber
\eea
which can be used to reproduce $F_i(n)$ as
\be
F_i(n)= \frac{d}{da} F_i(n,a) |_{a=0} \, .
\nonumber
\ee

The new functions $F_i(n,a)$ can be evaluated with the usage of the above procedure only one time. We will have
\be
\tilde{C_i}(n+1,a) \, \tilde{F_i}(n+1.a) - \tilde{C_i}(n,a) \, \tilde{F_i}(n,a) = \tilde{F_i}^{(1)}(n,a) ,
\nonumber
\ee
where the new functions $\overline{F_i}^{(1)}(n)$ can be summed by standard formulas from techbooks.

Here we will present the exact evaluation only the function $F_1(n,a)$. The other functions can be evaluated similiry and we will
show only the basic steps.

 \subsection{$F_1(n,a)$}

 Consider $F_1(n+1,a)$ and express it through $F_1(n,a)$ as
 \bea
 &&F_1(n+1,a) = \sum_{m=1}^{n} \, \frac{\Gamma(n+a+1)\Gamma(n-m+a+1)}{m\Gamma(2n-m+a+2)\Gamma(1+a)}
 = \frac{\Gamma(n+a+1)}{n\Gamma(n+a+2)\Gamma(1+a)} \nonumber \\
 &&+  \sum_{m=1}^{n-1} \, \frac{\Gamma(n+a)\Gamma(n-m+a)}{m\Gamma(2n-m+a)\Gamma(1+a)} \, \frac{(n+a)(n-m+a)}{m(2n-m+a)(2n+1-m+a)} \, .
\label{F1n1}
 \eea

Expanding the last factor, we have after some little algebra
\bea
&&\frac{(n-m+a)}{m(2n-m+a)(2n+1-m+a)}= \frac{(n+a)}{(2n+a)(2n+1+a)} \, \left[\frac{1}{m} + \frac{1}{2n-m+a}\right]
\nonumber \\ &&-
  \frac{n+1}{2n+1+a} \, \frac{1}{(2n-m+a)(2n+1-m+a)} \, ,
\nonumber
\eea
that leads for the r.h.s. of (\ref{F1n1})
\bea
&&\frac{1}{n(n+1+a)} + \frac{(n+a)^2}{(2n+a)(2n+1+a)} \, \left[F_1(n,a) + \sum_{m=1}^{n-1} \,
  \frac{\Gamma(n+1+a)\Gamma(n-m+a)}{\Gamma(2n+1-m+a)\Gamma(1+a)}  \right]
\nonumber \\ &&-
\frac{(n+1)(n+a)}{2n+1+a} \, \sum_{m=1}^{n-1} \,
  \frac{\Gamma(n+1+a)\Gamma(n-m+a)}{\Gamma(2n+2-m+a)\Gamma(1+a)} \, .
\label{F1n1.1}
  \eea

The last two series can be evaluated since
\bea
 &&\sum_{m=1}^{n-1} \,
  \frac{\Gamma(n+a)\Gamma(n-m+a)}{\Gamma(2n+\alpha-m+a)\Gamma(1+a)} =  \sum_{m=1}^{n-1} \,
  \frac{\Gamma(n+a)\Gamma(m+a)}{\Gamma(n+m+\alpha+a)\Gamma(1+a)}\nonumber \\
 && = \frac{\Gamma(n+a)}{\Gamma(1+a)} \, \hat{\Phi}_1(n-2,a+1,n+\alpha+a+1) \, ,
\label{F1n1a}
  \eea
  where the function $\hat{\Phi}_1(n,a,b)$ is defined and evaluated in Appendix A. Using these calculations
  \be
  \hat{\Phi}_1(n-2,a+1,n+\alpha+a+1) = \frac{1}{n+\alpha-1} \,  \left[\frac{\Gamma(1+a)}{\Gamma(n+\alpha+a)}
    -\frac{\Gamma(n+a)}{\Gamma(2n+\alpha+a-1)} \right] \, ,
  \label{Phin_2}
  \ee
  we have for the above series
\be
 \sum_{m=1}^{n-1} \, 
 \frac{\Gamma(n+a)\Gamma(n-m+a)}{\Gamma(2n+\alpha-m+a)\Gamma(1+a)} = \frac{1}{n+\alpha-1} \,
 \left[\frac{\Gamma(1+a)}{\Gamma(n+\alpha+a)}-\frac{\Gamma(n+a)}{\Gamma(2n+\alpha+a-1)} \right] \, \frac{\Gamma(n+a)}{\Gamma(1+a)} \, .
\label{Phin_2.1}
  \ee
  
%

Putting the result to (\ref{F1n1.1}), we have
\bea
&&\frac{1}{n(n+1+a)} + \frac{(n+a)^2}{(2n+a)(2n+1+a)} \, \Biggl[F_1(n,a) \nonumber \\ &&
  + \frac{1}{n} \, \left(
  \frac{\Gamma(n+a)}{\Gamma(n+1+a)}- \frac{\Gamma^2(n+a)}{\Gamma(2n+a)\Gamma(1+a)}  \right) \Biggr]
\nonumber \\ &&-
\frac{(n+a)}{2n+1+a} \, \left(
  \frac{\Gamma(n+a)}{\Gamma(n+12+a)}- \frac{\Gamma^2(n+a)}{\Gamma(2n+1+a)\Gamma(1+a)}  \right) \, .
\label{F1n1.2}
  \eea

  After some algebra, we have for $F_1(n+1,a)$:
  \bea
  && F_1(n+1,a) =  \frac{(n+a)^2}{(2n+a)(2n+1+a)} \, F_1(n,a) + \frac{3n-2a}{n(2n+a)(2n+1+a)}
 \nonumber \\ &&
  + \frac{a}{n} \,
  \frac{\Gamma(n+a)\Gamma(n+a+1)}{\Gamma(2n+2+a)\Gamma(1+a)} \, .
  \label{F1n1.3}
  \eea

  Consider
  \be
   F_1(n,a) =  C_1(n,a) \overline{F}_1(n,a)
\nonumber
\ee
with
\be
C_1(n,a) = \frac{\Gamma^2(n+a)}{\Gamma(2n+a)\Gamma(1+a)} \,.
\nonumber
\ee

Then, we have
 \bea
  &&  \overline{F}_1(n+1,a) =   \overline{F}_1(n,a) + \frac{3n-2a}{n} \, \frac{\Gamma(2n+a)\Gamma(1+a)}{\Gamma^2(n+1+a)} 
 + \frac{a}{n(n+a)} \, \nonumber \\ &&
 = \overline{F}_1(1,a) + \sum_{m=1}^n \left(\frac{3n-2a}{n} \, \frac{\Gamma(2m+a)\Gamma(1+a)}{\Gamma^2(m+1+a)} + \frac{a}{m(m+a)} \right) \, .
  \label{F1n1.4}
  \eea

  Since $F_1(1,a)=0$, then $\overline{F}_1(1,a)=0$. Thus, we obtain the final result for $F_1(n,a)$:
  \be
  F_1(n,a) =  \frac{\Gamma^2(n+a)}{\Gamma(2n+a)\Gamma(1+a)} \, \sum_{m=1}^{n-1} \left(\frac{3n-2a}{n} \, \frac{\Gamma(2m+a)\Gamma(1+a)}{\Gamma^2(m+1+a)} + \frac{a}{m(m+a)} \right)\, .
  \label{F1n1.5}
  \ee

  Then,
  \bea
  && F_1(n,0) = 3 \frac{\Gamma^2(n)}{\Gamma(2n)} \, \sum_{m=1}^{n-1} \, \frac{\Gamma(2m)}{\Gamma^2(m+1)} = 3 \frac{(n!)^2}{(2n)!} W_1 \nonumber \\
  && F_1(n) = \frac{\Gamma^2(n)}{\Gamma(2n)} \, \Biggl[S_2(2n-1) + \sum_{m=1}^{n-1} \, \frac{\Gamma(2m)}{\Gamma^2(m+1)} \, \nonumber \\
  && \cdot \left[3\left\{2S_2(m-1)-S_2(2m-1)-2S_2(n-1)+S_2(2n-1)\right\}+\frac{4}{m}\right]\Biggr]  \nonumber \\
  && =  \frac{\Gamma^2(n)}{\Gamma(2n)} \, \Biggl[\overline{S}_2 + 3 \left(\overline{S}_1 W_1 - \overline{W}_{1,1}\right)
    -6\left(S_1 W_1 - W_{1,1}\right)+ 4W_2\Biggr],
     \label{F1n1.6}
  \eea
  where
  $S_{\pm a,...} \equiv S_{\pm a,...}(n-1)$, $\overline{S}_{\pm a,...} \equiv S_{\pm a,...}(2n-1)$ (see definition
of $S_{\pm a,...}(n)$ in (\ref{Sin}), 
$W_{\pm a,...} \equiv W_{\pm a,...}(n-1)$ and $\overline{W}_{\pm a,...} \equiv W_{\pm a,...}(2n-1)$ with
\begin{eqnarray}
W_{a}(n)\ =\ \sum^n_{m=1} \,
\frac{\hat{C}_m^{-1}}{m^a},
\ \ W_{a,b,c,\cdots}(n)~=~ \sum^n_{m=1}  \,
\frac{\hat{C}_m^{-1}}{m^a}\, S_{b,c,\cdots}(m) \, . \label{FI5}
\end{eqnarray}

  \subsection{$F_i(n,a)$  ($i=2,3,4$)}

 Consider $F_i(n+1,a)$ $i=2,3,4$ and express it through $F_i(n,a)$ as in the previous subsection
 \be
 F_i(n+1,a) = K_i(n,a) 
 \, F_i(n,a) + f_i(n,a) \, ,
\label{Fi.1}
  \ee
  where
  \be
  K_2(n,a) =  \frac{(n+a)^2}{(2n+a)(2n+1+a)},~~K_3(n,a) = \frac{n(n-a)}{(2n-a)(2n+1-a)},~~
  K_4(n,a) = \frac{n+a}{n}
\label{Ki.1}
  \ee
  and
  \bea
&&f_2(n,a)= \frac{3n-2a}{n(2n+a)(2n+1+a)}
  - \frac{n[n(4n+3)+a(3n+2)]}{(2n+a)(2n+a+1)} \,
  \frac{\Gamma^2(n)}{\Gamma(2n+2)}, \nonumber \\
&&f_3(n,a)= \frac{3n^2-3an+a^2}{n(2n-a)(2n+1-a)}
  + \frac{a}{2(2n-a)(2n-a+1)} \,
  \frac{\Gamma^2(n)}{\Gamma(2n)}, \nonumber \\
&&f_4(n,a)= \frac{3n-2a}{n-a} \, \frac{\Gamma(2n)}{\Gamma^2(n+1)} -  \frac{1}{n} \, .
  \label{fi.1}
  \eea

  Consider
  \be
   F_i(n,a) =  C_i(n,a) \overline{F}_i(n,a),~~ f_i(n,a) =  C_i(n,a) \overline{f}_i(n,a)
\nonumber
\ee
with
\be
C_2(n,a) = \Gamma(1+a) C_2(n,a),~~  C_3(n,a) =
\frac{\Gamma(n)\Gamma(n-a)}{\Gamma(2n-a)},~~
C_4(n,a) = \frac{\Gamma(n-a)}{\Gamma(n)}\,.
\nonumber
\ee

Then, we have
 \be
 \overline{F}_i(n+1,a) =   \overline{F}_i(n,a) + \overline{f}_i(n,a)
 = \overline{F}_i(1,a) + \sum_{m=1}^n \overline{f}_i(m,a) \, ,
  \label{tFi.1}
  \ee
  where
  \bea
&&\overline{f}_2(n,a) = \frac{3n+2a}{n} \, \frac{\Gamma(2n+a)}{\Gamma^2(n+1+a)} 
  - \left(4n+3 + \frac{a}{n}(3n+2)\right)\, \frac{\Gamma^2(n+1)}{\Gamma(2n+2)}
  \, \frac{\Gamma(2n+a)}{\Gamma^2(n+1+a)} \, , \nonumber \\
  &&\overline{f}_3(n,a) = \frac{3n^2-3an+a^2}{n(n-a)} \, \frac{\Gamma(2n-a)}{\Gamma(n+1-a)\Gamma(n+1)} 
  + \frac{a}{2n}\, \frac{\Gamma(n)}{\Gamma(2n)}  \, \frac{\Gamma(2n+a)}{\Gamma(n+1+a)} \, ,
  \nonumber \\
  &&\overline{f}_4(n,a) = \frac{3n-a}{n-a} \, \frac{\Gamma(2n)}{\Gamma(n+1-a)\Gamma(n+1)} 
  - \frac{\Gamma(n)}{\Gamma(n+1-a)} \, .
\label{tfi.1}
  \eea
  
  Since $F_j(1,a)=0$ $(j=3,4)$ and $F_2(1,a)= (\Psi(1+a)-\Psi(1)-1)/(a-1)$, then
  \be
  \overline{F}_j(1,a)=0~~(j=3,4),~ \overline{F}_2(1,a)= \frac{1}{\Gamma(1-a)} \, \Bigl(1-\Psi(1+a)+\Psi(1)\Bigr) \, .
\label{tFi.2}
  \ee
  
 Thus, we obtain the final result for $F_i(n,a)$:
  \be
  F_i(n,a) = C_i(n,a) \, \left[ \overline{F}_j(1,a) +
  \sum_{m=1}^{n-1} \overline{f}_i(m,a)
\right] \, .
  \label{Fi.2}
  \ee

  Then,
  \be
  F_2(n,0) = \frac{\Gamma^2(n)}{\Gamma(2n)} \, \Bigl[\overline{S}_1 - 2S_1 + 3W_1 \Bigr],~~
  F_3(n,0) = 3\frac{(n!)^2}{(2n)!} W_1,~~
  F_4(n,0) = \frac{\Gamma^2(n)}{\Gamma(2n)} \,
  \Bigl[3W_1-S_1 \Bigr],  \nonumber \\
\label{Fi.3}
  \ee
  and
  \bea
  && F_2(n) =
  \frac{\Gamma^2(n)}{\Gamma(2n)} \, \Biggl[\zeta_2 - \frac{1}{2} \left(\overline{S}_2 - 2S_2 +
    (\overline{S}_1 - 2S_1)^2\right)
 - 3 \left(\overline{S}_1 W_1 - \overline{W}_{1,1}\right) \nonumber \\
 &&   +6\left(S_1 W_1 - W_{1,1}\right)- 4W_2\Biggr],
  \label{F2.1} \\
&& F_3(n) =
  \frac{\Gamma^2(n)}{\Gamma(2n)} \, \Biggl[\frac{1}{2} \overline{S}_2  
 + 3 \left(\overline{S}_1 W_1 - \overline{W}_{1,1}\right)
    -3\left(S_1 W_1 - W_{1,1}\right)+ 3W_2\Biggr],
  \label{F3.1} \\
&& F_4(n) =
 \frac{\Gamma^2(n)}{\Gamma(2n)} \, \Biggl[3\left(S_1 W_1 - W_{1,1}\right)- W_2
   - \frac{1}{2} \left(S^2_1 - S_2\right) \Biggr] \, .
  \label{F4.1}
  \eea
  
  Taking the results for $F_i(n)$ ($i=1,2,3,4$) together, we have for $P_{126}$:
  \bea
 P_{125} &=& - \frac{\hat{N}}{(q^{2})^2} \,
 \sum_{n=1} \, \frac{(-x)^n}{2n} \,
 \frac{(n!)^2}{(2n)!} \, \biggl\{
 \frac{1}{\ep^2} - \frac{1}{\ep} \Bigl(S_1 + \ln x) + 4\overline{S}_1S_1 - \frac{3}{2}
 \Bigl(5 S_1^2 + S_2 \Bigr) -\zeta_2
\nonumber \\
&& + \frac{2}{n} S_1 - S_1 \ln x + \frac{1}{2} \ln^2 x \biggr\} \, ,
\label{P126f}
\end{eqnarray}
 where all series
$W_{\pm a,...}$ and $\overline{W}_{\pm a,...}$ are canceled.

\subsection{Properties of series}

This scheme was successfully used to calculate the two-loop two-point \cite{Kotikov:1990zs,Kotikov:1990kg,Fleischer:1999hp}
and three-point diagrams \cite{Fleischer:1997bw,FleKoVe,Kniehl:2005bc}
with one nonzero mass. This procedure is very powerful, but rather complicated. However, there are some simplifications based
on representations of series of Feynman integrals.


Indeed, the inverse-mass expansion of two-loop two-point
and three-point diagrams
\footnote{The diagrams
  are complicated two-loop Feynman integrals that do not have cuts of
  three massive particles.  thus, their results should be expressed as combinations of Polylogarithms. 
Note that we consider only three-point diagrams with independent upward momenta $q_1$ and $q_2$, which satisfy 
the conditions $q_1^2=q_2^2=0$ and $(q_1+q_2)^2\equiv q^2 \neq 0$, where $q$ is downward momentum.}
with one nonzero mass (massless and massive propagators are shown by dashed and solid lines,
respectively), can be considered as
\begin{eqnarray}
\mbox{ FI} &=& 
  \, \frac{\hat{N}}{q^{2\alpha}} \,
\sum_{n=1} \, C_n \, {(\eta x)}^n
\, \biggl\{F_0(n) +
\biggl[\ln x
  \, F_{1,1}(n)  +
\frac{1}{\varepsilon} \, F_{1,2}(n) \biggr] 
\label{FI1} \\
&& + \biggl[\ln^2 x
  \, F_{2,1}(n)  + \frac{1}{\varepsilon} \,\ln x
  \, F_{2,2}(n) + \frac{1}{\varepsilon^2} \, F_{2,3}(n) + \zeta(2)
 \, F_{2,4}(n) \biggr]
+ \cdots \biggr\} \, ,
\nonumber
\end{eqnarray}
where $x=q^2/m^2$ (as it was shown in (\ref{P126}), $\eta =1$
or $-1$
and $\alpha=1$ and $2$ for
two-point and three-point cases, respectively. 
The normalization $\hat{N}$ is defined below Eq. (\ref{P126}).
Moreover,
\begin{eqnarray}
C_n  ~=~ \frac{(n!)^2}{(2n)!} ~\equiv ~ \hat{C}_n \, .
\label{FI1b}
\end{eqnarray}
for diagrams with
two-massive-particle-cuts ($2m$-cuts). For the diagrams with one-massive-particle-cuts ($m$-cuts)
$C_n = 1$.

For  $m$-cut
case,
the coefficients $F_{N,k}(n)$ should have the form
\begin{eqnarray}
F_{N,k}(n) ~ \sim ~ \frac{S_{\pm a,...}}{n^b}\, ,  \frac{\zeta(\pm a)}{n^b}\, , 
\label{FI1c}
\end{eqnarray}
where $S_{\pm a,...} \equiv S_{\pm a,...}(j-1)$
are harmonic sums defined in (\ref{Sin})
and $\zeta(\pm a)$ are the Euler-Zagier constants
\be
\zeta(\pm a)\ =\ \sum^{\infty}_{m=1} \frac{(\pm 1)^m}{m^a},
\ \ \zeta(\pm a,\pm b,\pm c,\cdots )~=~ \sum^{\infty}_{m=1}
\frac{(\pm 1)^m}{m^a}\, S_{\pm b,\pm c,\cdots}(m-1) \, .  \label{Euler}
\ee


For  $2m$-cut
case,
the coefficients $F_{N,k}(n)$ can be more complicated
\begin{eqnarray}
F_{N,k}(n) ~ \sim ~ \frac{S_{\pm a,...}}{n^b},  ~ \frac{V_{a,...}}{n^b}
,  ~ \frac{W_{a,...}}{n^b} \, ,
\label{FI1d}
\end{eqnarray}
where $S_{\pm a,...}\equiv S_{\pm a,...}(n-1)$ and $W_{\pm a,...}$ are defined in Eqs. (\ref{Sin}) and (\ref{FI5}),
respectively, and
$V_{\pm a,...} \equiv V_{\pm a,...}(j-1)$
with
\begin{eqnarray}
V_{a}(j)\ =\ \sum^j_{m=1}
\, \frac{\hat{C}_m}{m^a},
\ \ V_{a,b,c,\cdots}(j)~=~ \sum^j_{m=1}  \,
\frac{\hat{C}_m}{m^a}\, S_{b,c,\cdots}(m),  \label{FI4}
\end{eqnarray}

The terms $\sim V_{a,...}$ and $\sim W_{a,...}$
can come only in the $2m$-cut
case.
%
The origin of the appearance of these  terms 
is the product of series (\ref{FI1})
with the different 
coefficients $C_n =1$ and
$C_n = \hat{C}_n
$.

\subsection{Other examples}

As an example, consider two-loop two-point diagrams $I_1$
and $I_{12}$
studied in \cite{FleKoVe}.

\vskip 0.5cm
\be
I_1 \, =
 \hspace{3mm}
\raisebox{1mm}{{
    \begin{axopicture}(90,10)(0,4)
  \SetWidth{2.0}
\Arc(45,-7)(40,90,160)
\SetWidth{0.7}
\Arc(45,-7)(40,20,90)
\Arc(45,17)(40,200,270)
\Line(45,-25)(45,35)
\Arc(45,17)(40,270,340)
\Vertex(45,-23){2}
\Vertex(45,33){2}
\Vertex(5,5){2}
\Vertex(85,5){2}
\Line(5,5)(-5,5)
\Line(85,5)(95,5)
\Text(-3,-5)[b]{$\to$}
\Text(-3,-12)[b]{$q$}
\end{axopicture}
}}
\hspace{3mm},~~
I_{12} \, =
 \hspace{3mm}
\raisebox{1mm}{{
    \begin{axopicture}(90,10)(0,4)
 \SetWidth{2.0}
 \Arc(45,-7)(40,90,160)
 \Arc(45,17)(40,200,270)
 \SetWidth{0.7}
 \Arc(45,-7)(40,20,90)
\Line(45,-25)(45,35)
\Arc(45,17)(40,270,340)
\Vertex(45,-23){2}
\Vertex(45,33){2}
\Vertex(5,5){2}
\Vertex(85,5){2}
\Line(5,5)(-5,5)
\Line(85,5)(95,5)
\Text(-3,-5)[b]{$\to$}
\Text(-3,-12)[b]{$q$}
\end{axopicture}
}}
\hspace{3mm} \, .
\label{I1+2}
\ee
\vskip 1cm

Their results are
\begin{eqnarray}
I_1 &=& - \frac{\hat{N}}{q^{2}} \,
\sum_{n=1} \, \frac{(-x)^n}{n} \, \biggl\{
\frac{1}{2} \ln^2 x
- \frac{2}{n} \ln x
+ \zeta(2)
+2S_2 -2 \frac{S_1}{n} + \frac{3}{n^2} \biggr\} \, ,
\label{FI6a} \\
I_{12} &=& - \frac{\hat{N}}{q^{2}} \,
\sum_{n=1} \, \frac{(-x)^n}{n^2} \, \biggl\{\frac{1}{n} +
 \frac{(n!)^2}{(2n)!} \, \biggl(
 -2 \ln x
 -3 W_1 + \frac{2}{n} \biggr) \biggr\} \, .
\label{FI6c}
\end{eqnarray}

From (\ref{FI6a}) 
one can see that the corresponding functions
$F_{N,k}(n)$ have the form 
\begin{eqnarray}
F_{N,k}(n) ~ \sim ~ \frac{1}{n^{3-N}},~~~~(N\geq 2),
\label{FI8}
\end{eqnarray}
if we introduce the following complexity of the sums ($\overline{\Phi}=(S,V,W)$) 
\begin{eqnarray}
\overline{\Phi}_{\pm a} \sim \overline{\Phi}_{\pm a_1, \pm a_2}
\sim \overline{\Phi}_{\pm a_1,\pm a_{2},\cdots,\pm a_m}
\sim \zeta_{a} \sim \frac{1}{n^a},~~~~ (\sum_{i=1}^m a_i =a) \, .
\label{FI9}
\end{eqnarray}

The number $3-N$ determines the level of transcendentality (or complexity, or weight)
of the coefficients $F_{N,k}(n)$. The property greatly reduces  the number
of the possible elements in $F_{N,k}(n)$.
The level of transcendentality decreases if we consider the singular parts of diagrams and/or coefficients in front of
$\zeta$-functions and of logarithm powers.
Thus, finding the parts we can predict the rest  using the ansatz 
based on the results already obtained, but containing elements with a higher 
level  of transcendentality.

Other two-loop two-point
integrals in \cite{FleKoVe} have similar form. They were exactly calculated
by differential equation method \cite{Kotikov:1990zs,Kotikov:1990kg}.

Now we consider two-loop three-point diagrams, 
$P_5$
and $P_{12}$:
\vskip 0.7cm
\be
P_{5} \, =
 \hspace{3mm}
\raisebox{1mm}
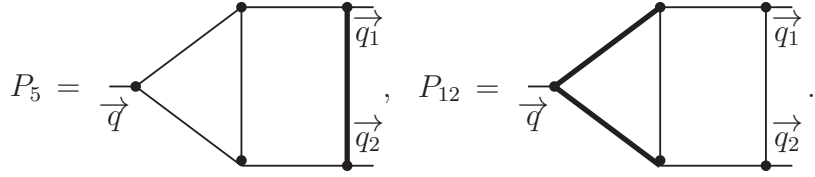
{{
    \begin{axopicture}(90,10)(0,4)
  \SetWidth{0.7}
  \Line(5,5)(45,35)
  \Line(5,5)(45,-25)
\Line(45,-25)(45,35)
\SetWidth{0.7}
\Line(45,35)(85,35)
\Line(45,-25)(85,-25)
\SetWidth{2.0}
\Line(85,35)(85,-25)
\SetWidth{0.7}
\Vertex(45,35){2}
\Vertex(45,-23){2}
\Vertex(5,5){2}
\Vertex(85,35){2}
\Vertex(85,-25){2}
\Line(5,5)(-5,5)
\Line(85,35)(95,35)
\Line(85,-25)(95,-25)
\Text(-3,-5)[b]{$\to$}
\Text(-3,-12)[b]{$q$}
\Text(93,28)[b]{$\to$}
\Text(93,20)[b]{$q_1$}
\Text(93,-12)[b]{$\to$}
\Text(93,-20)[b]{$q_2$}
\end{axopicture}
}}
\hspace{3mm}
,~~
P_{12} \, =
 \hspace{3mm}
\raisebox{1mm}{{
    \begin{axopicture}(90,10)(0,4)
  \SetWidth{2.0}
  \Line(5,5)(45,35)
  \Line(5,5)(45,-25)
\SetWidth{0.7}
  \Line(45,-25)(45,35)
\Line(45,35)(85,35)
\Line(45,-25)(85,-25)
\Line(85,35)(85,-25)
\Vertex(45,35){2}
\Vertex(45,-23){2}
\Vertex(5,5){2}
\Vertex(85,35){2}
\Vertex(85,-25){2}
\Line(5,5)(-5,5)
\Line(85,35)(95,35)
\Line(85,-25)(95,-25)
\Text(-3,-5)[b]{$\to$}
\Text(-3,-12)[b]{$q$}
\Text(93,28)[b]{$\to$}
\Text(93,20)[b]{$q_1$}
\Text(93,-12)[b]{$\to$}
\Text(93,-20)[b]{$q_2$}
\end{axopicture}
}}
\hspace{3mm} \, .
\nonumber
\ee
\vskip 1cm

Their results are (see \cite{FleKoVe}):
\begin{eqnarray}
P_5 &=& \frac{\hat{N}}{(q^{2})^2} \,
\sum_{n=1} \, \frac{x^n}{n} \, \biggl\{
-6\zeta_3 + 2S_1\zeta_2
+6S_3-2S_1S_2+ 4 \frac{S_2}{n}-\frac{S_1^2}{n}
+ 2\frac{S_1}{n^2} \nonumber \\
&&+ \biggl(-4S_2+S_1^2-2\frac{S_1}{n}\biggr) \, \ln x
+ S_1 \ln^2 x
\biggl\} \, ,
\label{FI7b} \\
P_{12} &=& \frac{\hat{N}}{(q^{2})^2} \,
\sum_{n=1} \, \frac{(-x)^n}{n^2} \,
 \frac{(n!)^2}{(2n)!} \, \biggl\{
 \frac{2}{\ep^2} + \frac{2}{\ep} \biggl(S_1 -3 W_1 + \frac{1}{n} -\ln x
\biggr) -6 W_2 -18 W_{1,1}
\nonumber \\ && -13S_2 + S_1^2- 6S_1W_1 +2 \frac{S_1}{n} +
\frac{2}{n^2}
-2 \bigg(S_1+\frac{1}{n}\biggr)\ln x
+ \ln^2 x 
\biggr\} \, .
\nonumber
\end{eqnarray}

 Now the coefficients
$F_{N,k}(n)$ have the form
\begin{eqnarray}
  F_{N,k}(n) ~ \sim ~ \frac{1}{n^{4-N}},~~~~(N\geq 3),
\label{FI12}
\end{eqnarray}

The diagram 
$P_5$ 
(and also $P_1$, $P_3$ and $P_6$ in \cite{FleKoVe})
was calculated
exactly by differential equation
method \cite{Kotikov:1990zs,Kotikov:1990kg}.
To find the results for  
$P_{12}$ (and also
all others
in \cite{FleKoVe}) we have used the knowledge of the several $n$ terms in the
inverse-mass expansion (\ref{FI1}) (usually less than $n=100$) and
the following arguments: 
\begin{itemize}
\item
If
a two-loop two-point diagram with a ``similar topology'' (for example, 
$I_{12}$ for $P_{12}$, etc.)
was already calculated, we should consider a similar set of basic elements for corresponding $F_{N,k}(n)$ of
two-loop three-point diagrams
but with a higher level of complexity.
\item
Let the diagram under consideration contain singularities and/or powers of logarithms.
Since the coefficients are very simple before the leading singularity, or the largest degree of the logarithm, or
the largest $\zeta$-function, they can often be predicted directly from the first few terms of the expansion.

Moreover, often we can calculate the singular part using a different technique
(see \cite{FleKoVe} for extraction of $\sim W_1(n)$ part). Then we should expand the
singular parts, find the main elements and try to use them
(with the corresponding increase in the level of complexity) in order to predict
the regular part of the diagram. If we need to find  $\ep$-suppressed terms, we should
increase the level of complexity of the corresponding basic elements.
\end{itemize}

Later, using the ansatz for $F_{N,k}(n)$ and several terms (usually less than 100) in the above
expression, which can be  exactly calculated, we obtain a system of
algebraic equations for the parameters of the ansatz. Solving the system, we
can obtain the analytical results for Feynman integrals without exact calculations.
To check the results, we only need to calculate a few more terms in the
above inverse-mass expansion (\ref{FI1}) and compare them with the
predictions of our anzatz with the fixed coefficients indicated above.

Thus, the considered arguments give a possibility to find results for many complicated
two-loop three-point diagrams without direct calculations.
Several process options have been successfully used to calculate
Feynman diagrams for many processes (see 
\cite{Fleischer:1997bw,FleKoVe,Kniehl:2005bc,Kniehl:2006bg}).

Note that properties similar to (\ref{FI8}) and (\ref{FI12}) were recently
observed \cite{Eden:2012rr} in the so-called double
operator-product-expansion limit of some four-point diagrams.\\

\subsection{Modern technique of massive diagrams}

Coefficients have the structure   (\ref{FI8}) and (\ref{FI12}) with the rule
(\ref{FI9}). Note that these conditions greatly reduce the number of
possible harmonic sums. In turn, the restriction is associated with a specific form of differential equations
for the Feynman integrals under consideration.
Differentials equations can be formally represented
as \cite{Kotikov:2010gf,Kotikov:2012ac}
\bea
\left((x+a)\frac{d}{dx} - \overline{k}\ep
\right) \, \mbox{ FI } \, = \,
 \mbox{ less complicated diagrams} (\equiv \rm{FI}_1),
\label{Int}
 \eea
with some number $a$ and some function $\overline{k}(x)$.
This form is generated by IBP procedure for diagrams including
an inner $n$-leg one-loop subgraph, which in turn contains the product $k^{\mu_1}...k^{\mu_m}$ of its internal momenta $k$ with $m=n-3$.

Indeed, for ordinary degrees 
  $\alpha_i=1+a_i\ep$ with arbitrary $a_i$ of subgraph propagators, the IBP relation (\ref{TreIBP}) gives the
  coefficient
  $d-2\alpha_1 - \sum^p_{i=2} \alpha_i+m \sim \ep$ for $m=n-3$. Important examples of applying the
  rule are the diagrams $I_1$, $I_{12}$ and $P_5$, $P_{12}$, $P_{126}$ (for the case $n=2$ and $n=3$)
  and also the diagrams in 
  in Ref.
  \cite{Gehrmann:2011xn} (for the case $n=3$ and $n=4$).
  However, we note that the results for the non-planar diagrams (see Fig. 3 of \cite{FleKoVe}) obey the Eq. (\ref{FI12}) but
  their subgraphs do not comply with the above rule. 
The disagreements may be related to the on-shall vertex of the subgraph, but this requires additional research.
 
Taking the set of less complicated Feynman integrals $\rm{FI}_1$ as diagrams having internal $n$-leg subgraphs,
we get their result stucture similar to the one given above (\ref{FI12}), but with  a lower level of complexity.

So, the integrals $\rm{FI}_1$ should obey to the following equation
\bea
\left((x+a_1)\frac{d}{dx} - \overline{k}_1\ep
\right) \, \mbox{ FI$_1$ } \, = \,
 \mbox{ less$^2$ complicated diagrams} (\equiv \rm{FI}_2) .
\label{Int.1}
 \eea

 Thus, we will have the set of equations for all Feynman integrals $\rm{FI}_n$ as
 \bea
\left((x+a_n)\frac{d}{dx} - \overline{k}_n\ep
\right) \, \mbox{ FI$_n$ } \, = \,
 \mbox{ less$^{n+1}$ complicated diagrams} (\equiv \rm{FI}_{n+1}),
\label{Int.n}
 \eea
 with the last integral $\rm{FI}_{n+1}$ contains only tadpoles.
  
 Moreover, following  \cite{Henn:2013pwa}, we can reconstruct the above set of inhomogeneous equations as 
the the homogeneous matrix equation
\footnote{For complicated diagrams, there is an extension in Ref. \cite{Adams:2018yfj}.}
(see Ref. \cite{Lee:2014ioa} containg methods to obtain the equation)

\begin{eqnarray} 
  \frac{d}{dx} \widehat{FI} - \ep  \widehat{K}  \widehat{FI}=0
\label{Henn}
\end{eqnarray}
 for the vector
 \begin{eqnarray}
   \widehat{FI} =
   \left(  \begin{array}{l}
     \rm{FI}\\  \rm{FI}_1/\ep \\   ... \\  \rm{FI}_n/\ep^n
\end{array} \right)
   \,,
   \nonumber
\end{eqnarray}
 where the matrix $ \widehat{K}$ contains
 the functions $\overline{k}_j/(x+a_j)$ as its elements.
 The form (\ref{Henn}) is called as the ``canonic basic''. It is now very popular 
 (see, for example, recent papers in Ref. \cite{Duhr:2020kzd}).
 
Please note that for real calculations of $\rm{FI}_{n}$ it is convenient to replace
 \begin{eqnarray}
   \rm{FI}_{n} =  \widetilde{\rm{FI}}_{n}  \overline{\rm{FI}}_{n},
   \nonumber
\end{eqnarray}
 where the term $\overline{\rm{FI}}_{n}$ obeys the corresponding homogeneous equation
 \bea
\left((x+a_n)\frac{d}{dx} - \overline{k}_n\ep
\right) \, \overline{\rm{FI}}_{n}
\, =\, 0 \, .
\label{IntBar.n}
\eea
 
The replacement simplifies the above equation (\ref{Int.n}) to the following form
\bea
(x+a_n)\frac{d}{dx}  \,  \widetilde{\rm{FI}}_{n} \, = \,   \widetilde{\rm{FI}}_{n+1}
\frac{ \overline{\rm{FI}}_{n+1}}{ \overline{\rm{FI}}_{n}} \, .
\label{Int.n.si}
 \eea
 having the solution
 \bea
 \widetilde{\rm{FI}}_{n}(x) = \int^x_0 \frac{dx_1}{x_1+a_{n}} \widetilde{\rm{FI}}_{n+1}(x_1)
\frac{ \overline{\rm{FI}}_{n+1}(x_1)}{ \overline{\rm{FI}}_{n}(x_1)}
 \label{Int.n.si1}
 \eea

Usually there are some cancellations in the ratio
$\overline{\rm{FI}}_{n+1}/ \overline{\rm{FI}}_{n}$
and sometimes it is equal to 1. In the last case, the
 equation (\ref{Int.n.si1}) coincides wuth the definition of Goncharov Polylogariths (see \cite{Duhr:2014woa} and
 the references therein).\\

 The series (\ref{FI6a}), (\ref{FI6c}) and (\ref{FI7b}) can be expressed as a combination of the Nilson \cite{Devoto:1983tc}
 and Remiddi-Vermaseren \cite{Remiddi:1999ew}
 polylogarithms with weight $4-N$ (see \cite{FleKoVe,Fleischer:1997bw}). More complicated cases were examined in
 \cite{Davydychev:2003mv}.

 \section{
  ${\mathcal N}=4$ SYM}

 Note that  both in the massless case and in the massive case the results of scalar diagrams have an important property:
 all elements have the form (see Eq. (\ref{FI9}))
 \be
 \frac{\overline{\Phi}_{a_1,a_2,...}(n+n_0)}{(n+n_0)^b} \, ~~ \mbox{or} ~~
 \, \frac{\overline{\Phi}_{a_1,a_2,...}(n+n_0)}{(n+n_0+1)^b} \, ,
   \label{S_a/b}
   \ee
where the quantity $\overline{a}+b$ ($\overline{a}=\sum_{i=1} |a_i|$) is fixed for each order of expansion in $\ep$.
 For integrals $I_1(1,n)$ and $I_2(1,n)$ (and also $I_1(n+1,n)$ and $I_2(n+1,n)$), this property is even more strict, since
 in this case $b=0$.

 Unfortunately, calculations in QCD (or another theory) mix orders of magnitude over $\ep$.
 Moreover, in expansions
 of diagrams containing propagators with non-trivial numerators (for example, as propagators of quarks
 and gluons) this property is also violated.
 
 It is an amazing fact that the property (\ref{S_a/b}) in its most strict form $b=0$ is restored for diagonal elements of
 anomalous dimensions \cite{KL,KoLiVe,KLOV} and DIS coefficient functions \cite{Bianchi:2013sta} within $N=4$ SYM.

 The anomalous dimensions  of the twist-2 Wilson operators 
govern the Bjorken scaling violation for parton distributions ($\equiv$ matrix elemens of the twist-2 Wilson operators)
in the framework of Quantum Chromodynamics.

Balitsky-Fadin-Kuraev-Lipatov
\cite{BFKL}
and Dokshitzer-Gribov-Lipatov-Altarelli-Parisi (DGLAP) \cite{DGLAP}
equations resum, respectively, the most important contributions
$\sim \alpha_s \ln(1/x_B)$ and $\sim \alpha_s \ln(Q^2/\Lambda^2)$ in different
kinematical regions of the Bjorken variable $x_B$ and the ``mass'' $Q^2$ of the
virtual photon in the lepton-hadron DIS
and, thus, they are the cornerstone in analyses of
the experimental data
from lepton-nucleon and nucleon-nucleon scattering.
In the supersymmetric generalization of QCD
the equations are simplified drastically \cite{KL00}.
In the ${\mathcal N}=4$ SYM the eigenvalues of the matrix of anomalous dimesion contain only one {\it universal} function
with shifted arguments \cite{LN4,KL}.\\

The  three-loop result
\footnote{
Note, that in an accordance with Ref.~\cite{next}
 our normalization of $\gamma (n)$ contains
the extra factor $-1/2$ in comparison with
the standard normalization (see~\cite{KL})
and differs by sign in comparison with one from Ref.~\cite{VMV}.}
for the universal anomalous dimension
$\gamma_{uni}(j)$
for ${\mathcal N}=4$ SYM is~\cite{KLOV}
\begin{eqnarray}
  \gamma_{uni}(n) ~=~ \hat a \gamma^{(0)}_{uni}(n)+\hat a^2
\gamma^{(1)}_{uni}(n) +\hat a^3 \gamma^{(2)}_{uni}(n) + ... , \qquad \hat a=\frac{\alpha N_c}{4\pi}\,,  \label{uni1}
\end{eqnarray}
where
\begin{eqnarray}
\frac{1}{4} \, \gamma^{(0)}_{uni}(j+2) &=& - S_1(n),  \label{uni1.1} \\
\frac{1}{8} \, \gamma^{(1)}_{uni}(j+2) &=& \Bigl(S_{3}(n) + \overline S_{-3}(n) \Bigr) -
2\,\overline S_{-2,1}(n) + 2\,S_1(n)\Bigl(S_{2}(n) + \overline S_{-2}(n) \Bigr),  \label{uni1.2} \\
\frac{1}{32} \, \gamma^{(2)}_{uni}(j+2) &=& 2\,\overline S_{-3}(n)\,S_2(n) -S_5(n) -
2\,\overline S_{-2}(n)\,S_3(n) - 3\,\overline S_{-5}(n)  +24\,\overline S_{-2,1,1,1}(n)\nonumber\\
&&\hspace{-2.5cm}+ 6\biggl(\overline S_{-4,1}(n) + \overline S_{-3,2}(n) + \overline S_{-2,3}(n)\biggr)
- 12\biggl(\overline S_{-3,1,1}(n) + \overline S_{-2,1,2}(n) + \overline S_{-2,2,1}(n)\biggr)\nonumber \\
&& \hspace{-2.5cm}  -
\biggl(S_2(n) + 2\,S_1^2(n)\biggr) \biggl( 3 \,\overline S_{-3}(n) + S_3(n) - 2\, \overline S_{-2,1}(n)\biggr)
- S_1(n)\biggl(8\,\overline S_{-4}(n) + \overline S_{-2}^2(n)\nonumber \\
&& \hspace{-2.5cm}  + 4\,S_2(n)\,\overline S_{-2}(n) +
2\,S_2^2(n) + 3\,S_4(n) - 12\, \overline S_{-3,1}(n) - 10\, \overline S_{-2,2}(n) + 16\, \overline S_{-2,1,1}(n)\biggr)
\label{uni1.5}
\end{eqnarray}
with
$S_{\pm a,\pm b,\pm c,...}(n)$ (see Eq. (\ref{Sin}))
and 
\begin{eqnarray}
\overline S_{-a,b,c,\cdots}(n) ~=~ (-1)^n \, S_{-a,b,c,...}(n)
+ S_{-a,b,c,\cdots}(\infty) \, \Bigl( 1-(-1)^n \Bigr).  \label{ha3}
\end{eqnarray}

The expression~(\ref{ha3}) is the analytical continuation (to real and 
complex $n$) \cite{AnalCont} of the harmonic sums $S_{-a,b,c,\cdots}(n)$
(see discussions andout the analytical continuation and its applications in Appendix C)

The results for $\gamma^{(3)}_{uni}(n)$ \cite{KLRSV,KoReZi}, $\gamma^{(4)}_{uni}(n)$ \cite{LuReVe}, $\gamma^{(5)}_{uni}(n)$
\cite{Marboe:2014sya} and  $\gamma^{(6)}_{uni}(n)$ \cite{Marboe:2016igj}
are obtained from the long-range asymptotic Bethe equations \cite{Staudacher:2004tk} 
for twist-two operators and the additional contribution
of the wrapping corrections.
The similar calculations for the anomalous dimension in the twist-three case can be found in \cite{Beccaria:2007pb}.
\\



Similary to the eqs. (\ref{FI8})  and  (\ref{FI12})
let us to introduce the
transcendentality level $i$ for the 
harmonic sums $S_{\pm a}(n)$ and
and Euler-Zagier constants $\zeta(\pm a)$
in the following way
\be
S_{\pm a,\pm b,\pm c,\cdots }(n) \sim \zeta(\pm a,\pm b,\pm c,\cdots )
\sim 1/n^i, ~~~~(i=a+b+c+ \cdots) \, . \label{Tran}
\ee

Then, the basic functions $\gamma_{uni}^{(0)}(n)$, $\gamma _{uni}^{(1)}(n)$ and $\gamma _{uni}^{(2)}(n)$ are
assumed to be of the types $\sim 1/n^{i}$ with the levels $i=1$, $i=3$ and
$i=5$, respectively. 
A violation of this property may be obtained from contributions of 
the terms appearing at a given order from previous orders of the perturbation theory. Such
contributions could be generated and/or removed using an appropriate finite
renormalization and/or redefinition of the coupling constant. But these terms do not appear in
the ${\overline{\mathrm{DR}}}$-scheme \cite{DRED}.

It is known, that at the first three orders of perturbation theory
(with the SUSY relation for the QCD color factors $C_{F}=C_{A}=N_{c}$) the
most complicated contributions (with $i=1,~3$ and $5$, respectively) are the
same as
in QCD~\cite{VMV}.
This property allows one to find the
universal anomalous dimensions 
$\gamma _{uni}^{(0)}(n)$, $\gamma _{uni}^{(1)}(n)$ and 
$\gamma_{uni}^{(2)}(n)$ without knowing all elements of the anomalous dimension matrix~\cite{KL},
which was verified for $\gamma _{uni}^{(1)}(j)$ by the exact 
calculations in~\cite{KoLiVe}.\\

Note that in ${\mathcal N}=4$ SYM some partial cases of anomalous dimension are also known for the large couplings
from string calculations and AdS/QFT correspondence \cite{Maldacena:1997re}. We would like to note
that if the property of  the maximal transcendentality exists at low coupling, then sometimes it
appeares at large couplings (see, for example, the results for the cusp  anomalous dimension at low \cite{Kotikov:2006ts}
and large  \cite{Basso:2007wd} couplings,
both of which are based on
the Beisert-Eden-Staudacher equation \cite{Beisert:2006ez}). However, this is not true for Pomeron intercept,
which results lose the property of  maximal transcendentality at large couplings (see \cite{KLOV,Brower:2006ea,Costa:2012cb}).
The reason of the difference in the results for the cusp anomalous dimension
and Pomeron intercept is currently not clear. More research is required.


\section{Conclusion}

In this review, we presented the results of the calculation of some massless and massive two-loop Feynman integrals.
In the massless case, we considered scalar two-point diagrams with a traceless product in the numerator of one
propagator and diagrams depending on two momenta $q$ and $p$ when $p^2=0$.
In the massive case, we studied the $1/m^2$-expansion  of Feynman integrals.
The similarity of the structure of the expansion coefficients of massless and massive diagrams is shown.

Evolutions of the most complicated parts: $\Phi_i(n)$ $(i=1,2)$ for massless diagrams and $F_j(n)$ $(j=1,2,3,4)$ for
massive diagrams (see Subsections 3.2.1, 3.3.1, 6.1 and 6.2, respectively), are shown in details.

The main calculation method for $\Phi_i(n)$ and $F_j(n)$ is to build recurrence relations (i.e. the relationships between
$\Phi_i(n)$ and $\Phi_i(n-1)$ and also $F_j(n)$ and $F_j(n-1)$), which ingomogeneous parts contains only simpler amounts.
The calculation of the series in the ingomogeneous parts allows us to get complete information about the recurrence
relations and, as a result, solve them and get the desired values for the studied $\Phi_i(n)$ $(i=1,2)$ and
$F_j(n)$ $(j=1,2,3,4)$.

Note that this approach is close to the method of differential equations for calculating massive diagrams.
This is not surprising, since the differential equations for any function correspond to recurrence relations for the
coefficients of its expansion.

For the massless and massive diagrams under consideration, the level of  transcendentality (or complexity, or weight)
remains unchanged in any order of $\ep$ (see Subsections 6.3 and 6.4).
Moreover, it decreases in the presence of logarithms or $\zeta$-functions.
It is called as  {\it transcendentality principle}.
Its application leads to the possibility to get the results for most of integrals
without direct calculations.

This property is violated in physical models, such as QCD, where propagators (both quarks and gluons) contain momenta in
their numerators, which are responsible for mixing levels of complexity.
However, this property is preserved (see Section 7) for diagonalized quantities, such as diagonal anomalous
dimensions and coefficient
functions in $N=4$ SYM, which is an amazing but poorly studied property.\\

Author 
thanks the Organizing Committee of Helmholtz International Summer School
``Quantum Field Theory at the Limits: from Strong Fields to Heavy Quarks''
for invitation. He is grateful to Andrei Pikelner for his help with the Axodraw2.


\section{Appendix A, Some useful formulas}
\label{sec:a}
\def\theequation{A\arabic{equation}}
\setcounter{equation}0

{\bf A.}~~ Consider firstly the simple series
\be
\hat{\Phi}_1(n;a,b) = \sum_{m=0}^n \, \frac{\Gamma(m+a)}{\Gamma(m+b)} \, .
\label{SimSe}
\ee

It is conveninet to rewrite it throught ${}_2F_1$-hypergeometric functions with the argument $1$,
which are exactly expressed as a product of $\Gamma$-functions
\be
   {}_2F_1(a,b;c;1) = \frac{\Gamma(c)\Gamma(c-a-b)}{\Gamma(c-a)\Gamma(c-b)}~~ \mbox{for}~~ c-a-b>0 \, .
\label{F21}
\ee   

Consider $\hat{\Phi}_1(a,b)$ as
\be
\hat{\Phi}_1(n;a,b) = \sum_{m=0}^{\infty} \, \frac{\Gamma(m+a)}{\Gamma(m+b)} -\sum_{m=n+1}^{\infty} \,
\frac{\Gamma(m+a)}{\Gamma(m+b)}
=  \sum_{m=0}^{\infty} \, \left( \frac{\Gamma(m+a)}{\Gamma(m+b)} - \frac{\Gamma(m+a+n+1)}{\Gamma(m+b+n+1)} \right) \, .
\label{SimSe.1}
\ee

Using (\ref{F21}), we have
\be
\hat{\Phi}_1(n;a,b) = \frac{1}{b-a-1} \, \left[\frac{\Gamma(a)}{\Gamma(b-1)} - \frac{\Gamma(a+n+1)}{\Gamma(b+n)} \right]
\, . 
\label{SimSe.2}
\ee

Now we consider another series
\be
\hat{\Phi}_2(n;\alpha,\beta) = \sum_{m=0}^n \, \frac{\Gamma(m+\beta+1)}{\Gamma(n-m+1+\alpha)} \, .
\label{NoSimSe}
\ee

Rewriting
\be
\frac{\Gamma(n-m+1+\alpha)}{\Gamma(1+\alpha)} = (-1)^{n-m} \, \frac{\Gamma(-\alpha)}{\Gamma(-\alpha-n+m)} \, ,
\label{ration}
\ee
we see that
\bea
\hat{\Phi}_2(n;\alpha,\beta) &=& \frac{(-1)^n}{\Gamma(-\alpha)\Gamma(1+\alpha)} \,  \hat{\Phi}_1(-n-\alpha,\beta)
\nonumber \\
&=&  \frac{(-1)^n}{\Gamma(-\alpha)\Gamma(1+\alpha)} \,  \frac{1}{\alpha+\beta+n} \,
\left[\frac{\Gamma(-n-\alpha)}{\Gamma(\beta)} - \frac{\Gamma(1-\alpha)}{\Gamma(\beta+n+1)} \right]
\, . 
\label{NoSimSe.2}
\eea

Since
\be
\Gamma(-n-\alpha) = (-1)^n \, \frac{\Gamma(-\alpha)\Gamma(1+\alpha)}{\Gamma(n+1+\alpha)} \, ,
\label{ration.1}
\ee
we obtain the useful relation
\be
\hat{\Phi}_2(n;\alpha,\beta) =  \frac{1}{\alpha+\beta+n} \,
\left[\frac{1}{\Gamma(n+1+\alpha)\Gamma(\beta)} + \frac{(-1)^n}{\Gamma(\alpha)\Gamma(\beta+n+1)} \right]
\, ,
\label{NoSimSe.3}
\ee
which has important particular cases
\be
\hat{\Phi}_2(n;\alpha=0,\beta) =  \frac{1}{\beta+n} \, \frac{1}{\Gamma(n+1)\Gamma(\beta)},~~
\hat{\Phi}_2(n;\alpha,\beta=0) =  \frac{(-1)^n}{\alpha+n} \, \frac{1}{\Gamma(n+1)\Gamma(\alpha)} \, . \label{NoSimSe.5} 
\ee

{\bf B.}~~ Consider $\overline{C}_2^{(4)}(0,n)$ considered in Eq. (\ref{oC24.0.2a}). It has the form
\bea
&&\overline{C}_2^{(4)}(0,n)= - \sum_{k=0}^{n+1} \,
\frac{(-1)^k n! S_1(k) S_1(n+1-k)}{k! (n-k+1)!} \nonumber \\
&&=- \frac{\partial}{\partial x} \, \frac{\partial}{\partial y} \, \sum_{k=0}^{n+1} \,
\frac{(-1)^k n! \Gamma(1-x) \Gamma(1-y)}{\Gamma(k+1-x) \Gamma(n-k+2-y)}\Biggr{|}_{x=0,y=0} \, .
\label{oC24.1A} 
\eea

Taking the results (\ref{NoSimSe.3}) we have
\bea
&&\overline{C}_2^{(4)}(0,n)= - \frac{\partial}{\partial x} \, \frac{\partial}{\partial x} \, \frac{n!}{n+2-x-y} \,
\left[\frac{(-x)\Gamma(1-y)}{\Gamma(n+3-y)} + \frac{(-1)^n \Gamma(1-x)}{\Gamma(n+2-x)}\right] \Biggr{|}_{x=0,y=0}
\nonumber \\
&&= - n! \,  \frac{\partial}{\partial x} \, \left[\frac{1}{(n+2-x)^2} \, \left(\frac{(-x)}{\Gamma(n+3)}
  + \frac{(-1)^n \Gamma(1-x)}{\Gamma(n+2-x)}\right) + \frac{(-x)}{(n+2-x)} \, \frac{S_1(n+2)}{\Gamma(n+3)}\right]
\Biggr{|}_{x=0} \, . \nonumber
\eea

Performing the derivative with respect of $x$, we can
parametrize the final result in the form
\be
\overline{C}_2^{(4)}(0,n)= - \frac{n!}{n+2} \, \Bigl[A_1(n) + (-1)^n A_2(n) \Bigr] \, ,
\label{oC24.2A} 
\ee
where 
\bea
&&A_1(n) = - \frac{1}{(n+2)^2} \, \frac{1}{\Gamma(n+2)} -\frac{S_1(n+2)}{\Gamma(n+3)} = - \frac{1}{\Gamma(n+3)}
\, \left(S_1(n+2) + \frac{1}{n+2} \right) \, , \label{A1n} \\
&&A_2(n) = \frac{2}{(n+2)^2} \, \frac{1}{\Gamma(n+2)} +\frac{S_1(n+1)}{\Gamma(n+3)} = \frac{1}{\Gamma(n+3)}
\, \left(S_1(n+1) + \frac{2}{n+2} \right) = - A_1(n) \, . \label{A1n}
\eea

So, we see that $\overline{C}_2^{(4)}(0,n)=0$ for even $n$ values.\\

{\bf C.}~~ Consider the series
\be
S_{1,1}(n) = \sum_{m=1}^n \, \frac{S_{1}(m)}{m},~~S_{1,2}(n) = \sum_{m=1}^n \, \frac{S_{2}(m)}{m},~~\sum_{m=1}^n \,
\frac{S^2_{1}(m)}{m}
\label{Sij.A}
\ee
and expess them trough standard sums $S_{i}(n)$ and the one $S_{2,1}(n)$.

For the first series $S_{1,1}(n)$ we have
\bea
S_{1,1}(n)=\sum_{m=1}^n \, \frac{1}{m} \,  \sum_{l=1}^m \, \frac{1}{l} = \sum_{l=1}^m \, \frac{1}{l} \,
\sum_{m=l}^n \, \frac{1}{m} = \sum_{l=1}^m \, \frac{1}{l} \, \left(S_1(n)-S_1(l) + \frac{1}{l} \right) \, .
\label{S11.A}
\eea
Comparing l.h.s. and r.h.s., we see
\be
S_{1,1}(n)=\frac{1}{2} \, \sum_{l=1}^m \, \frac{1}{l} \, \left(S_1(n) + \frac{1}{l} \right)
= \frac{1}{2} \, \Bigl(S_1^2(n) + S_2(n)\Bigr) \, .
\label{S11.A1}
\ee

By analogy with the case $S_{1,1}(n)$ we can obtain
\be
S_{1,2}(n)= \sum_{l=1}^m \, \frac{1}{l} \, \left(S_1(n)-S_1(l) + \frac{1}{l} \right)
= S_1(n)S_2(n) + S_{2,1}(n)+S_3(n) \, 
\label{S12.A}
\ee
and
\bea
&&\sum_{m=1}^n \, \frac{S^2_{1}(m)}{m} = \sum_{m=1}^n \, \frac{S_{1}(m)}{m} \,  \sum_{l=1}^m \, \frac{1}{l}
=\sum_{l=1}^n \, \frac{1}{l} \, \sum_{m=l}^n \, \frac{S_1(m)}{m} \nonumber \\
&&=
\sum_{l=1}^m \, \frac{1}{l} \, \left[ \sum_{m=1}^n \, \frac{S_1(m)}{m} -  \sum_{m=1}^l \, \frac{S_1(m)}{m}
+  \frac{S_1(l)}{l} \right] \, .
\label{S111.A}
\eea

Taking the results (\ref{S11.A1}), we have
\be
\sum_{m=1}^n \, \frac{S^2_{1}(m)}{m} =   \frac{1}{2} \, \Biggl[ \Bigl(S_1^2(n) + S_2(n)\Bigr) \, S_1(n) -
\sum_{l=1}^m \, \frac{1}{l} \, \Bigl(S_1^2(l) + S_2(l)\Bigr) +  2S_{2,1}(n) \Biggr] \, .
\label{S111.A1}
\ee

Comparing l.h.s. and r.h.s., we see
\be
\sum_{m=1}^n \, \frac{S^2_{1}(m)}{m} = \frac{1}{3} \, \left[
  \, \Bigl(S_1^2(n) + S_2(n)\Bigr) \, S_1(n) -
  \, S_{1,2}(n) + 2 S_{2,1}(n) \right] \, .
\label{S111.A2}
\ee

Taking the results (\ref{S12.A}), we have the final result
\be
\sum_{m=1}^n \, \frac{S^2_{1}(m)}{m} = \frac{1}{3} \, \left[ S_1^3(n) - S_3(n)\right] +  S_{2,1}(n) \, .
\label{S111.A3}
\ee

The evaluation of the more complicated sums can be found in Ref. \cite{Blumlein:2009cf}.

\section{Appendix B,
  method of ``projectors''}
\label{sec:a}
\def\theequation{B\arabic{equation}}
\setcounter{equation}0

In Refs. \cite{Kazakov:1986mu,Kotikov:1987mw}
we used a special case of the “projection” method \cite{Gorishnii:1983su,Tkachov:1983st} — is so-called
“differentiation” method,
which allows for a diagram depending on two momenta $p$ and $q$, where $p^2 = 0$,
to obtain the coefficients for powers of $[2(pq)]/q^2$.
\footnote{In the case of nonzero quark masses, similar canculations were done in Ref. \cite{Kotikov:2001ct}.}
These coefficients will be
call the ``moments''.

As a first example, we consider
the diagram $J_1(\alpha,q,p)$
%
\vskip 0.5cm
\bea
J_1(\alpha,q,p) =
\,
\raisebox{1mm}{{
    \begin{axopicture}(90,10)(0,4)
  \SetWidth{1.0}
\Arc(45,-7)(40,20,90)
\Arc(45,-7)(40,90,160)
\Line(45,-25)(45,35)
\Arc(45,17)(40,200,340)
\Vertex(45,-23){2}
\Vertex(45,33){2}
\Vertex(5,5){2}
\Vertex(85,5){2}
\Line(5,5)(-5,5)
\Line(85,5)(95,5)
\Line(45,-25)(45,-35)
\Line(85,5)(95,10)
\Text(75,-20)[t]{$\alpha$}
\Text(-3,-5)[b]{$\to$}
\Text(-3,-12)[b]{$q$}
\Text(93,-5)[b]{$\to$}
\Text(93,-12)[b]{$q$}
\Text(93,20)[b]{$\to$}
\Text(93,12)[b]{$p$}
\Text(52,-30)[b]{$\to$}
\Text(52,-38)[b]{$p$}
\end{axopicture}
}}
\hspace{3mm}
\underset{p^2=0}{=}
\sum_{k} J_(\alpha,k) \,
\frac{2^k p_{\lambda_1} ... p_{\lambda_k} q^{\lambda_1} ... q^{\lambda_k}}{q^{2(k+\alpha+2\ep)}} \, .
\nonumber
\eea
\vskip 1cm

We differentiate expression (6) on the right and on the left $n$ times with respect to $p$ and set
$p = 0$. On the left we have\\
%
\be
\frac{d}{dp_{\mu_1}} ... \frac{d}{dp_{\mu_n}} \, \biggl\{
\, \hspace{3mm}
\raisebox{1mm}{{
    \begin{axopicture}(90,10)(0,4)
  \SetWidth{1.0}
\Arc(45,-7)(40,20,90)
\Arc(45,-7)(40,90,160)
\Line(45,-25)(45,35)
\Arc(45,17)(40,200,340)
\Vertex(45,-23){2}
\Vertex(45,33){2}
\Vertex(5,5){2}
\Vertex(85,5){2}
\Line(5,5)(-5,5)
\Line(85,5)(95,5)
\Line(45,-25)(45,-35)
\Line(85,5)(95,10)
\Text(75,-20)[t]{$\alpha$}
\Text(-3,-5)[b]{$\to$}
\Text(-3,-12)[b]{$q$}
\Text(93,-5)[b]{$\to$}
\Text(93,-12)[b]{$q$}
\Text(93,20)[b]{$\to$}
\Text(93,12)[b]{$p$}
\Text(52,-30)[b]{$\to$}
\Text(52,-38)[b]{$p$}
\end{axopicture}
}}
\hspace{3mm}
\Biggr\}
\Biggr|_{p=0}
= \hat{S} \,
\frac{2^{n} \Gamma(n+\alpha)}{\Gamma(\alpha)} \, \hspace{3mm}
\raisebox{1mm}{{
    \begin{axopicture}(90,10)(0,4)
  \SetWidth{1.0}
\Arc(45,-7)(40,20,90)
\Arc(45,-7)(40,90,160)
\Line(45,-25)(45,35)
\Arc(45,17)(40,200,270)
\Arc[arrow](45,17)(40,270,340)
\Vertex(45,-23){2}
\Vertex(45,33){2}
\Vertex(5,5){2}
\Vertex(85,5){2}
\Line(5,5)(-5,5)
\Line(85,5)(95,5)
\Text(63,-10)[b]{$\scriptstyle n$}
\Text(75,-22)[t]{$\scriptstyle \alpha+n$}
\Text(-3,-5)[b]{$\to$}
\Text(-3,-12)[b]{$q$}
\end{axopicture}
}} 
\hspace{3mm} \, ,
\nonumber
\ee
\vskip 1cm
\hspace{-8mm} where $\hat{S}$ is a symmetrization on indeces: $\lambda_i$, $\mu_j$ ($i=1, ...,m$, $j=1, ...,n$).

On the right we get
\bea
\sum_k \, J_1(\alpha,k) \, \frac{2^k q^{\nu_1} ... q^{\nu_k}}{q^{2(k+\alpha+2\ep)}} \, \frac{d}{dp_{\mu_1}} ... \frac{d}{dp_{\mu_n}} \,
\Big( p^{\nu_1} ... p^{\nu_k} \Bigr)\biggl|_{p=0} = \hat{S} \, n! \,  J_1(\alpha,n) \, \frac{2^n q^{\nu_1} ... q^{\nu_n}}{q^{2(n+\alpha+2\ep)}} \, .
\nonumber
\eea

Therefore, for the moment $J_1(\alpha,n)$
of the diagram we have the following expression:
\vskip 0.5cm
\bea
J_1(\alpha,n) \, \frac{q^{\nu_1} ... q^{\nu_n}}{q^{2(n+\alpha+2\ep)}} =
\hat{S} \, 
\frac{
  \Gamma(n+\alpha)}{n!\Gamma(\alpha)} \, \hspace{3mm}
\raisebox{1mm}{{
    \begin{axopicture}(90,10)(0,4)
  \SetWidth{1.0}
\Arc(45,-7)(40,20,90)
\Arc(45,-7)(40,90,160)
\Line(45,-25)(45,35)
\Arc(45,17)(40,200,270)
\Arc[arrow](45,17)(40,270,340)
\Vertex(45,-23){2}
\Vertex(45,33){2}
\Vertex(5,5){2}
\Vertex(85,5){2}
\Line(5,5)(-5,5)
\Line(85,5)(95,5)
\Text(63,-10)[b]{$\scriptstyle n$}
\Text(75,-22)[t]{$\scriptstyle \alpha+n$}
\Text(-3,-5)[b]{$\to$}
\Text(-3,-12)[b]{$q$}
\end{axopicture}
}}
\hspace{3mm}
\overset{\alpha=1}{=}
\hat{S} \,
\hspace{3mm}
\raisebox{1mm}{{
    \begin{axopicture}(90,10)(0,4)
  \SetWidth{1.0}
\Arc(45,-7)(40,20,90)
\Arc(45,-7)(40,90,160)
\Line(45,-25)(45,35)
\Arc(45,17)(40,200,270)
\Arc[arrow](45,17)(40,270,340)
\Vertex(45,-23){2}
\Vertex(45,33){2}
\Vertex(5,5){2}
\Vertex(85,5){2}
\Line(5,5)(-5,5)
\Line(85,5)(95,5)
\Text(63,-10)[b]{$\scriptstyle n$}
\Text(75,-22)[t]{$\scriptstyle n+1$}
\Text(-3,-5)[b]{$\to$}
\Text(-3,-12)[b]{$q$}
\end{axopicture}
}}
\hspace{3mm} \, .
\nonumber
\eea
\vskip 1cm
\hspace{-8mm} Next, we will neglect
the symmetrizator $\hat{S}$.

Note that this transformation from the diagram to its moment remains valid for arbitrary indices of
the diagram lines, as well as the presence of additional momentums in the propagators of the diagram
(if the latter are located on a differentiable line, then some changes will be required).\\

The second consider diagram is
\vskip 1cm
\bea
\overline{J}_1(\alpha,q,p) =
\raisebox{1mm}{{
    \begin{axopicture}(90,10)(0,4)
  \SetWidth{1.0}
\Arc(45,-7)(40,20,90)
\Arc(45,-7)(40,90,160)
\Line(45,-25)(45,35)
\Arc(45,17)(40,200,340)
\Vertex(45,-23){2}
\Vertex(45,33){2}
\Vertex(5,5){2}
\Vertex(85,5){2}
\Vertex(65,-18){2}
\Line(5,5)(-5,5)
\Line(85,5)(95,5)
\Line(65,-18)(65,-28)
\Line(85,5)(95,10)
\Text(72,0)[t]{$\alpha$}
\Text(-3,-5)[b]{$\to$}
\Text(-3,-12)[b]{$q$}
\Text(93,-5)[b]{$\to$}
\Text(93,-12)[b]{$q$}
\Text(93,20)[b]{$\to$}
\Text(93,12)[b]{$p$}
\Text(72,-25)[b]{$\to$}
\Text(72,-32)[b]{$p$}
\end{axopicture}
}}
\hspace{3mm}
\, .
\nonumber
\eea
\vskip 1cm

By analogy with the previous diagram we have for its moments:
\vskip 0.5cm
\bea
\overline{J}_1(\alpha,n) \, \frac{q^{\nu_1} ... q^{\nu_n}}{q^{2(n+\alpha+2\ep)}} =
\hat{S} \, 
\frac{
  \Gamma(n+\alpha)}{n!\Gamma(\alpha)} \, \hspace{3mm}
\raisebox{1mm}{{
    \begin{axopicture}(90,10)(0,4)
  \SetWidth{1.0}
\Arc(45,-7)(40,20,90)
\Arc(45,-7)(40,90,160)
\Line(45,-25)(45,35)
\Arc(45,17)(40,200,270)
\Arc[arrow](45,17)(40,270,340)
\Vertex(45,-23){2}
\Vertex(45,33){2}
\Vertex(5,5){2}
\Vertex(85,5){2}
\Line(5,5)(-5,5)
\Line(85,5)(95,5)
\Text(63,-10)[b]{$\scriptstyle n$}
\Text(75,-22)[t]{$\scriptstyle \alpha+n+1$}
\Text(-3,-5)[b]{$\to$}
\Text(-3,-12)[b]{$q$}
\end{axopicture}
}}
\hspace{3mm}
\overset{\alpha=1}{=}
\hat{S} \,
\hspace{3mm}
\raisebox{1mm}{{
    \begin{axopicture}(90,10)(0,4)
  \SetWidth{1.0}
\Arc(45,-7)(40,20,90)
\Arc(45,-7)(40,90,160)
\Line(45,-25)(45,35)
\Arc(45,17)(40,200,270)
\Arc[arrow](45,17)(40,270,340)
\Vertex(45,-23){2}
\Vertex(45,33){2}
\Vertex(5,5){2}
\Vertex(85,5){2}
\Line(5,5)(-5,5)
\Line(85,5)(95,5)
\Text(63,-10)[b]{$\scriptstyle n$}
\Text(75,-22)[t]{$\scriptstyle n+2$}
\Text(-3,-5)[b]{$\to$}
\Text(-3,-12)[b]{$q$}
\end{axopicture}
}}
\hspace{3mm}
\, .
\nonumber
\eea
\vskip 1cm

A similar conclusion can be drawn for the following diagram
\vskip 1cm
\bea
\hat{J}_1(\alpha,\beta,q,p) =
\raisebox{1mm}{{
    \begin{axopicture}(90,10)(0,4)
  \SetWidth{1.0}
\Arc(45,-7)(40,20,90)
\Arc(45,-7)(40,90,160)
\Line(45,-25)(45,35)
\Arc(45,17)(40,200,340)
\Vertex(45,-23){2}
\Vertex(45,33){2}
\Vertex(5,5){2}
\Vertex(85,5){2}
\Vertex(22,-15){2}
\Line(5,5)(-5,5)
\Line(85,5)(95,5)
\Line(22,-15)(10,-25)
\Line(85,5)(95,10)
\Text(75,-20)[t]{$\alpha$}
\Text(35,-25)[t]{$\beta$}
\Text(-3,-5)[b]{$\to$}
\Text(-3,-12)[b]{$q$}
\Text(93,-5)[b]{$\to$}
\Text(93,-12)[b]{$q$}
\Text(93,20)[b]{$\to$}
\Text(93,12)[b]{$p$}
\Text(15,-30)[b]{$\to$}
\Text(15,-38)[b]{$p$}
\end{axopicture}
}}
\hspace{3mm}
\, .
\nonumber
\eea
\vskip 1cm

Its moment has the form
\bea
&&\hat{J}_1(\alpha,\beta,n) \, \frac{q^{\nu_1} ... q^{\nu_n}}{q^{2(n+\alpha+2\ep)}} =
\sum_{k=0}^{n} \, 
\frac{
  \Gamma(k+\beta) \Gamma(n-k+\alpha)}{(n-k)!k!\Gamma(\alpha)\Gamma(\beta)} \, \hspace{3mm}
\raisebox{1mm}{{
    \begin{axopicture}(90,10)(0,4)
  \SetWidth{1.0}
\Arc(45,-7)(40,20,90)
\Arc(45,-7)(40,90,160)
\Line(45,-25)(45,35)
\Arc[arrow](45,17)(40,200,270)
\Arc[arrow](45,17)(40,270,340)
\Vertex(45,-23){2}
\Vertex(45,33){2}
\Vertex(5,5){2}
\Vertex(85,5){2}
\Line(5,5)(-5,5)
\Line(85,5)(95,5)
\Text(25,-10)[b]{$\scriptstyle k$}
\Text(63,-10)[b]{$\scriptstyle n-k$}
\Text(75,-22)[t]{$\scriptstyle \alpha+n-k$}
\Text(25,-22)[t]{$\scriptstyle \beta+k+1$}
\Text(-3,-5)[b]{$\to$}
\Text(-3,-12)[b]{$q$}
\end{axopicture}
}}
\hspace{3mm}
\nonumber \\&&
\nonumber \\&&
\nonumber \\&&
\overset{\alpha=\beta=1}{=}
\, 
\sum_{k=0}^{n} \, \hspace{5mm} \,
\raisebox{1mm}{{
    \begin{axopicture}(90,10)(0,4)
  \SetWidth{1.0}
\Arc(45,-7)(40,20,90)
\Arc(45,-7)(40,90,160)
\Line(45,-25)(45,35)
\Arc[arrow](45,17)(40,200,270)
\Arc[arrow](45,17)(40,270,340)
\Vertex(45,-23){2}
\Vertex(45,33){2}
\Vertex(5,5){2}
\Vertex(85,5){2}
\Line(5,5)(-5,5)
\Line(85,5)(95,5)
\Text(25,-10)[b]{$\scriptstyle k$}
\Text(63,-10)[b]{$\scriptstyle n-m-k$}
\Text(75,-22)[t]{$\scriptstyle n-k+1$}
\Text(25,-22)[t]{$\scriptstyle k+2$}
\Text(-3,-5)[b]{$\to$}
\Text(-3,-12)[b]{$q$}
\end{axopicture}
}}
\hspace{3mm} \, .
\nonumber
\eea
\vskip 1cm

Note that there is another technique to calculate the considered diagrams: the “gluing” method \cite{Chetyrkin:1982zq}.
Using the orthogonality of traceless product, on can obtain the moment of the diagram by an additional integration on the
$q$ momentum with a
propagator,
which has an index $\delta$ and the
additional  tranceless product. This additional integration leads to very complicated three-loop diagrams. For the
considered Feynman integrals  $J_1(\alpha,q,p)$, $\overline{J}_1(\alpha,q,p)$, $\hat{J}_1(\alpha,\beta,q,p)$
these “gluing”  three-loop diagrams have the following form
\vskip 0.5cm
\bea
\raisebox{1mm}
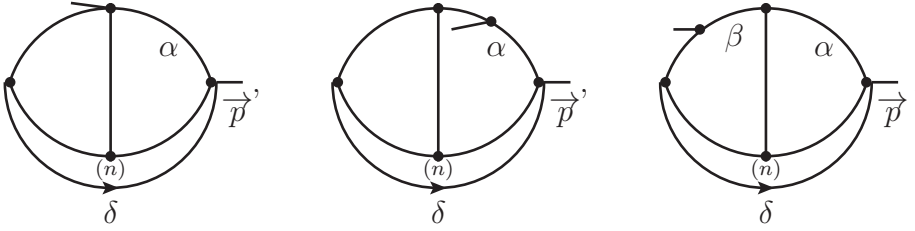
{{
    \begin{axopicture}(90,10)(0,4)
  \SetWidth{1.0}
\Arc(45,-7)(40,20,90)
\Arc(45,-7)(40,90,160)
\Line(45,-25)(45,35)
\Arc(45,17)(40,200,270)
\Arc(45,17)(40,270,340)
\Arc[arrow](45,5)(40,180,360)
\Vertex(45,-23){2}
\Vertex(45,33){2}
\Vertex(7,5){2}
\Vertex(83,5){2}
\Line(45,33)(30,35)
\Line(85,5)(95,5)
\Text(45,-32)[b]{$\scriptstyle (n)$}
\Text(67,20)[t]{$\alpha$}
\Text(45,-40)[t]{$\delta$}
\Text(93,-5)[b]{$\to$}
\Text(93,-12)[b]{$p$}
\end{axopicture}
}}
\hspace{3mm},
\, \hspace{5mm}
\raisebox{1mm}{{
    \begin{axopicture}(90,10)(0,4)
  \SetWidth{1.0}
\Arc(45,-7)(40,20,90)
\Arc(45,-7)(40,90,160)
\Line(45,-25)(45,35)
\Arc(45,17)(40,200,270)
\Arc(45,17)(40,270,340)
\Arc[arrow](45,5)(40,180,360)
\Vertex(45,-23){2}
\Vertex(65,28){2}
\Vertex(45,33){2}
\Vertex(7,5){2}
\Vertex(83,5){2}
\Line(65,28)(50,25)
\Line(85,5)(95,5)
\Text(45,-32)[b]{$\scriptstyle (n)$}
\Text(67,20)[t]{$\alpha$}
\Text(45,-40)[t]{$\delta$}
\Text(93,-5)[b]{$\to$}
\Text(93,-12)[b]{$p$}
\end{axopicture}
}}
\hspace{3mm},
\, \hspace{5mm}
\raisebox{1mm}{{
    \begin{axopicture}(90,10)(0,4)
  \SetWidth{1.0}
\Arc(45,-7)(40,20,90)
\Arc(45,-7)(40,90,160)
\Line(45,-25)(45,35)
\Arc(45,17)(40,200,270)
\Arc(45,17)(40,270,340)
\Arc[arrow](45,5)(40,180,360)
\Vertex(45,-23){2}
\Vertex(45,33){2}
\Vertex(7,5){2}
\Vertex(83,5){2}
\Vertex(20,25){2}
\Line(20,25)(10,25)
\Line(85,5)(95,5)
\Text(45,-32)[b]{$\scriptstyle (n)$}
\Text(67,20)[t]{$\alpha$}
\Text(33,27)[t]{$\beta$}
\Text(45,-40)[t]{$\delta$}
\Text(93,-5)[b]{$\to$}
\Text(93,-12)[b]{$p$}
\end{axopicture}
}}
\hspace{3mm} \, .
\nonumber
\eea
\vskip 1.5cm
\hspace{-8mm} The evaluation of these complicated diagrams is above of the slope of the paper.
Some example of application of the  “gluing” method can be found in Ref. \cite{Kazakov:1986mu}.

As a conclusion of Appendix B, we would like to note, when applying the method of ``projectors''
\cite{Gorishnii:1983su,Tkachov:1983st}, the expression obtained for the
$n$th moment of the initial diagram has alwais much
simpler form than using the “gluing” method \cite{Chetyrkin:1982zq}.

\section{Appendix C, about analytic continuation}
\label{sec:a}
\def\theequation{C\arabic{equation}}
\setcounter{equation}0

Here we demonstrate the direct calculation the particular case $C_2(1,n=0)$ from the general expression
$C_2(1,n)$ in (\ref{C2.1.a4}) by using an analytic continuation (from even $n$ values) of the sum
\be
S_{-2}(n) = \sum_{m=1}^n \, \frac{(-1)^m}{m^2} \, .
\label{S-2}
\ee

Indeed,
using the simple sum, it is very convenient to show the basic steps of the analytic continuation
itself. In the case of more general nested sums $S_{\pm a,\pm b,...}(n)$ formulas are more complex, which may
make it difficult to understand the procedure.

The basic idea of the analytic continuation is very simple: try to translate the argument $n$ from the upper
limit of the sum to the summed expression. After the procedure performed we have a possibility to expand and
to differenciate with respect of $n$ and so on.

Firstly we represent the sum $S_{-2}(n)$ in (\ref{S-2}) as
\be
S_{-2}(n) = \Bigl(\sum_{m=1}^{\infty} - \sum_{m=n+1}^{\infty}\Bigr) \, \frac{(-1)^m}{m^2}
= S_{-2}(\infty) - (-1)^n \,  \sum_{m=1}^{\infty} \, \frac{(-1)^m}{(m+n)^2}
\, .
\label{S-2.2}
\ee
and we see the unpleasant factor $(-1)^n$ in the front of the last term in r.h.s. 

Considering the variable $(-1)^n S_{-2}(n)$:
\be
(-1)^n S_{-2}(n) 
= (-1)^n S_{-2}(\infty) - \sum_{m=1}^{\infty} \, \frac{(-1)^m}{(m+n)^2}
\, ,
\label{S-2.3}
\ee
we move the factor $(-1)^n$ to the front of the first term.

Now we introduce the new $n$-dependent function $\overline{S}_{-2}(n)$ as
\be
\overline{S}_{-2}(n)= (-1)^n S_{-2}(n) + (1-(-1)^n) S_{-2}(\infty)
\, ,
\label{oS-2.1}
\ee
which coincides with the initial one $S_{-2}(n)$ at even $n$ values and have no  the unpleasant factor $(-1)^n$
\be
\overline{S}_{-2}(n)= S_{-2}(\infty) - \sum_{m=1}^{\infty} \, \frac{(-1)^m}{(m+n)^2}
\, .
\label{oS-2.2}
\ee

So, the function $\overline{S}_{-2}(n)$ can be considered as the analytic continuation (from even $n$ values)
of the sum $S_{-2}(n)$.

Now it is possible to consider the small-$n$ limit of $C_2(1,n)$ using the corresponding limit of
$\overline{S}_{-2}(n)$ as
\bea
\overline{S}_{-2}(n=\delta \to 0)&=& S_{-2}(\infty) - \sum_{m=1}^{\infty} \, \frac{(-1)^m}{m^2}
\, \left[1-2\frac{\delta}{m} + O(\delta^2)\right] \nonumber \\
&=& 2\delta \, \sum_{m=1}^{\infty} \, \frac{(-1)^m}{m^3} + O(\delta^2) = 2\delta \,  S_{-3}(\infty)
+ O(\delta^2)
\, .
\label{oS-2.3}
\eea

In r.h.s the function $S_{-3}(\infty)$ coincides with the Euler number $\overline{\zeta}_3$, where
\be
\overline{\zeta}_a= \sum_{m=1}^{\infty} \, \frac{(-1)^m}{m^a} = \left(\frac{1}{2^a}-1\right) \, \zeta_a
= - \frac{3}{4}\, \zeta_3~~\mbox{for}~ a=3 \, .
\label{ozeta3}
\ee

Thus, in the small-$n$ limit we have for $\overline{S}_{-2}(n)$:
\be
\overline{S}_{-2}(n=\delta \to 0) = - \frac{3}{4}\, \zeta_3 \delta + O(\delta^2) \, 
\label{oS-2.4}
\ee
and for the coefficient function $C_2(1,n=0)$ of (\ref{C2.1.a4}) 
\be
C_2(1,n=0) = - \frac{4}{\delta (1+\delta)} \, \overline{S}_{-2}(n=\delta \to 0) = 6\zeta_3 +  O(\delta) \, ,
\label{C21n=0}
\ee
that exactly coincides with $C_1(1,n=0)$.

This analytic continuation has many important uses. For example, to analyze the evolution of parton distributions and DIS
structure functions, there is a popular approach \cite{Parisi:1978jv}
which is based on the Gegenbauer polynomials, which  in-turn are associated with the moments of
parton distributions and structure functions.
Using in the analysis a simple evolution for moments, which is determined here by the simple differential DGLAP equations, at
the last step the parton distributions and/or structure functions are restored by summing (to a certain value $N_{\rm MAX}$) the
Gegenbauer polynomials.

In this analysis, evolution should be performed for both even and odd moments, so the analytic continuation is necessary.
Using this analytical continuation, a lot of analysis of experimental data was performed (see
a review in \cite{Krivokhizhin:2009zz})
by the method described here.

Another important application \cite{Kotikov:1998qt}
is the study of parton distributions and structure functions in the region of small values of
the Bjerken variable $x$, which directly relates with above study of the nested sums at $n \to 0$. The approch includes,
in particular,
an extraction of gluon density and longitudinal structure function from data for structural function $F_2$,
the evolution of parton distributions at low $x$ in nucleon and in nuclei, an ultrahigh-energy asymptotics 
of the neutrino-hadron interaction cross section.
Some review of these studies can be found in Ref. \cite{Kotikov:2007ua}.


\begin{thebibliography}{0}

\bibitem{Peterman:1978tb} 
  A.~Peterman,
  Phys.\ Rept.\  {\bf 53}, 157 (1979).

\bibitem{tHooft:1972tcz}
  G.~'t Hooft and M.~J.~G.~Veltman,
  Nucl.\ Phys.\ B {\bf 44} (1972) 189;
  C.~G.~Bollini and J.~J.~Giambiagi,
  Nuovo Cim.\ B {\bf 12} (1972) 20;
  G.~M.~Cicuta and E.~Montaldi,
  Lett.\ Nuovo Cim.\  {\bf 4} (1972) 329;
  G.~'t Hooft,
  Nucl.\ Phys.\ B {\bf 61} (1973) 455.
  

\bibitem{DEramo:1971hnd}
  M.~D'Eramo {\it et al.},
  Lett.\ Nuovo Cim.\  {\bf 2} (1971) no.17,  878.

\bibitem{Vasiliev:1981dg}
  A.~N.~Vasiliev {\it et al.},
  Theor.\ Math.\ Phys.\  {\bf 47} (1981) 465.

\bibitem{Kazakov:1984km}
  D.~I.~Kazakov,
  Phys.\ Lett.\  {\bf 133B} (1983) 406;
  Theor.\ Math.\ Phys.\  {\bf 58} (1984) 223
   [Teor.\ Mat.\ Fiz.\  {\bf 58} (1984) 343];
  N.~I.~Usyukina,
  Theor.\ Math.\ Phys.\  {\bf 54} (1983) 78
   [Teor.\ Mat.\ Fiz.\  {\bf 54} (1983) 124].
  V.~V.~Belokurov and N.~I.~Usyukina,
  J.\ Phys.\ A {\bf 16} (1983) 2811;
  Theor.\ Math.\ Phys.\  {\bf 79} (1989) 385
   [Teor.\ Mat.\ Fiz.\  {\bf 79} (1989) 63].
  
   
\bibitem{Kazakov:1983pk}
  D.~I.~Kazakov,
  Theor.\ Math.\ Phys.\  {\bf 62} (1985) 84
   [Teor.\ Mat.\ Fiz.\  {\bf 62} (1984) 127].


\bibitem{Kotikov:1995cw}
  A.~V.~Kotikov,
  Phys.\ Lett.\ B {\bf 375} (1996) 240.

 
\bibitem{Kotikov:2018uat}
  A.~V.~Kotikov and S.~Teber,
  Theor.\ Math.\ Phys.\  {\bf 194} (2018) no.2,  284.
  [Teor.\ Mat.\ Fiz.\  {\bf 194} (2018) no.2,  331]

\bibitem{Kotikov:2018wxe}
  A.~V.~Kotikov and S.~Teber,
  Phys.\ Part.\ Nucl.\  {\bf 50} (2019) no.1,  1

  
 \bibitem{Ryder}
   L. H. Ryder, ``Quantum Field Theory'', Cambridge University Press, 1996.

\bibitem{Davydychev:1995mq}
  A.~I.~Davydychev and J.~B.~Tausk,
  Phys.\ Rev.\ D {\bf 53} (1996) 7381
   
\bibitem{Chetyrkin:1982zq}
  K.~G.~Chetyrkin  {\it et al.},
  Phys.\ Lett.\  {\bf 119B} (1982) 407.

\bibitem{Gorishnii:1983su}
  S.~G.~Gorishnii  {\it et al.},
  Phys.\ Lett.\  {\bf 124B} (1983) 217;
  S.~G.~Gorishnii and S.~A.~Larin,
  Nucl.\ Phys.\ B {\bf 283} (1987) 452.
  
\bibitem{Tkachov:1983st}
  F.~V.~Tkachov,
  Phys.\ Lett.\  {\bf 124B} (1983) 212;
  K.~G.~Chetyrkin,
  Phys.\ Lett.\  {\bf 126B} (1983) 371.

\bibitem{Kazakov:1986mu}
  D.~I.~Kazakov and A.~V.~Kotikov,
  Theor.\ Math.\ Phys.\  {\bf 73} (1988) 1264;
  Nucl.\ Phys.\ B {\bf 307} (1988) 721
  [Nucl.\ Phys.\ B {\bf 345} (1990) 299].
  
\bibitem{Kotikov:1987mw}
  A.~V.~Kotikov,
  Theor.\ Math.\ Phys.\  {\bf 78} (1989) 134.



\bibitem{KK92}  D.~I.~Kazakov and A.~V.~Kotikov, 
Phys.\ Lett.\ B \textbf{291} (1992) 171;




\bibitem{Kotikov:1990zs}
  A.~V.~Kotikov,
  Mod.\ Phys.\ Lett.\ A {\bf 6} (1991) 677:
  Int.\ J.\ Mod.\ Phys.\ A {\bf 7} (1992) 1977.

\bibitem{Henn:2014yza}
  J.~M.~Henn and J.~C.~Plefka,
  Lect.\ Notes Phys.\  {\bf 883} (2014) 1.

\bibitem{Tarasov:1996br}
  O.~V.~Tarasov,
  Phys.\ Rev.\ D {\bf 54} (1996) 6479;
  Nucl.\ Phys.\ B {\bf 502} (1997) 455

\bibitem{Lee:2012cn}
  R.~N.~Lee,
  arXiv:1212.2685 [hep-ph];
  J.\ Phys.\ Conf.\ Ser.\  {\bf 523} (2014) 012059;
  R.~N.~Lee, A.~V.~Smirnov and V.~A.~Smirnov,
  JHEP {\bf 1004} (2010) 020
  
\bibitem{KL00}
A.~V.~Kotikov and L.~N.~Lipatov, Nucl.\ Phys.\ \textbf{B582} (2000) 19.


  
\bibitem{BFKL}
L.~N.~Lipatov, Sov.\ J.\ Nucl.\ Phys.\ \textbf{23} (1976) 338;
V.~S.~Fadin {\it et al.},
Phys.\ Lett.\ B \textbf{60} (1975) 50;
E.~A.~Kuraev,{\it et al.},
Sov.\ Phys.\ JETP \textbf{44} (1976) 443;
Sov.\ Phys.\ JETP 
\textbf{45} (1977) 199;
I.~I.~Balitsky and L.~N.~Lipatov,
Sov.\ J.\ Nucl.\ Phys.\ \textbf{28} (1978) 822;
JETP\ Lett.\ \textbf{30} (1979) 355.


\bibitem{next}
V.~S.~Fadin and L.~N.~Lipatov, Phys.\ Lett.\ B \textbf{429} (1998) 127;
G.~Camici and M.~Ciafaloni, Phys.\ Lett.\ B \textbf{430} (1998) 349.

\bibitem{BSSGSO}  L.~Brink {\it et al.},
Nucl.\ Phys.\ B \textbf{121} (1977) 77;
F.~Gliozzi {\it et al.},
Nucl.\ Phys.\ B \textbf{122} (1977) 253.

   
  \bibitem{KL}
A.~V.~Kotikov and L.~N.~Lipatov, Nucl.\ Phys.\ B \textbf{661}  (2003) 19;
in: {\it Proc. of the XXXV
Winter School}, Repino, S'Peterburg, 2001 (hep-ph/0112346).



\bibitem{KoLiVe}
  A.~V.~Kotikov {\it et al.},
Phys.\ Lett.\ B \textbf{557} (2003) 114.

\bibitem{KLOV}
  A.~V.~Kotikov {\it et al.},
Phys.\ Lett.\ B {\bf 595} (2004) 521.

\bibitem{Bianchi:2013sta}
  L.~Bianchi {\it et al.},
  Phys.\ Lett.\ B {\bf 725} (2013) 394


\bibitem{VMV}
  S.~Moch {\it et al.},
Nucl.\ Phys.\ B {\bf 688} (2004) 101;
A.~Vogt {\it et al.},
Nucl.\ Phys.\ B {\bf 691} (2004) 129;
J.~A.~M.~Vermaseren {\it et al.},
  Nucl.\ Phys.\ B {\bf 724} (2005) 3.

\bibitem{Broadhurst:1987ei}
  D.~J.~Broadhurst,
  Z.\ Phys.\  C {\bf 47} (1990) 115.

\bibitem{FleKoVe}  J. Fleischer {\it et al.},
Nucl. Phys. B \textbf{547} (1999) 343;
Acta Phys. Polon. B \textbf{29} (1998) 2611.



\bibitem{Eden:2011we}
  B.~Eden {\it et al.},
  Nucl.\ Phys.\ B {\bf 862} (2012) 193;
  L.~J.~Dixon,
  J.\ Phys.\ A {\bf 44} (2011) 454001;
  L.~J.~Dixon {\it et al.},
  JHEP {\bf 1201} (2012) 024;
  T.~Gehrmann {\it et al.},
  JHEP {\bf 1203} (2012) 101;
  A.~Brandhuber {\it et al.},
  JHEP {\bf 1205} (2012) 082;
  J.~M.~Henn {\it et al.},
  JHEP {\bf 1112} (2011) 024



\bibitem{Schlotterer:2012ny}
  O.~Schlotterer and S.~Stieberger,
  J.\ Phys.\ A {\bf 46} (2013) 475401;
  J.~Broedel {\it et al.},
  Fortsch.\ Phys.\  {\bf 61} (2013) 812;
  S.~Stieberger and T.~R.~Taylor,
  Phys.\ Lett.\ B {\bf 716} (2012) 236.

\bibitem{Eden:2012rr} 
  B.~Eden,
  arXiv:1207.3112 [hep-th];
  R.~G.~Ambrosio {\it et al.},
  JHEP {\bf 1501} (2015) 116;
D.~Chicherin {\it et al.},
  JHEP {\bf 1603} (2016) 031;
  B.~Eden and A.~Sfondrini,
  JHEP {\bf 1602} (2016) 165

\bibitem{Basso:2020xts}
  B.~Basso {\it et al.},
  Phys.\ Rev.\ Lett.\  {\bf 124} (2020) no.16,  161603;
  S.~Caron-Huot  {\it et al.},
  JHEP {\bf 1908}, 016 (2019);
  JHEP {\bf 1807}, 170 (2018);
  Phys.\ Rev.\ Lett.\  {\bf 117} (2016) no.24,  241601;
  S.~Abreu {\it et al.},
  JHEP {\bf 1903} (2019) 123;
  Phys.\ Rev.\ Lett.\  {\bf 122} (2019) no.12,  121603;
  L.~J.~Dixon {\it et al.},
  JHEP {\bf 1702} (2017) 137;
  JHEP {\bf 1702} (2017) 112;
  JHEP {\bf 1601} (2016) 053;
  JHEP {\bf 1406} (2014) 116.

\bibitem{Badger:2019djh}
  S.~Badger {\it et al.},
  Phys.\ Rev.\ Lett.\  {\bf 123} (2019) no.7,  071601;
  D.~Chicherin {\it et al.},
  JHEP {\bf 1903} (2019) 115;
  Phys.\ Rev.\ Lett.\  {\bf 122} (2019) no.12,  121602;
  JHEP {\bf 1903} (2019) 042;
  JHEP {\bf 1805} (2018) 164;
  T.~Gehrmann {\it et al.},
  JHEP {\bf 1810} (2018) 103
  
\bibitem{Bianchi:2018rrj}
  L.~Bianchi  {\it et al.},
  JHEP {\bf 1902} (2019) 134;
  JHEP {\bf 1902} (2019) 182;
  A.~Brandhuber  {\it et al.},
  Phys.\ Rev.\ Lett.\  {\bf 119} (2017) no.16,  161601;
  JHEP {\bf 1611} (2016) 143;
  JHEP {\bf 1608} (2016) 134;
  Phys.\ Rev.\ Lett.\  {\bf 115} (2015) no.14,  141602

\bibitem{Caron-Huot:2020vlo}
  S.~Caron-Huot   {\it et al.},
  arXiv:2003.03120 [hep-th];
  J.~Henn {\it et al.},
  JHEP {\bf 2002} (2020) 019;
  JHEP {\bf 1809} (2018) 012;
  JHEP {\bf 1810} (2018) 059;
  J.~M.~Henn and B.~Mistlberger,
  JHEP {\bf 1905} (2019) 023.

\bibitem{Fotopoulos:2019vac}
  A.~Fotopoulos  {\it et al.},
  JHEP {\bf 2003} (2020) 130;
  S.~Stieberger and T.~R.~Taylor,
  Phys.\ Lett.\ B {\bf 793} (2019) 141;
  Nucl.\ Phys.\ B {\bf 913} (2016) 151;
  Phys.\ Lett.\ B {\bf 750} (2015) 587;
  G.~Puhlfürst and S.~Stieberger,
  Nucl.\ Phys.\ B {\bf 902} (2016) 186

  
\bibitem{deLeeuw:2019qvz}
  M.~De Leeuw {\it et al.},
  arXiv:1912.12231 [hep-th];
  B.~Eden {\it et al.},
  JHEP {\bf 1811} (2018) 097;
  JHEP {\bf 1802} (2018) 170;
  JHEP {\bf 1709}, 156 (2017);
  B.~Eden and A.~Sfondrini,
  JHEP {\bf 1710} (2017) 098;
  B.~Eden and V.~A.~Smirnov,
  JHEP {\bf 1610} (2016) 115.



  
\bibitem{Kotikov:2010gf}
  A.~V.~Kotikov,
  In *Diakonov, D. (ed.): Subtleties in quantum field theory* 150-174
  [arXiv:1005.5029 [hep-th]];
  Phys.\ Part.\ Nucl.\  {\bf 44} (2013) 374;
  A.~V.~Kotikov and A.~I.~Onishchenko,
  arXiv:1908.05113 [hep-th].


\bibitem{Kotikov:2012ac}
  A.~V.~Kotikov,
  Theor.\ Math.\ Phys.\  {\bf 176} (2013) 913;
  Theor.\ Math.\ Phys.\  {\bf 190} (2017) no.3,  391

  
\bibitem{Chetyrkin:1980pr}
  K.~G.~Chetyrkin {\it et al.},
  Nucl.\ Phys.\ B {\bf 174} (1980) 345.

\bibitem{Broadhurst:1996yc}
  D.~J.~Broadhurst and A.~V.~Kotikov,
  Phys.\ Lett.\ B {\bf 441} (1998) 345;
  S.~Teber,
  Phys.\ Rev.\ D {\bf 86} (2012) 025005;
  A.~V.~Kotikov and S.~Teber,
  Phys.\ Rev.\ D {\bf 87}, no. 8, 087701 (2013)
  
\bibitem{Kotikov:1989nm} 
  A.~V.~Kotikov,
  JETP Lett.\  {\bf 58}, 731 (1993)
  [Pisma Zh.\ Eksp.\ Teor.\ Fiz.\  {\bf 58}, 785 (1993)];
  Phys.\ Atom.\ Nucl.\  {\bf 75} (2012) 890;
  A.~V.~Kotikov {\it et al.},
  Phys.\ Rev.\ D {\bf 94} (2016) no.5,  056009;
   Erratum: [Phys.\ Rev.\ D {\bf 99} (2019) no.11,  119901];
  A.~V.~Kotikov and S.~Teber,
  Phys.\ Rev.\ D {\bf 94} (2016) no.11,  114011
   Addendum: [Phys.\ Rev.\ D {\bf 99} (2019) no.5,  059902];
  Phys.\ Rev.\ D {\bf 89} (2014) no.6,  065038


   
\bibitem{Teber:2016unz}
  S.~Teber and A.~V.~Kotikov,
  Theor.\ Math.\ Phys.\  {\bf 190} (2017) no.3,  446

  
\bibitem{Chetyrkin:1981qh}
  K.~G.~Chetyrkin and F.~V.~Tkachov,
  Nucl.\ Phys.\  B {\bf 192} (1981) 159;
  F.~V.~Tkachov,
  Phys.\ Lett.\  B {\bf 100} (1981) 65;




\bibitem{Gorishnii:1984te}
  S.~G.~Gorishnii and A.~P.~Isaev,
  Theor.\ Math.\ Phys.\  {\bf 62} (1985) 232
   [Teor.\ Mat.\ Fiz.\  {\bf 62} (1985) 345];
  D.~J.~Broadhurst,
  Z.\ Phys.\ C {\bf 32} (1986) 249.

\bibitem{Kazakov:1984bw}
  D.~I.~Kazakov,
JINR preprint  JINR-E2-84-410.


\bibitem{Vermaseren:1998uu}
  J.~A.~M.~Vermaseren,
  Int.\ J.\ Mod.\ Phys.\ A {\bf 14} (1999) 2037


\bibitem{Broadhurst:1999xk}
  D.~J.~Broadhurst,
  hep-th/9909185.

 
  
\bibitem{Kotikov:2019bqo}
  A.~V.~Kotikov and S.~Teber,
  Phys.\ Rev.\ D {\bf 100} (2019) no.10,  105017;
  arXiv:1912.10957 [hep-th].




\bibitem{Kotikov:1990kg}
  A.~V.~Kotikov,
  Phys.\ Lett.\  B {\bf 254} (1991) 158:
  Phys.\ Lett.\  B {\bf 259} (1991) 314;
  Phys.\ Lett.\  B {\bf 267} (1991) 123;
  Mod.\ Phys.\ Lett.\ A {\bf 6} (1991) 3133;
  E.~Remiddi,
  Nuovo Cim.\  A {\bf 110} (1997) 1435.


\bibitem{Kniehl:2005bc}
  B.~A.~Kniehl {\it et al.},
  Nucl.\ Phys.\  B {\bf 738} (2006) 306;
  Nucl.\ Phys.\ B {\bf 948} (2019) 114780

\bibitem{Kniehl:2005yc}
  B.~A.~Kniehl and A.~V.~Kotikov,
  Phys.\ Lett.\ B {\bf 638} (2006) 531;
  Phys.\ Lett.\ B {\bf 712} (2012) 233.



\bibitem{Fleischer:1999hp}
  J.~Fleischer {\it et al.},
  Phys.\ Lett.\  B {\bf 462} (1999) 169.


\bibitem{Fleischer:1997bw}
  J.~Fleischer {\it et al.},
  Phys.\ Lett.\  B {\bf 417} (1998) 163

\bibitem{Kotikov:2007vr}
  A.~Kotikov {\it et al.},
  Nucl.\ Phys.\ B {\bf 788} (2008) 47.



\bibitem{Kniehl:2006bg}
  B.~A.~Kniehl {\it et al.},
  Phys.\ Rev.\ Lett.\  {\bf 97} (2006) 042001;
  Phys.\ Rev.\  D {\bf 79} (2009) 114032;
  Phys.\ Rev.\ Lett.\  {\bf 101} (2008) 193401;
Phys. Rev. A {\bf 80} (2009) 052501;


  
\bibitem{Gehrmann:2011xn}
  T.~Gehrmann  {\it et al.},
  JHEP {\bf 1203} (2012) 101.
  
  
\bibitem{Henn:2013pwa}
  J.~M.~Henn,
  Phys.\ Rev.\ Lett.\  {\bf 110} (2013) 251601;
  J.\ Phys.\ A {\bf 48} (2015) 153001

\bibitem{Adams:2018yfj}
  L.~Adams and S.~Weinzierl,
  Phys.\ Lett.\ B {\bf 781} (2018) 270

\bibitem{Lee:2014ioa}
  R.~N.~Lee,
  JHEP {\bf 1504} (2015) 108;
  JHEP {\bf 1810} (2018) 176;
  R.~N.~Lee and A.~I.~Onishchenko,
  JHEP {\bf 1912} (2019) 084;


\bibitem{Duhr:2020kzd}
  C.~Duhr {\it et al.},
  arXiv:2004.04752 [hep-ph];
  arXiv:2001.07717 [hep-ph];
  J.~Henn {\it et al.},
  arXiv:2002.09492 [hep-ph];
  H.~Frellesvig {\it et al.},
  arXiv:2002.07776 [hep-ph];
  A.~Bissi {\it et al.},
  arXiv:2002.04604 [hep-th];
  C.~Dlapa {\it et al.},
  arXiv:2002.02340 [hep-ph];
  M.~Misiak {\it et al.},
  arXiv:2002.01548 [hep-ph];
  C.~Anastasiou {\it et al.},
  arXiv:2001.06295 [hep-ph];
  M.~L.~Czakon and M.~Niggetiedt,
  arXiv:2001.03008 [hep-ph].
  
\bibitem{Duhr:2014woa}
  C.~Duhr,
  arXiv:1411.7538 [hep-ph].





\bibitem{Devoto:1983tc}
  A.~Devoto and D.~W.~Duke,
  Riv.\ Nuovo Cim.\  {\bf 7N6} (1984) 1.
  

\bibitem{Remiddi:1999ew}
  E.~Remiddi and J.~A.~M.~Vermaseren,
  Int.\ J.\ Mod.\ Phys.\ A {\bf 15} (2000) 725
  

\bibitem{Davydychev:2003mv}
  A.~I.~Davydychev and M.~Y.~Kalmykov,
  Nucl.\ Phys.\  B {\bf 699} (2004) 3.


  
\bibitem{DGLAP}
V.~N.~Gribov and L.~N.~Lipatov, Sov.\ J.\ Nucl.\ Phys.\ \textbf{15} (1972) 438;
\textbf{15} (1972) 675;
L.~N.~Lipatov, Sov.\ J.\ Nucl.\ Phys.\ \textbf{20} (1975) 94;
G.~Altarelli and G.~Parisi, Nucl.\ Phys.\ B \textbf{126} (1977) 298;
Yu.~L. Dokshitzer, Sov.\ Phys.\ JETP \textbf{46} (1977) 641.


  

\bibitem{LN4}
L.N. Lipatov, 
Nucl. Phys. Proc. Suppl. \textbf{99A} (2001) 175.


\bibitem{AnalCont}
A.V. Kotikov, Phys. At. Nucl. \textbf{57} (1994) 133;
A.~V.~Kotikov and V.~N.~Velizhanin,
in: {\it Proc. of the XXXIX
Winter School}, Repino, S'Peterburg, 2005 (hep-ph/0501274).


\bibitem{KLRSV}
  A.V. Kotikov {\it et al.},
J. Stat. Mech. {\bf 0710} (2007) P10003;
Z. Bajnok,
R.A. Janik, and T. Lukowski,
Nucl. Phys. B {\bf 816} (2009) 376.



\bibitem{KoReZi}
  A.V. Kotikov {\it et al.},
 Nucl. Phys. B {\bf 813} (2009) 460;
 M. Beccaria {\it et al.},
Nucl. Phys. B {\bf 827} (2010) 565.


\bibitem{LuReVe} 
  T.~Lukowski {\it et al.},
  Nucl.\ Phys.\ B {\bf 831}, 105 (2010).

\bibitem{Marboe:2014sya}
  C.~Marboe {\it et al.},
  JHEP {\bf 1507} (2015) 084
  
\bibitem{Marboe:2016igj}
  C.~Marboe and V.~Velizhanin,
  JHEP {\bf 1611} (2016) 013

\bibitem{Staudacher:2004tk}
M.~Staudacher,
JHEP {\bf 0505} (2005) 054;
N. Beisert and M.~Staudacher,
Nucl. Phys. B {\bf 727} (2005) 1.

\bibitem{Beccaria:2007pb}
  M.~Beccaria,
  JHEP {\bf 0709} (2007) 023;
  M.~Beccaria {\it et al.},
  JHEP {\bf 0903} (2009) 129;
  V.~N.~Velizhanin,
  JHEP {\bf 1011} (2010) 129


\bibitem{DRED}  W.~Siegel, Phys.\ Lett.\ B \textbf{84} (1979) 193.

\bibitem{Maldacena:1997re}
  J.~M.~Maldacena,
  Int.\ J.\ Theor.\ Phys.\  {\bf 38} (1999) 1113
   [Adv.\ Theor.\ Math.\ Phys.\  {\bf 2} (1998) 231];
   S.~S.~Gubser {\it et al.},
  Phys.\ Lett.\ B {\bf 428} (1998) 105;
  E.~Witten,
  Adv.\ Theor.\ Math.\ Phys.\  {\bf 2} (1998) 253.

\bibitem{Kotikov:2006ts}
  A.~V.~Kotikov and L.~N.~Lipatov,
  Nucl.\ Phys.\ B {\bf 769} (2007) 217;
  M.~K.~Benna {\it et al.},
  Phys.\ Rev.\ Lett.\  {\bf 98} (2007) 131603.

  
\bibitem{Basso:2007wd}
  B.~Basso {\it et al.},
  Phys.\ Rev.\ Lett.\  {\bf 100} (2008) 091601;
  B.~Basso and G.~P.~Korchemsky,
  Nucl.\ Phys.\ B {\bf 807} (2009) 397.
  
  
\bibitem{Beisert:2006ez}
  N.~Beisert {\it et al.},
  J.\ Stat.\ Mech.\  {\bf 0701} (2007) P01021

\bibitem{Brower:2006ea}
  R.~C.~Brower {\it et al.},
  JHEP {\bf 0712} (2007) 005.

  
\bibitem{Costa:2012cb}
  M.~S.~Costa {\it et al.},
  JHEP {\bf 1212} (2012) 091;
  A.~V.~Kotikov and L.~N.~Lipatov,
  Nucl.\ Phys.\ B {\bf 874} (2013) 889;
  N.~Gromov {\it et al.},
  JHEP {\bf 1407} (2014) 156.






\bibitem{Blumlein:2009cf}
  J.~Blumlein, D.~J.~Broadhurst and J.~A.~M.~Vermaseren,
  Comput.\ Phys.\ Commun.\  {\bf 181} (2010) 582
  
 

\bibitem{Kotikov:2001ct}
  A.~V.~Kotikov {\it et al.},
  Eur.\ Phys.\ J.\ C {\bf 26} (2002) 51

  
\bibitem{Parisi:1978jv}
  G.~Parisi and N.~Sourlas,
  Nucl.\ Phys.\ B {\bf 151} (1979) 421.

\bibitem{Krivokhizhin:2009zz}
  V.~G.~Krivokhizhin and A.~V.~Kotikov,
  Phys.\ Part.\ Nucl.\  {\bf 40} (2009) 1059.

\bibitem{Kotikov:1998qt}
  A.~V.~Kotikov and G.~Parente,
  Nucl.\ Phys.\ B {\bf 549} (1999) 242;
   A.~Y.~Illarionov {\it et al.},
  Phys.\ Part.\ Nucl.\  {\bf 39} (2008) 307.

  

    

  
\bibitem{Kotikov:2007ua}
  A.~V.~Kotikov,
  Phys.\ Part.\ Nucl.\  {\bf 38} (2007) 1
   Erratum: [Phys.\ Part.\ Nucl.\  {\bf 38} (2007) 828].
  

   

\end{thebibliography}
\end{document}